\documentclass[12pt,preprint]{aastex}
\makeatletter
\renewcommand{\fps@figure}{thbp}
\renewcommand{\fps@table}{thbp}

\makeatother

\def\la{\mathrel{\hbox{\rlap{\hbox{\lower4pt\hbox{$\sim$}}}\hbox{$<$}}}}
\def\sun{\hbox{$\odot$}}

\renewcommand{\mag}{\mbox{$\;$mag}}

\begin{document}

\title{CEPHEID DISTANCES TO SNe\,Ia HOST GALAXIES BASED 
       ON A REVISED PHOTOMETRIC ZERO-POINT OF THE HST-WFPC2 AND NEW
       P-L RELATIONS AND METALLICITY CORRECTIONS} 
\author{A. Saha\altaffilmark{1}, F. Thim\altaffilmark{1,2}}
\affil{NOAO, P.O. Box 26732, Tucson, AZ 85726}
\email{saha@noao.edu, thim@noao.edu}
\author{G. A. Tammann, B. Reindl}
\affil{Astronomisches Institut der Universit\"at Basel,\\
       Venusstrasse 7, CH-4102 Binningen, Switzerland}
\email{G-A.Tammann@unibas.ch, reindl@astro.unibas.ch}
\and
\author{A. Sandage}
\affil{The Observatories of the Carnegie Institution of Washington,\\
       813 Santa Barbara Street, Pasadena, CA 91101}

\altaffiltext{1}{NOAO is operated by the Association of Universities
  for Research in Astronomy, Inc. (AURA) under cooperative agreement
  with the National Science Foundation}
\altaffiltext{2}{present address: {\tt thim@brandenburg-gmbh.de}:
  Brandenburg GmbH, Technologiepark 19, 33100 Paderborn, Germany}

\begin{abstract}
With this paper we continue the preparation for a forthcoming summary
report of our experiment with the Hubble Space Telescope (HST) to
determine the Hubble constant using type Ia supernovae as standard
candles. Two problems are addressed.
(1) We examine the need for, and determine the value of, the
corrections to the apparent magnitudes of our program Cepheids in the
eleven previous calibration papers due to sensitivity drifts and
charge transfer effects of the HST WFPC2 camera over the life time of
the experiment from 1992 to 2001.
(2) The corrected apparent magnitudes are applied to all our previous
photometric data from which revised distance moduli are calculated for
the eight program galaxies that are parents to the calibrator Ia
supernovae. Two different Cepheid P-L relations are used; one for the
Galaxy and one for the LMC. These differ both in slope and zero-point
at a fixed period.

     The procedures for determining the absorption and reddening
corrections for each Cepheid are discussed. Corrections
for the effects of metallicity differences between the program
galaxies and the two adopted P-L relations are derived and
applied. The distance moduli derived here for the eight supernovae
program galaxies, and for 29 others, average $0.20\mag$ fainter (more
distant) than those derived by Gibson et~al. and Freedman et~al. in
their 2000 and 2001 summary papers for reasons discussed in this paper. 
The effect on the Hubble constant
is the subject of our forthcoming summary paper.
\end{abstract}

\setcounter{footnote}{2}
\section{INTRODUCTION}
\label{sec:intro}
This is the fourth paper of a set 
of preparations for a final summary paper of our experiment to
obtain the Hubble constant using type Ia supernovae as standard
candles. 
The first \citep*[][hereafter Paper~I]{Tammann:etal:03} is a 
re-calibration of the Cepheid period-luminosity relation for Galactic
Cepheids. The second 
\citep*[][hereafter Paper~II]{Sandage:etal:04} 
is the same, but for Cepheids in the LMC together with a small
correction to the Galactic relations from \citeauthor{Tammann:etal:03}.
The third \citep*[][hereafter Paper~III]{Reindl:etal:05} is a
re-determination of the reddening and absorption corrections for a
complete sample of modern type Ia supernovae, leading to a revised
Hubble diagram that is corrected to the cosmic kinematic velocity
frame.

     In this paper we examine the need for, and
determine the values of, corrections to the apparent magnitudes
of Cepheids that are listed in the original papers of our HST
Cepheid discovery program over the period from 1992 to 2001:
\citet{Sandage:etal:92,Sandage:etal:94,Sandage:etal:96},
\citet{Saha:etal:94,Saha:etal:95,Saha:etal:96a,Saha:etal:96b,Saha:etal:97,Saha:etal:99,Saha:etal:01a,Saha:etal:01b}.
These revisions are based on re-calibrations of the photometric
properties of the Hubble Space Telescope (HST) over that interval.

     The organization of the paper is as follows. The known problem of
the charge transfer efficiency (CTE) of the CCD chips of the 
Wide Field Planetary Camera 2 (WFPC2) of HST, and the
discovery of the effect is set out in \S~\ref{sec:redetermination}. 
Also, we discuss here the
determination of non-linearities in the responses of the WFPC2 to
exposure time (the long and short exposure problem) and to the
background and crowding levels, as well as a determination of
zero-point differences of the WFPC2 photometry from those we had
adopted earlier, which were based on the 
\citet{Holtzman:etal:95a,Holtzman:etal:95b} zero-points. 
These differences are calculated over the time interval from 
1994 to 2000, to look for changes as the detectors have aged  
and degraded. All these effects are determined by the direct
comparison of HST WFPC2 data with ground based photometry of stars in
the globular cluster NGC\,2419. These comparisons lead to the adopted
corrections to the original photometry, which are presented in
Table~\ref{tab:final_delta_mags}. The final corrections to our
original Cepheid apparent magnitudes that were listed in the original
11 papers cited above are made from this table. The discussion in 
\S~\ref{sec:redetermination} is rather detailed, and we hope of general 
interest to any one doing photometry with the WFPC2.  However, the
reader who is uninterested in the specifics may wish to skip this
section.   

     Strictly speaking, our re-calibration method, which directly
compares the results of our WFPC2 photometry to a ground based
sequence of stars of similar brightness, should need no further 
validation.  However, there are several issues that can give rise to  
confusion, and we 
devote space to highlight our photometry procedure
for WFPC2 data and contrast it with those of others. 
We also discuss how some apparent discrepancies between our photometry
and  those reported by others are really inconsequential.

     The corrected magnitudes determined in \S~\ref{sec:newadopted}
are set out in Table~\ref{tab:cep:complete} for 
the eight program galaxies (IC\,4182, NGC\,5253, NGC\,4536, NGC\,4496A,
NGC\,4639, NGC\,3527, NGC\,4527, and NGC\,3982). They refer to a
variety of mean metallicities, galaxy-to-galaxy. 
If the known difference in the P-L relations
between the Galaxy and the LMC (Paper~I\,\&\,II) is due to mean
metallicity differences between these galaxies, then different Cepheid
period-luminosity relations must be applied to the individual program
galaxies.  
The P-L relations in $V$ and $I$ for the Galaxy and LMC from
\citeauthor{Sandage:etal:04} are repeated in
\S~\ref{sec:distance}. They are applied to the Cepheids of the eight
program galaxies to yield two sets of apparent and
absorption-corrected distance moduli $\mu^{0}$(Gal) and $\mu^{0}$(LMC)
listed in Table~\ref{tab:cep:complete}. The differences of the two
absorption-corrected moduli, $\mu^{0}$(Gal) and $\mu^{0}$(LMC), are
investigated further in \S~\ref{sec:metal} and found to be a strong
function of $\log P$. On the assumption that the differences are a
linear function of the metallicity, period-dependent metallicity
corrections to the distance moduli of the program galaxies are
derived and tested against external evidence. It is shown in 
\S~\ref{sec:analysis} that Cepheid distances depend in general on the 
periods of the Cepheids considered, particularly if the P-L relations in 
different pass-bands are used to solve for the distance and the reddening.

     The multiple evidence for the validity of the adopted
zero-point of the present distance scale is in
\S~\ref{sec:zeropoint}. The adopted distances are compared with the
results of several previous authors in \S~\ref{sec:comparison}.  

  The conclusions in \S~\ref{sec:conclusions} summarize the nine
principle research points made in the paper.

\section{THE ZERO-POINT RE-DETERMINATION}
\label{sec:redetermination}
%
\subsection{Problems with Charge Transfer Efficiencies}
\label{sec:redetermination:problem}
It has been known since the discovery by Stetson in 1994, as reported
by \citet{Kelson:etal:96}, that photometry of stars observed on HST
WFPC2 frames taken of the globular clusters Pal 4 and NGC\,2419
did not agree at the few percent level between frames taken with short
exposures and those taken with long exposures. In the course of study
of the reason for this exposure-time effect it was also discovered
that there is a non-linear dependence of the electron counts per pixel
on the incident light level, and that the effect is more prominent
when the position of the object is farther from the read-out amplifier
of the CCD. We now know that the charge transfer efficiencies of the
WFPC2 CCDs depend on the level of the background counts and on the net
counts from an object (\citealt*{Saha:etal:00} (SLP);
\citealt{Dolphin:02}; \citealt{Whitmore:Heyer:02}).  
Long exposures on faint targets where the ``blank'' sky levels are
high compared with short exposures on standards have quite different
background levels. 

     We were aware of the problem in 1994 and, with the limited
knowledge then available, we adopted the {\em ad hoc\/} procedure to
correct the initial seminal \citet{Holtzman:etal:95a}
and \citet{Holtzman:etal:95b} ``pipe-line'' zero-points that had been
determined on a short exposure basis to a long exposure basis by
adding $0.05\mag$ to the Holtzman zero-points (making the adopted
magnitudes of the faint targets fainter). We continued to use 
this correction procedure throughout our series of papers (even as
progressively better understanding of the problem began to emerge)
with the intention of eventually correcting everything retrospectively
once the ultimate understanding was in hand. One purpose of the
present paper is to affect this understanding and to correct our
initial photometry to our ``final'' system using the corrections
derived in this paper.\footnote{As detailed later, we were consistent
  in all but one of our original papers in reporting uncorrected mean
  $V$ and $I$ magnitudes for the Cepheids, and applying the
  $0.05\mag$ {\em ad hoc\/} correction only for the derived distance
  modulus. The exception was for NGC\,3982 where the referee insisted
  we break with our tradition. Thus the mean magnitudes of Cepheids in
  NGC\,3982 in \citet{Saha:etal:01b} are already on the long exposure
  basis. This is accordingly accounted for when making our correction
  in \S~5.} 

\subsubsection{The Background Problem}
\label{sec:redetermination:problem:back}
Studies of the role of varying background levels on photometry
systematics of the WFPC2 have been presented by various
authors; the most important for our purpose are 
\citet{Stetson:98}, \citet{Saha:etal:00}, \citet{Dolphin:00} and
\citet{Dolphin:02}. 
The conclusion of these studies is that when there is sufficient
background exposure (few hundred electrons per pixel), charge
transfers are essentially loss-less. At progressively lower exposure
levels, the lost charge increases, and there is more fractional loss
from fainter objects than from brighter ones. The discussions in these
papers show that the corrections needed are procedure dependent,
particularly in the details of aperture correction and what procedure 
is used to determine the background level. This is because the far
wings of the HST/WFPC2 PSFs contain a significant fraction of the
light, but have low S/N. Different procedures handle the S/N optimization
vs flux normalization problem differently, thus placing different relative 
weights on the contribution from the far wings.
To complicate matters, the {\it fractional\/} loss of charge due to
CTE problems differs from near the core of the PSF than from the far
wings: they are larger for the latter.
Thus the prescription to correct for CTE effects that is derived for 
one procedure of extracting photometry from
WFPC2 images may not be applicable to a different procedure. 
This point is key to the discussion in  
\S~\ref{sec:redetermination:Discussion:Comparison}, 
when comparing the results of CTE corrections from different authors.
We had made the fortuitous decision to follow {\em exactly\/} the same
prescription for DoPHOT based WFPC2 photometry, as described in
\citet{Saha:etal:96a} throughout the series of papers. Also, the 
\citet{Saha:etal:00} empirical derivation of charge transfer
corrections using exposures of various durations of NGC\,2419 was done
with the exact same procedure. 

     The \citet{Saha:etal:00} study provides a prescription for
converting the measured instrumental magnitudes from exposures with a
small background level, to the value that would have been measured in
the event that charge transfer were absent, i.e.\ as in an image with
sufficiently high exposure of the background to avoid the CTE problem.  
Hence, we can directly use the results from the
\citeauthor{Saha:etal:00} study to correct for the effects of the
charge transfer problem in all of our previous papers on the
Cepheid-supernovae program where the problem exists.

\subsubsection{A Summary of Our Photometry Procedure}
\label{sec:redetermination:summary:photometry}
To place the specific revisions presented here in the context of other
efforts to obtain the best possible photometric calibration of WFPC2,
one must first understand how our methodology differs with
others. There are two aspects to this: 
1) procedural differences in measuring the brightness of a stellar
object, i.e. how the details of the DoPHOT based photometry as
implemented by us throughout, and 
2) the use of the recently calibrated standard stars to establish the
WFPC2 photometric zero-points that reduce the `lever arm' of
corrections necessary to account for non-linearity and CTE anomalies
in WFPC2. The new standard sequence is known to differ systematically
from some sequences in use in the past, as noted in
\citet*{Saha:etal:05} (SDTW). We elaborate on each of the above
issues. 

     The DoPHOT based photometry procedure is fully described in 
\citet{Saha:etal:96a}. It is summarized as follows. 
The PSF incident on a WFPC2 CCD has a sharp core with flared wings. 
The shape of the central core changes slowly with position in the
field of view (FOV), whereas beyond a few times 0.1 arc-seconds, the
extended low level wings are due to scattering from micro-roughness in
the mirrors, and do not change with position. Our procedure asserts
that the incident PSF beyond a radius of 0.5 arcsec (5 pixels on the
three WF chips) is invariant with position on the FOV.  
This assertion has been verified on well exposed (high sky background)
images with many well exposed stars (such as on a field of the Leo~I
dwarf galaxy) by comparing the light within the above radii (using
aperture photometry) to the light within a 1 arc-second radius
aperture, and further noting that the difference is not a function of
position on the FOV. 
DoPHOT itself is tuned to measure the peak height of an analytic
PSF by fitting to the pixels in a $ 9 \times 9 $ pixel box centered  
on the star being measured. The resulting instrumental magnitude 
is first corrected for PSF variation with FOV by using a procedure
that is fully described in \citet{Saha:etal:96a}. The resulting
instrumental magnitude is then normalized (via an aperture correction)
to the value that would be measured in an aperture of 0.5 arc-second
radius. This is the same aperture size as in  
\citep{Holtzman:etal:95a,Holtzman:etal:95b}, and thus their
photometric calibration and color equations were directly applied. 
However, while our Cepheid data were deep exposures with high enough
background to eliminate the non-linear effects of poor CTE, the
Holtzman calibration itself was based on shallow exposures that were
beset with CTE problems. 
As an {\it ad hoc\/} correction to account for this, we added 0.05 mag
to our final Cepheid magnitudes, and waited for a better understanding
and characterization of the corrections. 
 
     In many applications where well exposed isolated stars are not
plentifully distributed over the entire FOV, determining the aperture
correction from DoPHOT fitted values to 0.5 arcsec radius aperture
magnitudes is noisy. Since \citet{Saha:etal:96a} it was  
realized that both the zero-point and the position dependent variations
in conditions where the sky and stars are well exposed are stable: that 
applying fixed values determined from suitable data (high sky
background where CTE effects are minimal, and well exposed bright
stars distributed well over the field) agree over a range of epochs
with rms magnitude errors less than 0.02 mag. Applying such a fixed
correction is thus less error prone than determining even just the
constant term in aperture corrections from non-optimal data.

     The DoPHOT procedure is thus tantamount to measuring a fitted
magnitude and correcting it to the aperture equivalent of Holtzman's
``flight system''. Since the correction was calculated from images
with sky exposures high enough to mitigate CTE problems, they apply to
objects whose wings out to 5 pixels are unaffected by CTE problems.  

     Let us contrast this with the case where direct aperture
measurements are made. When the background is not high enough to
mitigate the CTE effects, the CTE anomaly will begin to modify the
wings of the observed PSF. Specifically, fainter stars will be
affected more and brighter stars less so, even at the same sky
background. Hence, {\it for direct aperture measurements in
insufficient background levels, the measured magnitude can be expected
to be a function of the brightness of the object, with stronger
dependence as the aperture size is increased}.  
Indeed, this dependence is seen in the aperture based studies, such as 
\citet{Dolphin:00} and \citet{Dolphin:02}.  
This implies that {\it the CTE corrections are procedure dependent}.  
Corrections appropriate for one photometry prescription are not
generally transferable to another. 

     It is extremely important to understand that due to the way our
photometry procedure is constructed (as outlined above), the
dependence of the CTE correction on brightness of the object is 
{\it not\/} explicitly seen for our procedure, as demonstrated in
\citeauthor{Saha:etal:00}. The reason is two-fold: 
\begin{enumerate}
\item 
The DoPHOT procedure fits a profile which is heavily weighted by the
core, and the fitted ``sky'' is determined with respect to a profile's
own wings rather than to pixel values in the far wings. Further, the
aperture correction is derived for the case where CTE problems are
mitigated, and applied even to data that do not have sufficient
background levels.  
So unless CTE problems eat into the wings within the $9 \times 9$
pixel fit box, which happens only in very faint objects, the effect is
muted from what one would measure directly with a 0.5 arcsec radius
aperture. 

\item
In \citeauthor{Saha:etal:00}, a CTE correction is made not with
respect to the real background, but to the background reported by the
PSF fit. The analytic PSF used does not pretend to trace the far
wings, but goes to zero within 1 arcsec radius.  Thus the fitted
`background' is really the star's own wings (effectively somewhere
between 0.5 and 1.0 arcsec radius).  
In practice, brighter stars have brighter fitted backgrounds, and so 
smaller CTE corrections per the correction equations in
\citeauthor{Saha:etal:00}. Fortuitously, the net result is to cancel
any brightness dependence, as demonstrated empirically from the
analysis of photometry in \citeauthor{Saha:etal:00}. 
Actually it is not just a happy accident that this
is so: way before we understood what CTE problems were doing, we had
noticed non-linearities when comparing the same stars observed in
exposures of differing durations. These non-linearities were 
{\it empirically\/} removed when the analytic PSF was given the form
that we used, and when the fitting box in DoPHOT was set to $9 \times
9$ or smaller.  
\end{enumerate}

     In our prior series of Cepheid papers, we equated the 0.5 arcsec
radius aperture magnitudes from our DoPHOT based procedure described
above, to Holtzman's magnitudes on his ``flight system'', and then
used his color equations to reach $V$ and $I$. 
However, recall that Holtzman's measurements of the relatively bright
stars with WFPC2 used short exposures with very low background levels,
and bore the full brunt of the CTE problem. Holtzman used a ramp to
make a correction for position dependence of the CTE, but could not
correct for stellar brightness or background value using the knowledge
then available. 
His derived zero-points, and perhaps even his color dependencies must
therefore be re-evaluated.

     Given this context, the \citeauthor{Saha:etal:00} paper
established a prescription to correct for CTE anomalies at a
particular epoch (1994) in the life of WFPC2, that is 
{\it specific to our reduction procedure}.   
It left open two questions:
\begin{enumerate}
\item
How well does this prescription work for other epochs, i.e.\ as further 
degradation of CTE happened with time, and
\item
How must the zero-point calibration be revised, to account for the CTE
issues with the data used by Holtzman. 
\end{enumerate}

     This paper addresses both these questions. 
First by testing the \citeauthor{Saha:etal:00} prescription at
different epochs and looking for variations, and second, by comparing
archived WFPC2 observations of photometric sequences newly established
from the ground.  

     Note that the conclusions here can only be applied to our
specific photometry procedure, since, as discussed above, corrections
due to CTE problems are procedure specific. Other studies for CTE
corrections of other procedures have been done, notably by
\citet{Dolphin:00}, \citet{Dolphin:02} and \citet{Heyer:etal:04}. 

     The tests presented in this paper, along with the analysis in
\citeauthor{Saha:etal:00} and the ground based photometric sequences
presented in \citeauthor{Saha:etal:05} form a closed system: they are
fully self-consistent, and in principle need no further reference. 
However, much has been said and written about WFPC2 calibrations, and
our referee has pointed out several points of confusion. Disagreement
with other calibrations could (mistakenly) suggest that there is some
defect or mistake in the execution of our method.  
These points of confusion are discussed and clarified later, 
in \S~\ref{sec:redetermination:Discussion:Comparison}.

\subsubsection{The Test for a Variation of the WFPC2 Sensitivities
  over the Six Year Duration of our HST Observations}
\label{sec:redetermination:problem:var}
As mentioned above, two remaining issues must be accounted for,
both related to the CTE problem: 
(1) A number of studies by others have shown that the charge
transfer problem in the HST WFPC2 camera has worsened over
time, probably due to the progressive damage due to the high levels of
incident cosmic ray flux. Because the Cepheid data were all
obtained on deep HST frames where the CTE effects are very small, it 
is expected that any corrections necessary for these did not evolve
over the six year duration of our observations, even though the
detectors have deteriorated. 
(2) However, this is not the case for the conversion of standard star
observations taken with short exposures (and insignificant background
levels in such exposures) to what would have been if the exposure
times were long. Hence, the zero-point calibration is expected to be a
function of time (on the time scale of a year, because change is slow)
when the observations for the standards were made, because CTE
corrections are significant for these standard star observations. 

     We have studied both problems by repeatedly 
observing (with WFPC2) a field of stars over time, 
in which we now 
have a ground calibrated sequence of stars.  A
standard-star sequence in the globular cluster NGC\,2419 was observed
with HST over the six years of our Cepheid observations from 
1994 to 2000 (\citeauthor{Saha:etal:05}). 

We have compared the WFPC2 instrumental magnitudes on deep
exposures (or for short exposures, corrected for CTE losses as
prescribed by \citeauthor{Saha:etal:00}) with the ground-based values
for the standard stars in NGC\,2419.  
The data over the six year period were used to revise the Holtzman
zero-point so that they give the CTE-corrected instrumental magnitudes
over the six year period. We describe the results in the remainder of
this section. 

     Magnitudes on the Landolt \citep{Landolt:92} $B,V,R$, and $I$ system 
(Johnson for $B$ and $V$; Cousins for $R$ and $I$) of a faint sequence of
stars in NGC\,2419 were measured down to faint levels (to $V\sim23$)
with the WIYN $3.5\;$m telescope at several epochs from 2001 and 2003
(\citeauthor{Saha:etal:05}). This calibration was specifically
designed to enable the retro-active calibration of HST WFPC2
data, by targeting fields that have been observed repeatedly by 
WFPC2 over its lifetime.

     Armed with this ``faint star ground truth'', we have compared
magnitudes of the same stars measured on HST WFPC2 archival
frames of NGC\,2419 made in 1994, 1997, and 2000 with the ground based
magnitudes. It turns out that the WFPC2 observations were 
fortuitously made with different pointings by centering the cluster
in different chips, hence the zero-point corrections could not only be
made for each of the four chips (the standard-star field was large
enough to cover the four-chip area), but, because of the pointing 
differences, any systematic zero-point difference could be studied as
a function of crowding. 

      The frames of the individual WFPC2 images of NGC\,2419 are listed
in Tables~\ref{tab:ZP_diff_V} and \ref{tab:ZP_diff_I}, along with
their archival dataset names and their individual exposure times. 
The exposure times range between 20 and 1400 s in $V$ and 10 and 1300
s in $I$. The observations in year 2000 were made at two distinct
epochs, which have been distinguished as 2000a and 2000b. 

     Object identification and photometry of the WFPC2
observations was done with the DoPHOT program
\citep{Schechter:Mateo:Saha:93} as modified for use with HST/WFPC2
\citep{Saha:etal:96a}.  
DoPHOT was run with the same tuning of parameters as in
\citet{Saha:etal:00}. The exact same settings were also used in all
our Cepheid discovery and photometry papers. 
The individual WFPC2 $V$ and $I$ magnitudes for the NGC\,2419
comparisons have been corrected for CTE losses by using equations (24)
to (27) given in \citet{Saha:etal:00}. 
The magnitude correction, $\Delta m$, is a function of background level 
in electrons ($B$), and the ordinate $y$ in pixels.\footnote{Other 
  studies have required additional terms for defining the correction,
  in particular, the instrumental magnitude (or ``counts'') of the
  objects themselves. However, the DoPHOT procedure described in
  \citet{Saha:etal:96a} measures ``background'' in a way that 
  fortuitously cancels the dependence on incident brightness of an
  object, as is definitively demonstrated in the \citet{Saha:etal:00}
  study.}  

     The $y$ axis is the direction along which charge is read out
through the parallel registers of the CCD. As the charge from any
pixel travels from ``top to bottom'' during readout, some fraction of
the charge is lost in traps. Charge from a pixel near the ``bottom''
travels only a short distance through the parallel registers, and so
loses a smaller fraction than charge from a pixel near the ``top'',
which has more opportunities to lose charge. This is the reason for
the $y$ dependence. The lost charge also depends in a non-linear way
on the charge being carried, and on whether the traps have already
been filled by charge from preceding pixels. Semi-empirical models to
reconstruct the loss-free image have not been successful however, and
one must thus resort to purely empirical methods, such as that 
demonstrated by \citet{Saha:etal:00}.

     When the background ``sky'' levels are high enough, the lost
charge comes essentially from the background. Also the traps are
probably quickly filled by the first few pixels worth of charge that
traverse it. The overall observed effect is that as the background
increases, the amplitude of the correction to the photometry required
becomes progressively smaller, until for high enough backgrounds it is
no longer noticeable. For exposures with the {\em F555W\/} 
and {\em F814W\/} filters that exceed 1000s in duration, the
accumulated blank sky exposure is large enough to mitigate the effects
from the bad CTE. Thus the exposures of the galaxies for finding and
measuring Cepheids, which are all longer than 1000s, are not affected
by the CTE problem to any measurable degree. This conclusion too, was
demonstrated in \citet{Saha:etal:00}. 
However, the standard star photometry is affected, and the standard
star measurements must be corrected for both $y$ and background level
dependent CTE effects in order to be on the same footing as the target
Cepheid exposures. The overall correction is thus an offset to the
magnitudes of the Cepheids reported earlier in the papers on their 
discovery. However the calculation of the appropriate offset requires
the application of the detailed CTE corrections worked out in 
\citet{Saha:etal:00} to the observations of the standard stars. 
The standards used here are stars in NGC\,2419: 
observations of NGC\,2419 have been made with a range of exposure
times and even pre-flashes, to span a range of background levels.

\begin{figure}[h]
   \epsscale{0.75}
   \plotone{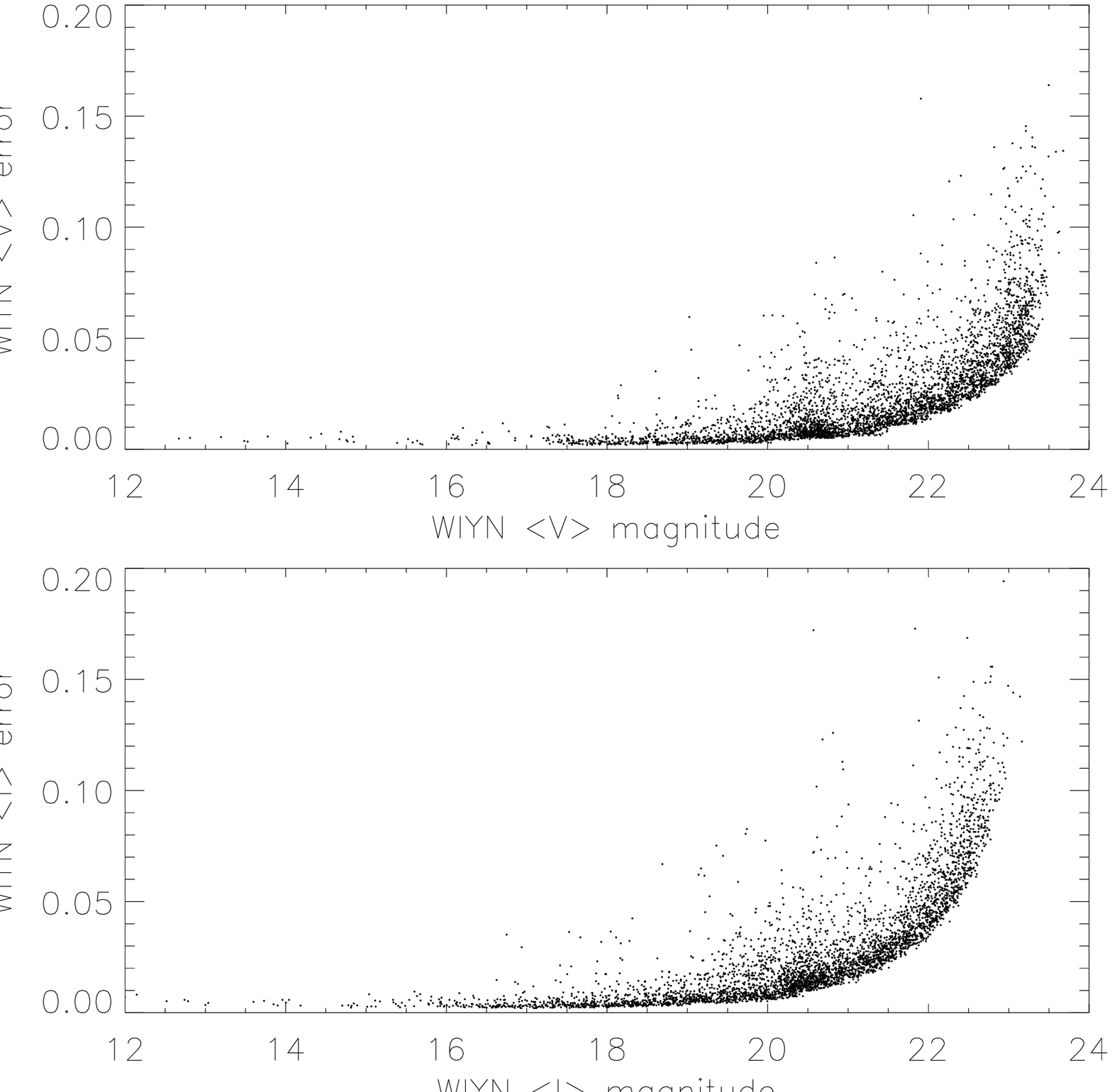}
   \caption{The error in the mean (rms/$\sqrt{N}$) from all of the
   various WIYN exposures on the various nights are plotted against
   the respective $\langle V\rangle$ and $\langle I\rangle$ mean
   magnitudes. These define the fidelity of the ground based standard
   sequence.}  
\label{fig:V_I_error}
\end{figure}

     In Figure~\ref{fig:V_I_error}, we plot the error in the mean which
are calculated from the rms variation in the magnitudes from all of
the various WIYN exposures on the various nights against the
respective $\langle V\rangle$ and $\langle I\rangle$ mean
magnitudes. These plots define the random error estimates for a single 
star for the ground based photometric sequence, as a function of 
brightness (over and above any {\it systematic\/} errors (which are 
given in Table~2 of \citeauthor{Saha:etal:05}, and for $V$ and $I$ are
smaller than 0.01 mag). 
We shall see that these uncertainties are much smaller than the
measuring errors on WFPC2 data.

\begin{figure}[h]
   \epsscale{0.60}
   \plotone{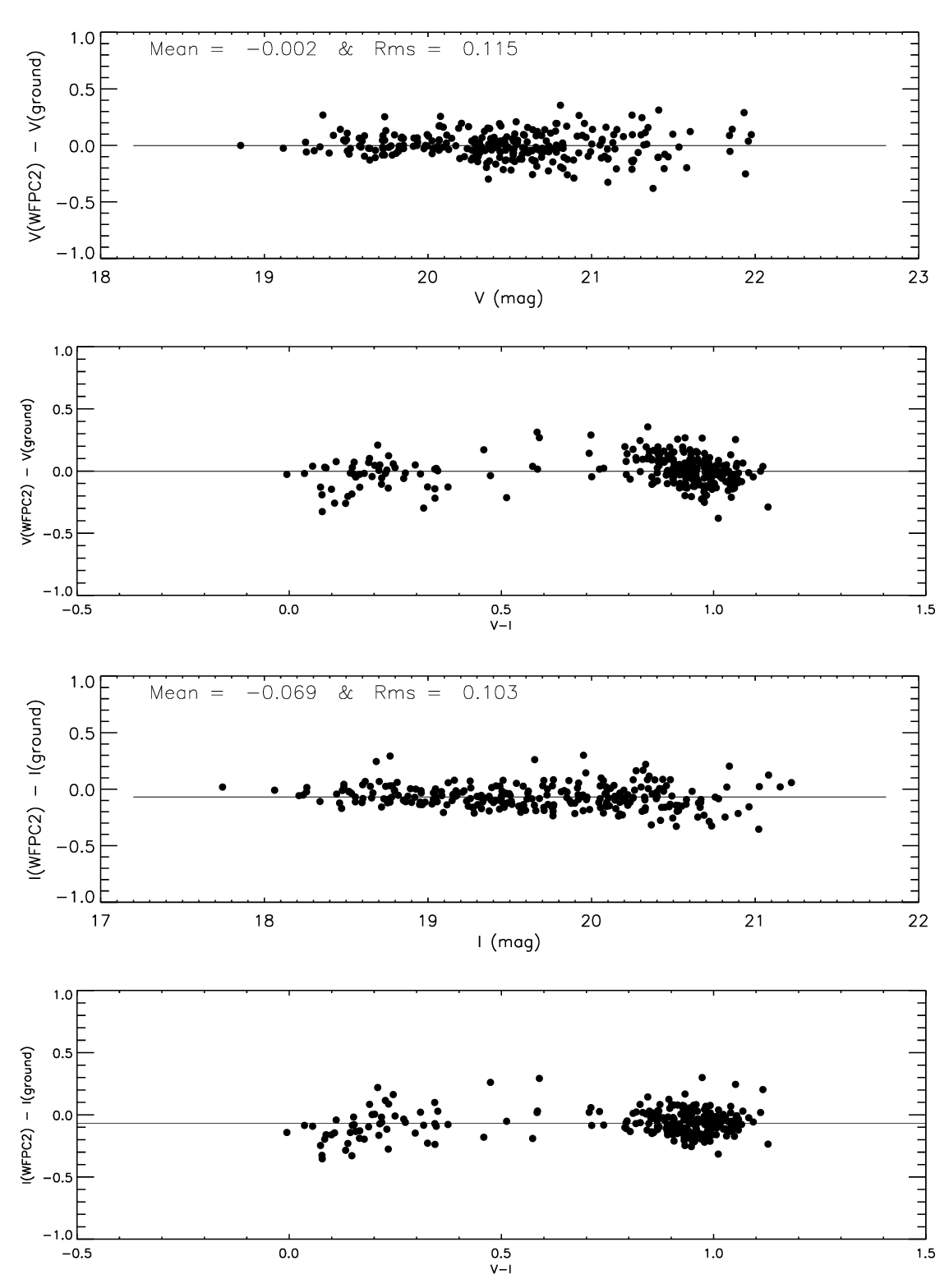}
   \caption{The comparison of stars on WF3 in one particular epoch,
   measured with our DoPHOT based procedure, corrected for CTE effects
   according to the prescription in \citeauthor{Saha:etal:00}, and
   calibrated with the Holtzman color equations and zero-points.  
   The abscissae show magnitudes and colors of the stars from the
   ground based sequence. This figure shows how faint the comparison
   stars are, the absence of any obvious non-linearity on brightness,
   and on the colors of the stars.}   
\label{fig:Comp_V_I}
\end{figure}

     The number of stars on the individual WFPC2 chips that
can be matched to stars in the ground based sequence (and therefore the
accuracy with which the mean corrections, mindful of the individual
accuracies from Figure~\ref{fig:V_I_error}) can be determined, are a
function of exposure time and the pointing with respect to the center
of NGC\,2419. The numbers range between a few to several hundreds
in each filter per chip. 
In Figure~\ref{fig:Comp_V_I}, we show such a comparison of stars in one
chip (WF3) with CTE corrections applied according to the
\citeauthor{Saha:etal:00} prescription and with zero-points from
\citet{Holtzman:etal:95b}. The actual rms is about 0.1 mag, nearly a
factor of 2 higher than the uncertainty from the combined error
estimates from the WIYN based sequence and the measuring errors on the
WFPC2 frame (one must allow for a minimum of 0.03 mag measuring error
due to the acute under-sampling of the stellar PSF). We are uncertain about 
the source of the extra scatter, but the figure demonstrates that the 
scatter is not from any obvious systematic effect with brightness or color. 

The number of stars that could be matched
between the individual WFPC2 epochs and the ground sequence is
given in the Tables~\ref{tab:ZP_diff_V} and \ref{tab:ZP_diff_I} for
the $V$ and $I$ frames respectively. 
The zero-point differences in the 
sense WIYN minus WFPC2 magnitudes for objects with Observed
deviations $|\Delta m| < 0.4 $ are also given in the
Tables~\ref{tab:ZP_diff_V} and \ref{tab:ZP_diff_I}. The rms scatter
and the resulting uncertainty in the means (rms/$\sqrt{N}$), are also
in the tables. 

\begin{deluxetable}{ccrcrcrcrcrcrcrcrc}
\rotate
\tabletypesize{\scriptsize}
\tablecaption{$V$ Zero-Point differences WIYN $-$ WFPC2}
\tablewidth{0pt}
\tablehead{
\colhead{Epoch} & \colhead{Exp.time [s]} &
\colhead{PC}    & \colhead{rms} & \colhead{N} & \colhead{err$^{\rm\,a}$} & 
\colhead{WF2}   & \colhead{rms} & \colhead{N} & \colhead{err$^{\rm\,a}$} & 
\colhead{WF3}   & \colhead{rms} & \colhead{N} & \colhead{err$^{\rm\,a}$} & 
\colhead{WF4}   & \colhead{rms} & \colhead{N} & \colhead{err$^{\rm\,a}$} }
\startdata
\tableline
\noalign{\smallskip}
\multicolumn{18}{c}{1994} \\
\noalign{\smallskip}
\tableline
u2dj0a01p & 1400 &$-0.08$ &\nodata & 1  &\nodata &$-0.02$ &  0.08  & 2   &  0.06  &  0.00  & 0.12 & 125 & 0.01 &  0.05  & 0.13 & 302 & 0.01 \\
u2dj0a02p & 1400 &$-0.08$ &  0.04  & 3  &  0.02  &\nodata &\nodata & 0   &\nodata &$-0.01$ & 0.11 & 123 & 0.01 &  0.03  & 0.14 & 309 & 0.01 \\
u2dj0a03p & 1400 &$-0.07$ &  0.07  & 3  &  0.04  &$-0.02$ &  0.17  & 3   &  0.10  &$-0.01$ & 0.12 & 135 & 0.01 &  0.03  & 0.15 & 341 & 0.01 \\
u2dj0a04p & 1400 &$-0.06$ &  0.05  & 3  &  0.03  &$-0.11$ &  0.11  & 3   &  0.06  &  0.02  & 0.13 & 125 & 0.01 &  0.03  & 0.14 & 348 & 0.01 \\
u2dj0a05t & 1400 &$-0.08$ &  0.12  & 56 &  0.02  &$ 0.00$ &  0.10  & 100 &  0.01  &$-0.01$ & 0.10 & 124 & 0.01 &  0.05  & 0.14 & 301 & 0.01 \\
u2dj0a06t & 1400 &$-0.03$ &  0.13  & 43 &  0.02  &$-0.01$ &  0.09  & 106 &  0.01  &  0.00  & 0.10 & 122 & 0.01 &  0.04  & 0.13 & 306 & 0.01 \\
u2dj0a07t & 1400 &$-0.06$ &  0.13  & 57 &  0.02  &$-0.02$ &  0.10  & 102 &  0.01  &$-0.01$ & 0.11 & 121 & 0.01 &  0.03  & 0.15 & 277 & 0.01 \\
u2dj0a08p & 1400 &$-0.05$ &  0.12  & 57 &  0.02  &$ 0.01$ &  0.11  & 103 &  0.01  &  0.00  & 0.10 & 120 & 0.01 &  0.03  & 0.14 & 291 & 0.01 \\
\tableline
\noalign{\smallskip}
\multicolumn{18}{c}{1997} \\
\noalign{\smallskip}
\tableline
u4ct0207r & 300  &$-0.28$ &  0.12  & 2  &  0.08  &  0.01  &  0.13  & 320 &  0.01  &  0.01  & 0.13 & 226 & 0.01 &  0.00  & 0.11 & 147 & 0.01 \\ 
u4ct0206r & 300  &$-0.08$ &  0.09  & 5  &  0.04  &  0.02  &  0.13  & 333 &  0.01  &  0.02  & 0.12 & 222 & 0.01 &  0.02  & 0.10 & 150 & 0.01 \\
u4ct0208r & 300  &$-0.15$ &  0.33  & 2  &  0.23  &  0.01  &  0.13  & 407 &  0.01  &  0.02  & 0.12 & 223 & 0.01 &  0.03  & 0.11 & 148 & 0.01 \\
u4ct0202r & 40   &$-0.20$ &  0.26  & 2  &  0.18  &  0.02  &  0.11  & 489 &  0.01  &  0.01  & 0.13 & 245 & 0.01 &  0.00  & 0.25 & 10  & 0.08 \\
u4ct0205r & 40   &  0.02  &  0.11  & 2  &  0.08  &  0.01  &  0.14  & 501 &  0.01  &  0.01  & 0.14 & 173 & 0.01 &  0.06  & 0.11 & 65  & 0.01 \\
u4ct0204r & 40   &  0.02  &\nodata & 1  &\nodata &  0.00  &  0.13  & 547 &  0.01  &  0.02  & 0.14 & 276 & 0.01 &  0.01  & 0.14 & 152 & 0.01 \\
\tableline
\noalign{\smallskip}
\multicolumn{18}{c}{2000a} \\
\noalign{\smallskip}
\tableline
u6ah0304r & 400  &$-0.06$ &  0.10  & 28 &  0.02  &$-0.01$ &  0.11  & 213 &  0.01  &  0.07  & 0.03 & 2   & 0.02 &$-0.04$ & 0.09 & 31  & 0.02 \\
u6ah0305r & 100  &$-0.09$ &  0.08  & 15 &  0.02  &  0.00  &  0.11  & 187 &  0.01  &  0.02  & 0.01 & 2   & 0.01 &  0.00  & 0.03 & 15  & 0.01 \\
u6ah0306r &  20  &$-0.17$ &  0.14  &  5 &  0.06  &  0.02  &  0.03  & 69  &  0.01  &$-0.04$ & 0.07 & 2   & 0.05 &$-0.05$ & 0.07 & 6   & 0.03 \\
\tableline
\noalign{\smallskip}
\multicolumn{18}{c}{2000b} \\
\noalign{\smallskip}
\tableline
u6ah030ar & 400  &  0.01  &  0.06  & 3  &  0.03  &  0.01  &  0.12  & 356 &  0.01  &$-0.01$ & 0.11 & 228 & 0.01 &$-0.03$ & 0.11 & 223 & 0.01 \\
\enddata
\tablenotetext{a}{err is the error in the mean (rms/$\sqrt N$). The
  lower limit for the error in the mean is set to 0.01.}
\label{tab:ZP_diff_V}
\end{deluxetable}

\begin{deluxetable}{ccrcrcrcrcrcrcrcrc}
\rotate
\tabletypesize{\scriptsize}
\tablecaption{$I$ Zero-Point differences WIYN $-$ WFPC2}
\tablewidth{0pt}
\tablehead{
\colhead{Epoch} & \colhead{Exp.time [s]} &
\colhead{PC}    & \colhead{rms} & \colhead{N} & \colhead{err$^{\rm\,a}$} & 
\colhead{WF2}   & \colhead{rms} & \colhead{N} & \colhead{err$^{\rm\,a}$} & 
\colhead{WF3}   & \colhead{rms} & \colhead{N} & \colhead{err$^{\rm\,a}$} & 
\colhead{WF4}   & \colhead{rms} & \colhead{N} & \colhead{err$^{\rm\,a}$} }
\startdata
\tableline
\noalign{\smallskip}
\multicolumn{18}{c}{1994} \\
\noalign{\smallskip}
\tableline
u2dj0b01p & 1300 &  0.02  &\nodata & 1  &\nodata &$-0.03$ &  0.16  & 2   &  0.11  &  0.06  & 0.10 & 124 & 0.01 &  0.08  & 0.12 & 305 & 0.01 \\
u2dj0b02p & 1300 &  0.03  &  0.22  & 5  &  0.10  &\nodata &\nodata & 0   &\nodata &  0.07  & 0.10 & 125 & 0.01 &  0.07  & 0.13 & 310 & 0.01 \\
u2dj0b03p & 1300 &$-0.04$ &  0.24  & 6  &  0.10  &$-0.02$ &  0.06  & 2   &  0.04  &  0.06  & 0.10 & 141 & 0.01 &  0.06  & 0.14 & 358 & 0.01 \\
u2dj0b04p & 1300 &  0.03  &  0.25  & 5  &  0.11  &$-0.01$ &  0.00  & 2   &  0.01  &  0.07  & 0.11 & 134 & 0.01 &  0.06  & 0.14 & 368 & 0.01 \\
u2dj0b05p & 1300 &  0.01  &  0.11  & 57 &  0.01  &  0.03  &  0.10  & 102 &  0.01  &  0.06  & 0.12 & 126 & 0.01 &  0.07  & 0.14 & 306 & 0.01 \\
u2dj0b06t & 1300 &$-0.01$ &  0.11  & 61 &  0.01  &  0.04  &  0.09  & 106 &  0.01  &  0.06  & 0.10 & 121 & 0.01 &  0.06  & 0.13 & 308 & 0.01 \\
u2dj0b07t & 1300 &  0.01  &  0.10  & 56 &  0.01  &  0.02  &  0.10  & 115 &  0.01  &  0.06  & 0.10 & 139 & 0.01 &  0.07  & 0.13 & 309 & 0.01 \\
u2dj0b08t & 1300 &  0.02  &  0.10  & 56 &  0.01  &  0.03  &  0.09  & 105 &  0.01  &  0.07  & 0.08 & 120 & 0.01 &  0.07  & 0.13 & 295 & 0.01 \\
\tableline
\noalign{\smallskip}
\multicolumn{18}{c}{1997} \\
\noalign{\smallskip}
\tableline
u4ct010or & 1000 &$-0.02$ &  0.07  & 6  &  0.03  &  0.03  &  0.12  & 309 &  0.01  &  0.07  & 0.10 & 229 & 0.01 &  0.07  & 0.08 & 146 & 0.01 \\
u4ct010pr & 1000 &$-0.03$ &  0.08  & 9  &  0.03  &  0.03  &  0.12  & 332 &  0.01  &  0.06  & 0.11 & 226 & 0.01 &  0.05  & 0.09 & 153 & 0.01 \\
u4ct010lr & 300  &$-0.07$ &  0.07  & 6  &  0.03  &  0.01  &  0.12  & 407 &  0.01  &  0.07  & 0.11 & 229 & 0.01 &  0.07  & 0.10 & 153 & 0.01 \\
u4ct0107r & 40   &$-0.15$ &\nodata & 1  &\nodata &  0.04  &  0.12  & 497 &  0.01  &  0.10  & 0.11 & 245 & 0.01 &  0.06  & 0.13 & 7   & 0.05 \\
u4ct0101r & 10   &$-0.03$ &  0.37  & 2  &  0.26  &  0.03  &  0.15  & 516 &  0.01  &  0.09  & 0.15 & 169 & 0.01 &  0.12  & 0.12 & 66  & 0.01 \\
u4ct0108r & 40   &$-0.06$ &  0.07  & 3  &  0.04  &  0.03  &  0.13  & 572 &  0.01  &  0.09  & 0.13 & 282 & 0.01 &  0.08  & 0.12 & 150 & 0.01 \\
\tableline
\noalign{\smallskip}
\multicolumn{18}{c}{2000a} \\
\noalign{\smallskip}
\tableline
u6ah0301r & 400  &$-0.03$ &  0.09  & 28 &  0.02  &  0.04  &  0.11  & 212 &  0.01  &  0.11  & 0.08 & 2   & 0.06 &  0.05  & 0.11 & 30  & 0.02 \\
u6ah0302r & 100  &$-0.02$ &  0.07  & 15 &  0.02  &  0.05  &  0.10  & 187 &  0.01  &$-0.02$ & 0.17 & 3   & 0.10 &  0.06  & 0.06 & 15  & 0.02 \\
u6ah0303r &  20  &$-0.08$ &  0.10  & 6  &  0.04  &$-0.02$ &  0.09  & 70  &  0.01  &$-0.05$ & 0.19 & 3   & 0.11 &  0.06  & 0.06 & 6   & 0.02 \\
\tableline
\noalign{\smallskip}
\multicolumn{18}{c}{2000b} \\
\noalign{\smallskip}
\tableline
u6ah0307r & 400  &  0.04  &  0.11  & 3  &  0.06  &  0.04  &  0.12  & 355 &  0.01  &  0.06  & 0.11 & 229 & 0.01 &  0.06  & 0.11 & 225 & 0.01 \\
\enddata
\tablenotetext{a}{err is the error in the mean (rms/$\sqrt N$). The
  lower limit for the error in the mean is set to 0.01.}
\label{tab:ZP_diff_I}
\end{deluxetable}

\clearpage

%
\subsubsection{Discussion of Corrections and Comparison with CTE 
               corrections by Others}
\label{sec:redetermination:Discussion:Comparison}
At first glance, the zero-point differences in
Tables~\ref{tab:ZP_diff_V} and \ref{tab:ZP_diff_I} appear very large.  
However, the \citet{Holtzman:etal:95b} zero-points themselves are
based on short exposure data with shallow background, and require
correction for CTE effects. A large part of the differences are thus
just that such a correction would be to make the WFPC2 measurements 
fainter, i.e.\ subtract a few hundredths of a magnitude from the
values in these tables (akin to the 0.05 mag {\it ad hoc} correction
we were using earlier). By comparing directly to ground based
magnitudes, we circumvent the problem of CTE correcting Holtzman's
photometry. 

     Unfortunately there is room for confusion when reading our
results in the context of other material in the literature. In
particular, in a discussion of the accuracy of WFPC2 photometric
zero-points, \citet{Heyer:etal:04} show the comparison of zero-points
from five different sources:  
from \citet{Holtzman:etal:95b}, \citet{Dolphin:00}, \citet{Dolphin:02} 
and from WFPC2 instrument handbooks in '95 and '02. 
The \citet{Dolphin:00} and \citet{Dolphin:02} 
zero-points are based on measurements that are fully and 
self-consistently corrected for CTE correction with a prescription
suitable for his data reduction scheme. 
It is not made clear in the report whether and by what means the
remaining zero-points were adjusted for the latest CTE correction
schema. It is inconceivable that they did not make some adjustments
for CTE, else, going by the size of corrections in \citep{Dolphin:02},
the scatter in their data from one method to another should have been
much larger.   
Note that of the five sets, the \citet{Dolphin:02} yields the
faintest zero-points. This data set makes self-consistent CTE
corrections that are fully documented. 
Qualitatively at least, \citet{Dolphin:02} differs from the other
studies in the same sense as our corrections to the zero-point in this
paper. Since the details of any CTE corrections to the Holtzman 
zero-points are not made in \citet{Heyer:etal:04}, we are unable to
make any quantitative comparison of our zero-points with the five
others shown. 

     In a second part to their study, \citet{Heyer:etal:04} show the
comparison of CTE corrected aperture photometry (according to
prescription \citep{Dolphin:02}, which is suitable for direct 0.5
arc-sec aperture photometry)  vs. the ground based sequences in
NGC\,2419 of \citet{Stetson:00}\footnote{this paper describes a 
continually updated data base at \\
http://cadcwww.dao.nrc.ca/cadcbin/wdb/astrocat/stetson/query}
and an early (before publication) version of the ground based NGC\,2419
sequence in \citeauthor{Saha:etal:05}. The zero-point used is the mean
of the five studies mentioned earlier. The comparisons in $F555W$ are 
within acceptable margins for both the Stetson and
\citeauthor{Saha:etal:05} sequences. 
However, in $F814W$, the Stetson sequence is in better agreement 
in this comparison. The residuals against \citeauthor{Saha:etal:05}
are as large as 0.07 mag. 
The residuals against the Stetson sequence are smaller (only 
as large as 0.03 mag), but have the same sense in all four chips. 
The \citeauthor{Saha:etal:05} sequence in the $I$ band  
is known to differ from the Stetson sequence used 
(as discussed extensively in \citeauthor{Saha:etal:05}), 
and the difference is maximal (by up to 0.05 mag) for the color of the
giants in NGC\,2419, which are the stars in use for this comparison. 
If instead of comparing to the mean zero-point of the five 
methods, the comparison is made to only the \citet{Dolphin:02} results
(which, along with \citet{Dolphin:02}, have a well recorded CTE 
correction pedigree) the residuals for the $F814W$ passband in all the
chips for the comparison with \citeauthor{Saha:etal:05} are
substantially reduced, to 0.04 mag at most. 
The comparison with Stetson's sequence in $F814W$ have even smaller
residuals. However, there is more to this comparison, and we must bear
two things in mind: 
\begin{enumerate}
\item 
   The primary difference between \citet{Dolphin:00} and
   \citet{Dolphin:02} is that the former is based on both, the
   \citet{Walker:94} sequence in $\omega$ Cen, {\it and} Stetson's
   sequence in NGC\,2419, whereas the latter is based on {\it only\/}
   the Walker zero-points. The \citet{Heyer:etal:04} text misleads the
   reader into believing that both sequences are used in both cases.
   In $F814W$, the \citet{Dolphin:02} zero-points shown in
   \citet{Heyer:etal:04} are systematically fainter compared with
   those from \citet{Dolphin:00}, which is exactly what one would
   expect if the Stetson sequence is too bright, as alleged in
   \citeauthor{Saha:etal:05}. A {\it direct\/} comparison of NGC\,2419
   photometry by Dolphin using only the Walker zero-points is shown in
   Fig.~10 of \citeauthor{Saha:etal:05}, and used as one of several
   arguments by them for the \citeauthor{Saha:etal:05} sequence to be
   preferred over Stetson's sequence. To close the loop on
   consistency, the \citet{Heyer:etal:04} WFPC2 photometry which is
   compared to the Stetson and \citeauthor{Saha:etal:05} sequences
   must be systematically in error by $\approx 0.03\mag$ in $F814W$. 
\item
   In their analysis, \citet{Heyer:etal:04} use an aperture correction
   to correct to infinity. They say: ``Aperture photometry was
   performed on each data set using a 0.5 arcsec radius, and the
   values were corrected to infinity by subtracting 0.1 magnitudes''. 
   Such corrections to `infinity' are inherently uncertain, especially
   when CTE effects can have relatively large effect on the far
   wings. The correction used by \cite{Heyer:etal:04} was originally
   derived in \citet{Holtzman:etal:95b}, using short exposure data,
   where the far wings would have been muted by CTE
   effects.\footnote{One must make this correction to infinity when
     calibrating synthetically from spectrophotometry and instrument
     response, but it is better to circumvent it whenever possible
     when comparing to standard sequences.}
   This is especially likely in $F814W$, in which the PSF has more
   flared wings than for bluer bands. We should expect this correction
   to depend on the level of background, and on the exposure level
   (brightness) of each star. It would hardly be surprising if this
   correction thus results in an error by a few hundredths of a
   magnitude, in the sense that on Holtzman's short exposures, too
   little light in the wings would have been measured compared to one
   with no CTE effects. The correct zero-point for CTE corrected data
   would thus be {\it fainter\/} by a similar amount. This would
   further reduce the residuals in the \citet{Heyer:etal:04}
   comparison with the \citeauthor{Saha:etal:05} sequence. 
\end{enumerate}

Taken together, these clarifications explain what
\citet{Heyer:etal:04} find, and why they apparently favor the Stetson
sequences over \citeauthor{Saha:etal:05}. While one cannot be certain
if we have the correct explanation, it is certainly a plausible one. 
The point is that there are too many things to untangle 
and \citet{Heyer:etal:04} is short on specifics to allow us to resolve 
all the apparent discrepancies. In contrast, our approach taken here is 
direct and fully self-consistent. The only decision made has been to 
prefer the \citeauthor{Saha:etal:05} sequence over that of Stetson,
and that for reasons that are argued in \citeauthor{Saha:etal:05}.

     The noteworthy difference between Stetson's sequence and of
\citeauthor{Saha:etal:05} is the color-dependent discrepancy in
$I$. Comparisons of $B$, $V$ and $R$ passbands are wholly within the
bounds of expected uncertainties. 
In the $I$ band, \citeauthor{Saha:etal:05} get magnitudes that agree
with Stetson's at $(V\!-\!I) = 0$, but get progressively fainter as one
goes redder till $(V\!-\!I) \approx 1.0$, where it is as much as 0.04
mag fainter than Stetson's. 
At even redder colors the differences do not increase further 
(see Fig~8 in \citeauthor{Saha:etal:05}), but there are very few stars
to trace subtle changes.  
The majority of stars being compared in NGC\,2419 have colors near 
$(V\!-\!I) = 1.0$, and the net discrepancy is also thus 0.04 mag. This
is also the color region of interest for Cepheids.

     In a response of sorts to \citeauthor{Saha:etal:05},
\citet{Stetson:05} presents an extensive discussion and re-evaluation
of his NGC\,2419 sequence using the same image data used in
\citeauthor{Saha:etal:05}, as well as the many more observation sets
he has accumulated over the years. In \S~5.1 of that paper, Stetson
reports finding an error, which when corrected, removes the color
signature. However, there is still a discrepancy with the zero-point
of his new magnitudes for NGC\,2419 stars: the net 
result is that the color dependence in the discrepancy is
fully rectified, but a zero-point discrepancy of $\approx 0.02\mag$
(\citeauthor{Saha:etal:05} being fainter) is still present in $I$ for
stars of all colors in the range $ 0.0 \leq V\!-\!I \leq 2.0 $, which
is the color range of interest (Fig.~6 of \citep{Stetson:05}) to us.  

     In his \S~5.3.1, \citet{Stetson:05} speculates whether some
peculiarities of the WIYN data are responsible for the difference --
such as errors in shutter timing. Short exposures are shown to yield
systematically different magnitudes than longer ones by amounts of
order $0.02\mag$, but the sign of the difference changes from night to
night. In \citeauthor{Saha:etal:05}, these differences are captured in
the calculation of the systematic errors. No definite flaw in the WIYN
data are identified. The differences are comparable to the scatter in
the exposure to exposure residuals of the standard stars, and could
well be just random errors from one exposure to the next, possibly
reflecting imperfections in how photometric the prevailing conditions
were. 

     One must understand that Stetson's photometric sequence is the
sum total of all of the data he has acquired. In other words, he
calibrates the WIYN data to his extant primary and secondary standards 
that include Landolt standards as well as stars he has already
calibrated, in this case in NGC\,2419 itself. 
In reading  para 6 of the same subsection in \citet{Stetson:05}, it
appears that when he reduces the WIYN data of
\citeauthor{Saha:etal:05} with only the \citet{Landolt:92} standards,
he sees the same discrepancy of order $0.02\mag$ as we do with  
his (now modified and corrected for the color signature)
sequence. This he then reconciles using the shutter error
hypothesis. What is interesting here is that when Stetson does the
same experiment as us, i.e.\ compare against \citet{Landolt:92}
standards, he gets the same answer from the same data. 
This could mean, as he suggests, that there is something strange 
about the WIYN data. It could also mean that there is still some
residual systematic in his secondary standards, and in his values for
the Landolt stars.  Only an independent study can resolve the issue.
However, with the color signature corrections already made, the new
Stetson sequence is only $0.02\mag$ discrepant with
\citeauthor{Saha:etal:05}.

\subsection{Test for Corrections for Epoch}
\label{sec:redetermination:test}
\begin{figure}[t]
   \epsscale{0.65}
   \plotone{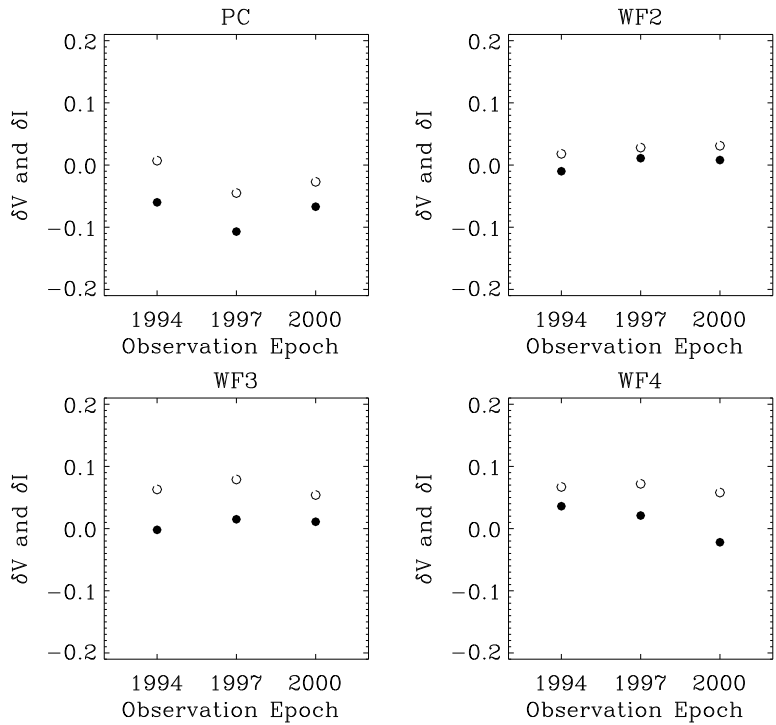}
   \caption{The zero-point differences WIYN $-$ WFPC2 for $V$ (filled
   circles) and $I$ (open circles) on an chip to chip
   basis. Zero-point dependencies with observation time are at most a
   few hundredths of a mag over 6 years, but larger zero-point
   systematic differences exist that are independent of time.}  
\label{fig:ZP_differences_time_mean}
\end{figure}

We have made several iterations in the tests for changes in the 
WFPC2 camera sensitivities (CTE changes with time and any other
effects) with observing time (the year). In a first attempt, we
combined all epochs in a given year to calculate the mean zero-point 
differences as a function of the epoch in that year. Separating the
years gives mean zero-point differences for 1994, 1997, and 2000 for
$V$ and $I$ separately.  
The conclusions from this first iteration are shown in
Figure~\ref{fig:ZP_differences_time_mean} for each chip separately (The
PC is the higher  resolution ``Planetary Camera'' chip). These
zero-point differences are in the sense WIYN minus WFPC2 magnitudes,
where the WFPC2 magnitudes are on the
\citet{Holtzman:etal:95b} system which we used throughout our original
11 paper series. Note again that the Holtzman magnitude zero-points
are on the ``short exposure'' scale. Filled circles in
Figure~\ref{fig:ZP_differences_time_mean} are the zero-point
differences in $V$ (Johnson system); open circles are the differences in
$I$ (Cousins system).

In Figure~\ref{fig:ZP_differences_time_mean} all data have been used in
a given year regardless of the number of stars that went into the
comparisons, or the exposure times (but they are corrected for the CTE 
effect using \citeauthor{Saha:etal:00}), or the crowding differences. 
We found no significant dependence on the observation epoch in a given
year from these data and all such epochs are averaged at the level of
$\sim\!0.02\mag$.\footnote{The mean magnitudes were calculated as a
  weighted mean, weighted by the inverse square of the standard
  error. This has the advantage that the scatter as well as the number
  of stars have been taken into account. If only one star was
  available, the standard error was estimated as the highest
  individual rms error of another individual epoch with the same
  exposure time. This is obviously an overestimate of the standard
  error, but it makes sure that epochs with just one matched star are
  given a low weight.} 

     However, Figure~\ref{fig:ZP_differences_time_mean} does show a
significant zero-point offset for the PC chip in $V$ (filled 
circles) and in $I$ (open circles) in chips 3 and 4, and a possible
secular change in $V$ in chip~4 between 1994 and 2000, all at a level
of about $0.06\mag$. 

\begin{figure}[t]
   \plotone{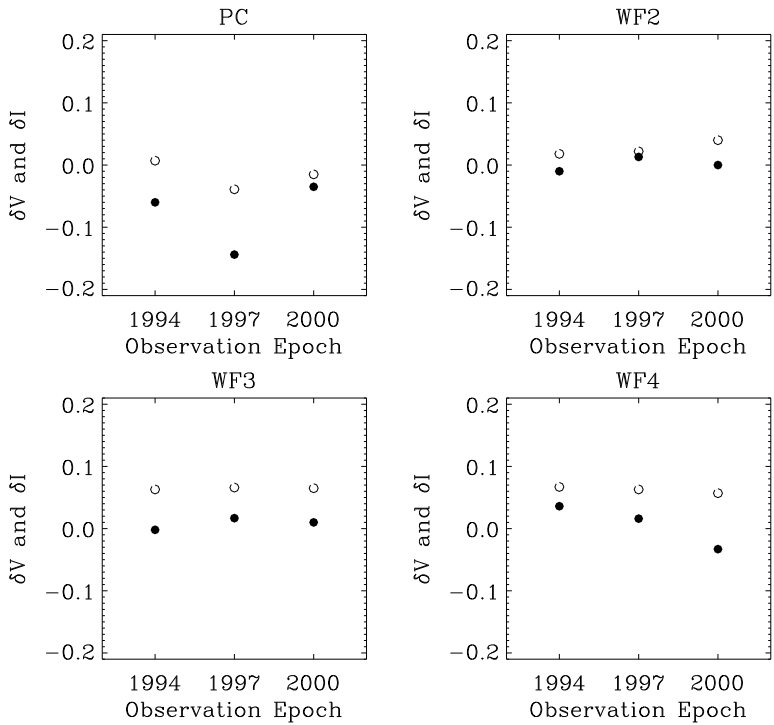}
   \caption{The zero-point differences WIYN $-$ WFPC2 for $V$ (filled
   circles) and $I$ (open circles) using only exposures longer than
   300s. No significant zero-point dependencies with observation time
   are seen, proving that our corrections for the CTE problem using
   the method of \citeauthor{Saha:etal:00} have worked.} 
\label{fig:ZP_differences_time_mean_long}
\end{figure}

In Figure~\ref{fig:ZP_differences_time_mean_long} we show the next
iteration using only long exposures of more than 300 seconds. 
No significant differences from Figure~\ref{fig:ZP_differences_time_mean}
are seen (with the exception of $V$ in the PC, which is almost
certainly an artifact of the small numbers of stars used), proving
that our corrections for the CTE by the method of
\citeauthor{Saha:etal:00} have worked.  

\subsection{Tests for Possible Crowding Problems}
\label{sec:redetermination:crowding}
\begin{figure}[t]
   \plotone{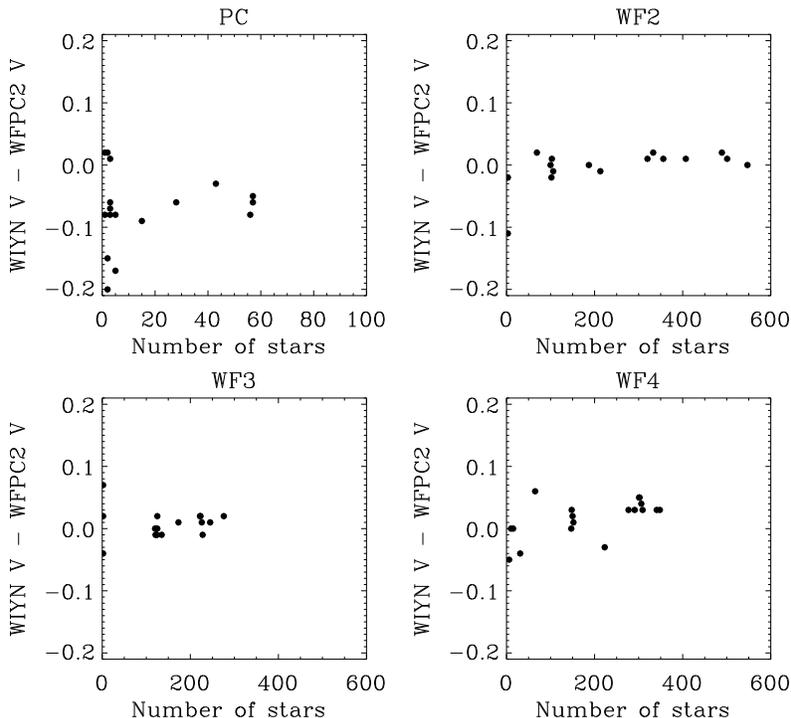}
   \caption{The zero-point differences WIYN $-$ WFPC2 for $V$ as a
   function of number of matched stars. The number of matched stars is
   used as a crowding index.} 
\label{fig4}
\end{figure}

\begin{figure}[t]
   \plotone{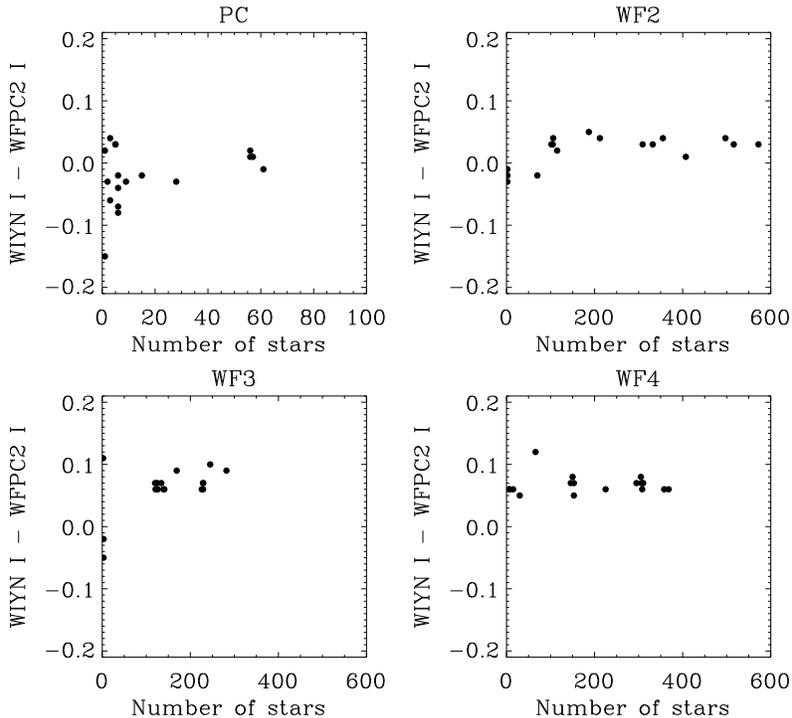}
   \caption{The zero-point differences WIYN $-$ WFPC2 for $I$ as a
   function of number of matched stars.} 
\label{fig5}
\end{figure}

We next investigated any possible zero-point differences as a function
of crowding, caused by a possible inability of the DoPHOT reduction
procedure to deal with and correct for closely adjacent images. The
number of matched pairs between the WIYN and the WFPC2 frames
on a given chip is used as a measure of the crowding density. The
results are shown in Figures~\ref{fig4} and \ref{fig5} for $V$ and $I$
respectively using the data from Tables~\ref{tab:ZP_diff_V} and
\ref{tab:ZP_diff_I}. No systematic zero-point dependence on the
crowing index (the number of stars) is present but, of course the
scatter in the figures is larger for the smaller number of stars,
showing the $\sqrt{N}$ dependence of the error in the mean with
respect to the rms scatter. 

\subsection{The Final Correction Table to the Holtzman et al. Zero-Points}
\label{sec:redetermination:holtz}
\begin{figure}[t]
   \plotone{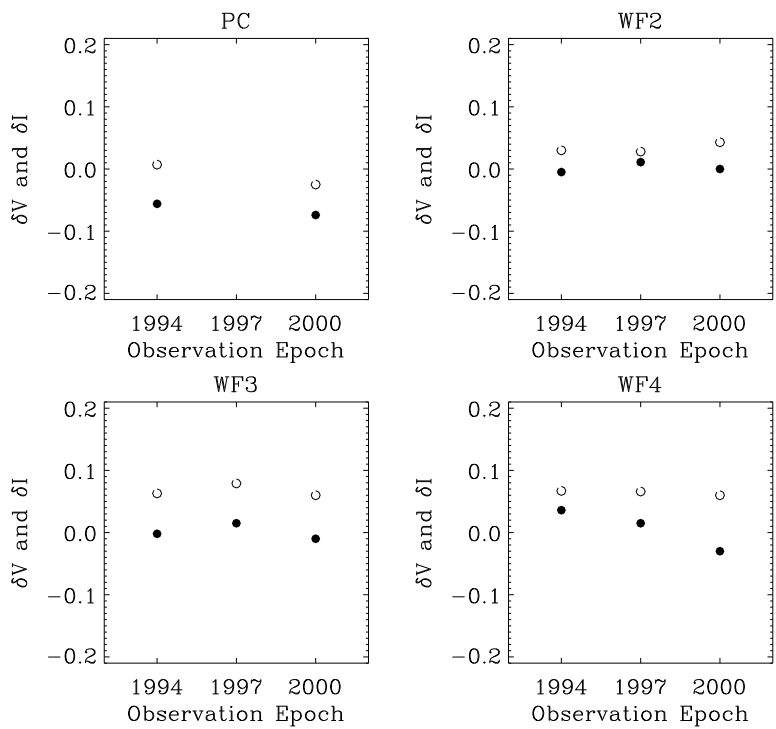}
   \caption{The zero-point differences WIYN $-$ WFPC2 for $V$ (filled
   circles) and $I$ (open circles) on an chip to chip basis for epochs
   with more than 10 stars in the PC and more than 100 stars in the WF
   chips.} 
\label{fig6}
\end{figure}

As a consequence of the increase in scatter for small $N$ seen in 
Figures~\ref{fig4} and \ref{fig5}, we have re-analyzed the data that
generated Figure~\ref{fig:ZP_differences_time_mean} using the comparison
data for the magnitude differences between the 
WIYN and the WFPC2
matched stars using mean values calculated only when the number of
such matched stars was 10 or greater. The data for
Figure~\ref{fig:ZP_differences_time_mean} that  were re-worked with this
restriction is shown in Figure~\ref{fig6}. For the 1997 $V$ data for the
PC, no observations with 10 or more matching stars can be
found. Figure~\ref{fig6} shows that there are no significant differences
with Figure~\ref{fig:ZP_differences_time_mean}. The systematic trend for
the $V$ data in chip 4 still remain, which could be interpreted as a
time-dependent zero-point change. 
However, the reality of the gradient depends on only the single 2000
epoch in $V$. There is no gradient in $I$, which should be present if
there is a real secular trend from changes in detector response. 
We therefore take the average of the zero-point differences in the 3
epochs shown in Figure~\ref{fig6}, rather than argue for a zero-point
change for just one chip in just one filter. 
We do not imply that no changes to the CTE took place over this period of 
time, we conclude only that over the six year interval early in the life 
of WFPC2, the effect of any such changes on our photometry are 
not noticeable at the 0.02 mag level.

     The final mean magnitude differences are again weighted means,
weighted by the inverse variance of the scatter of the individual
measurements. These values are shown in
Table~\ref{tab:final_delta_mags}, and are adopted as the zero-point
adjustments to the Holtzman et~al. zero-points (and therefore to our
initial magnitude system in the cited individual papers). 

     The corrections to the Holtzman scale are negative for the PC
chip (i.e.\ the corrected magnitudes found by applying the values in
Table~\ref{tab:final_delta_mags} to our originally listed values are 
brighter than these earlier published values) but are positive for the
WF chips (hence the corrected magnitudes on the WF chips here will be
fainter than the originals). Furthermore, the corrections are always
more positive in $I$ than in $V$ in all four chips. This means that the
corrected magnitudes of the Cepheids will be bluer than originally
published by a few hundredths of a magnitude, according to the values
in Table~\ref{tab:final_delta_mags}. This is an important change
because these small systematic effects in the $V\!-\!I$ colors can
drive significant (in the range of 10\%) differences in the derived
extinction and therefore in the distance modulus. These effects are
factored into the new distances derived in \S~\ref{sec:distance}. 
\begin{deluxetable}{lcr}
\tablecaption{$V$ and $I$ Mean Zero-Point 
differences WIYN $-$ WFPC2}
\tablewidth{0pt}
\tablehead{
\colhead{Chip} & 
\colhead{Filter} & 
\colhead{$\Delta$ m}} 
\startdata
PC   & $V$ & $-$0.062$\pm$0.021 \\
WF2  & $V$ &    0.004$\pm$0.012 \\
WF3  & $V$ &    0.004$\pm$0.012 \\
WF4  & $V$ &    0.025$\pm$0.021 \\
PC   & $I$ & $-$0.003$\pm$0.020 \\
WF2  & $I$ &    0.032$\pm$0.010 \\
WF3  & $I$ &    0.070$\pm$0.013 \\
WF4  & $I$ &    0.067$\pm$0.009 \\
\enddata
\label{tab:final_delta_mags}
\end{deluxetable}

\clearpage

\section{NEW ADOPTED CEPHEID APPARENT MAGNITUDES 
         FOR THE EIGHT PROGRAM GALAXIES}
\label{sec:newadopted}
%
\subsection{The New Magnitudes in $V$ and $I$}
\label{sec:newadopted:VI}
We have applied the corrections in Table~\ref{tab:final_delta_mags} to
the original data that were listed in each of the papers for the eight
parent galaxies cited in the Introduction. 
These corrected apparent magnitudes in $V$ and $I$ are listed in
column 2\&3 of Table~\ref{tab:cep:complete}, discussed later in
\S~\ref{sec:distance:individual}.
Recall that in the original papers of target galaxies observed with
WFPC2, the mean $V$ and $I$ reported magnitudes for Cepheids were 
{\em not\/} corrected for the ``long vs. short'' exposure effect. 
We refer to these as being on the ``Holtzman scale''. Rather,
$0.05\mag$ was added to the final distance modulus to account for this
effect. 
The exception, as stated in the footnote in
\S~\ref{sec:redetermination:problem}, was for the case of 
NGC\,3982, where, at the referee's insistence, the ad hoc $0.05\mag$
correction was applied to the individual Cepheid magnitudes. Thus the
corrections in Table~\ref{tab:final_delta_mags} are applied directly
to the $V$ and $I$ mean magnitudes of Cepheids in NGC\,4536, 
NGC\,4496A, NGC\,4639, NGC\,3627, and NGC\,4527, as reported in the
respective original discovery papers. 
For NGC\,3982, however, we must first subtract 0.05 magnitude from
the reported mean $V$ and $I$ mean magnitudes to return them to the
Holtzman scale, and then apply the corrections from
Table~\ref{tab:final_delta_mags}. Note also that for NGC\,4527, two
different and independent sets of photometry were presented in the
original paper \citep{Saha:etal:01a}. 
The values (DoPHOT based) in Table 4 of that paper are on the 
uncorrected Holtzman scale, and the corrections from
Table~\ref{tab:final_delta_mags} are applied directly to those
values. The values in Table 5 of the same paper are the magnitudes
from the ROMAFOT based analysis, which were adjusted by adding
$0.05\mag$, and must be backed off to the Holtzman scale before 
applying the corrections from Table~\ref{tab:final_delta_mags}.

     Note emphatically that the corrections in
Table~\ref{tab:final_delta_mags} do not apply to 
our original magnitudes for the Cepheids in IC\,4182 and in NGC\,5253
which were measured with the original 
{\em Wide Field and Planetary Camera (WFPC)\/} aboard HST. 
Because of the way those data were reduced, the listed magnitudes in
the two original papers for these two galaxies  
are already on the ground based $V$ and $I$ Landolt
zero-points. Hence, Table~\ref{tab:cep:complete} in this paper list
the magnitudes as originally given.

\subsection{Selection of the Cepheids Used}
\label{sec:newadopted:selection}
The Cepheids listed in Table~\ref{tab:cep:complete} {\em are only a
subset of the total Cepheid population we actually discovered}. They
are the same selection from the complete data that were used in the
original papers to derive the distance moduli. The exact criteria for
selection are more fully described in the original papers for each of
the galaxies: the goal of the selection being to choose objects least
affected by crowding, and those without abnormal colors so as to
obtain a control over separating the effects from absorption and
reddening from the measurement errors in color which, when they are
large, have an abnormally large and spurious effect on the
absorption-corrected mean modulus. 
We explain below the principles applied to select a suitable
sub-sample of Cepheids for distance determination. The detailed case
by case analyses are in the original papers, and are not repeated
here. One of the reasons our distances differ from those of
\citet{Freedman:etal:01} at the 0.2-0.3$\mag$ level (our distances are
larger; \S~\ref{sec:comparison} items 3 \& 6 and Appendix~A) is that
they included Cepheids which we discarded for reasons that we discuss
below and because they approached the absorption problem differently
than we do, as narrated for the case of NGC\,5253 at the end of this
section. 

     When apparent moduli in $V$ and $I$ ($\mu_{V}$ and $\mu_{I}$) are
obtained from some version of a P-L relation (e.g.\
equations~\ref{eq:PLgal:V}\,-\,\ref{eq:PLlmc:I:lt1} later here), they
will differ due to the wavelength dependence of reddening. 
As we show later (\S~\ref{sec:distance:absorption},
equations~\ref{eq:mu_R}\&\ref{eq:mu_0}), the true modulus, $\mu^{0}$,
is given by equation~(\ref{eq:mu_0}) as 
$\mu^{0} = 2.52\mu_I - 1.52\mu_V$ where the ratios of absorption to
reddening, taken from \citeauthor{Sandage:etal:04}, are 
${\cal R}_{V}=3.23$ and ${\cal R}_{I}=1.95$.

     Each Cepheid yields a value of $\mu^0$. Often the observational
errors in photometry due to crowding and faintness are significant,
and the individual $\mu^0$ values differ for different Cepheids 
in the same galaxy, often by substantial amounts (see the individual
listings in many tables in the original series of papers and in
Table~\ref{tab:cep:complete}).

     The procedure to obtain the most likely mean value of $\mu^0$
requires caution to avoid a biased final answer. The errors in
measurement propagate strongly into the de-reddened true modulus: the
coefficients in equation~(\ref{eq:mu_0}) amplify any observational
errors in $V$ and particularly in $I$, where the effects of crowding
are also more severe. 

     There are cases among the eight data sets and using all the
Cepheid data (not just the subsets listed in
Table~\ref{tab:cep:complete} later) where the 
observed period-color relation has a large scatter. The cause is a
combination of the large measuring errors (primarily in $I$, often due 
to the smaller number of epochs than in $V$), and differential
absorption (i.e.\ differences in the real reddening from
Cepheid-to-Cepheid). The bias effect is this: if the scatter is due to
measuring error where the errors are not symmetrically distributed,
then interpreting the color scatter as due to differential reddening
can produce a skewed (bias) mean true modulus. While currently
available photometric reduction programs do produce very 
realistic errors from fitting residuals, the estimate of errors from
confusion noise is more uncertain. Artificial star experiments to
model confusion errors do not work reliably for HST/WFPC2
data, since the PSFs are acutely undersampled (and we do not have
sufficient knowledge of the intrinsic PSF to simulate it well enough
in artificial data). 

     However, in Paper~V of the series \citep{Saha:etal:96a} we
devised a diagnostic to learn if there are large measuring errors as
opposed to differential extinction alone. In case of LMC it is well
known that the slope of the reddening line ${\cal R}_{V,V-I}=2.52$ is
nearly the same as that of the lines of constant period
($\beta_{V,V-I}=2.43$; see \citeauthor{Sandage:etal:04}, eq.~24) that
cut across the instability strip in the HR diagram 
(i.e. there is an intrinsic change of color with true
brightness that mimics the color change due to reddening). Hence, in
the plane of $\mu_{V}$ vs. $\mu_{V} - \mu_{I}$ (the latter is the
color excess $E(V\!-\!I)$), true Cepheids can occupy only a small
strip along the reddening vector of slope 2.52.  

     Cepheids whose data have insignificant measuring errors, and
therefore whose excess colors are due solely to differential
reddening, {\em can occupy only a small strip along the 
reddening line\/} (cf. Fig.~11 of \citealt{Saha:etal:96a} and
Figure~9 of \citealt{Saha:etal:96b}). Excess scatter about this
line {\em must\/} therefore be {\em due to measuring errors}, whose 
magnitude can therefore be estimated from the scatter. 
{\em This method of recognizing the presence of observational error 
rather than differential reddening was used to restrict our complete
lists of discovered Cepheids\/} 
in each of the host galaxies to the subset of
Cepheids in Table~\ref{tab:cep:complete} that are reliable for
measuring a dereddened distance modulus. In contrast, 
\citet{Gibson:etal:00}, who re-did our Cepheid analysis, do not make
such a test, and their results, we believe, are vulnerable to the
propagation of skewed photometric errors through the dereddening
procedure, due to measuring errors, not correctable reddening. 

     We now realize that our diagnostic must be employed with caution
because (metal-rich) Galactic Cepheids exhibit very shallow
constant-period lines of slope $\beta_{V,V-I}\approx0.6$
(\citeauthor{Sandage:etal:04}) in the HR diagram.  
Changing the intrinsic color of a Cepheid with fixed period has
therefore much less effect on $M_{V}$ than interstellar reddening. 
Consequently Galactic Cepheids and their counterparts have
considerably more scatter in the $\mu_{V}$ vs. $\mu_{V}-\mu_{I}$ plane 
than LMC-like Cepheids. 
Nevertheless, even this increased allowance does not accommodate the
much larger empirical scatter that we see in some of the galaxies we
have studied, and so our prior conclusion that the dominant source of
this excess scatter seen in some of our Cepheid data comes from
measurement errors from excessive crowding of objects. 

     Each of our eight cases of the SNe\,Ia host galaxies present
different absorption and reddening situations, because the severity of
differential extinction and confusion noise (measurement errors from
crowding) differ from galaxy to galaxy. 
We are aware that this requires us to handle each case individually,
so that the uncertainty in the derived SNIa absolute magnitude is
minimized in each individual case. 
The specific details for each case can be found in the original
papers cited in \S~\ref{sec:intro}.
We disagree with \citet{Gibson:etal:00} in their criticism of our
different treatment of the absorption problem from
galaxy-to-galaxy. We believe they have ignored the complex interplay
between reddening and measuring errors.

     The problem can be illustrated by considering our decisions
regarding optimal procedure for the case of NGC\,5253, as detailed in
\citet{Saha:etal:95}. In this case, the Cepheids are all outside the
central region of this amorphous (Am) galaxy, and show 
remarkable consistency in the apparent $V$-band moduli $\mu_{V}$, but
wide scatter in $\mu_{I}$. The source of the range in observed
$\mu_{V}-\mu_{I}$ cannot then be from differential extinction, since
that would cause a larger range in $\mu_{V}$ than in $\mu_{I}$,
contrary to what is observed. The culprit must be the measurement
errors in $I$, likely due to the higher level of object confusion in
$I$ than in $V$ (the mean color of the unresolved or quasi-resolved
fainter stars is redder than the mean color of Cepheids), 
which is exacerbated because we have only a few epochs available in
$I$.

\section{DISTANCE MODULI OF EIGHT SNe\,Ia HOST GALAXIES 
         BASED ON NEW CEPHEID P-L RELATIONS}
\label{sec:distance}
%
\subsection{The Adopted P-L Relations and Corresponding Apparent
  Distance Moduli}
\label{sec:distance:individual}
The Cepheids in the Galaxy and in LMC define {\em different\/} P-L
relations (Paper~I\&II; \citealt{Ngeow:Kanbur:04}; \citealt{Ngeow:etal:05}). 
The most likely reason for the difference are the different
metallicities of the two galaxies and their Cepheids. In fact the
metallicity-dependent line blanketing effect {\em must\/} have some
effect on the period-color (P-C) relations and hence on the P-L
relations. But the major cause for the different P-L relations is that
LMC Cepheids at given period or luminosity are hotter (and therefore
also bluer) than their Galactic counterparts
(\citeauthor{Sandage:etal:04}). The latter cannot be proved yet to be
a metallicity effect, but in the absence of an alternative
explication, we accept it as a working hypothesis.
The P-L relations from \citeauthor{Sandage:etal:04} for the here
relevant $V$ and $I$ magnitudes, not repeating the errors of the
coefficients, are  
\\
1) for the Galaxy
\begin{eqnarray}
\label{eq:PLgal:V}
   M_{V} &=& -3.087 \log P - 0.914, \\
\label{eq:PLgal:I}
   M_{I} &=& -3.348 \log P - 1.429, 
\end{eqnarray}
2a) for LMC ($P>10\;$days)
\begin{eqnarray}
\label{eq:PLlmc:V}
   M_{V} &=& -2.567 \log P - 1.634, \\
\label{eq:PLlmc:I}
   M_{I} &=& -2.822 \log P - 2.084, 
\end{eqnarray}
2b) for LMC ($P<10\;$days)
\begin{eqnarray}
\label{eq:PLlmc:V:lt1}
   M_{V} &=& -2.963 \log P - 1.335, \\
\label{eq:PLlmc:I:lt1}
   M_{I} &=& -3.099 \log P - 1.846. 
\end{eqnarray}
The P-L relation of LMC has a break of slope at $P=10\;$days.

     A tacit assumption in Paper~I \& II should be justified here. It
was assumed that the open clusters containing the calibrating
Cepheids, which in turn carry almost 50\% of the weight in
equations~(\ref{eq:PLgal:V}\,\&\,\ref{eq:PLgal:I}), have {\em solar
[Fe/H] on average}, because their metal-dependent main sequences were
fitted to the ZAMS of the Pleiades. Since very few relevant cluster
abundances are available the metallicity of their Cepheids is taken as
representative. Values of [Fe/H] are available for 14 of these
Cepheids from \citet{Fry:Carney:97}, \citet{Andrievsky:etal:02}, and
\citet{Luck:etal:03}; their mean is $[\mbox{Fe/H}] = -0.02\pm0.02$. 
The close to solar value follows also from the mean galactocentric
distance of the calibrating clusters of $\langle R\rangle=7.8\;$kpc,
which is based on an assumed value of $R_{\odot}=7.9\;$kpc. 
Open clusters near the solar circle are indeed known to have solar
[Fe/H]s on average \citep{Chen:etal:03}. 

     If one applies the Galactic
equations~(\ref{eq:PLgal:V}\,\&\,\ref{eq:PLgal:I}) to the individual
Cepheids with $V$ {\em and\/} $I$ magnitudes in
Table~\ref{tab:cep:complete} one obtains absolute 
magnitudes $M_{VI}$(Gal) which, if combined with the appropriate
apparent magnitudes, yield individual {\em apparent\/} distance moduli 
$\mu_{VI}$(Gal) not corrected for absorption. Analogously the LMC P-L
relations in equations~(\ref{eq:PLlmc:V}-\ref{eq:PLlmc:I:lt1}) yield
{\em apparent\/} moduli $\mu_{VI}$(LMC), again not corrected for
absorption. The resulting apparent moduli are listed in
Table~\ref{tab:cep:complete}, column~5 \& 6 and 8 \& 9, respectively.

     The selection of the Cepheids is explained in the original papers
and in \S~\ref{sec:newadopted:selection}. Cepheids with $P<10\;$days are
excluded except in NGC\,5253. Some of the Cepheids with the shortest
periods but being suspiciously bright were excluded in addition,
because they appear biased at the expense of fainter fellows
\citep{Sandage:88}. The excluded Cepheids are marked with an asterisk
in Table~\ref{tab:cep:complete}. Finally Cepheids are left out whose
$\mu_{V}$ or $\mu_{I}$ deviate by more than $2\sigma$ from the mean
$\langle\mu_{V}\rangle$ or $\langle\mu_{I}\rangle$; they are marked
with two asterisks in Table~\ref{tab:cep:complete}.

\begin{deluxetable}{clcrrcccccc}
\tablewidth{0pt}
\tabletypesize{\scriptsize}
\tablecaption{Magnitudes and Distance Moduli of Cepheids.\label{tab:cep:complete}}
\tablehead{
       \multicolumn{5}{c}{} &
       \multicolumn{3}{c}{P-L from Gal.} &
       \multicolumn{3}{c}{P-L from LMC} \\
 \colhead{} & \colhead{Cepheid} & 
 \colhead{$\log P$} &
 \colhead{$m_{V}$} &
 \colhead{$m_{I}$} &
 \colhead{$\mu_{V}$} &
 \colhead{$\mu_{I}$} &
 \colhead{$\mu_{0}$} &
 \colhead{$\mu_{V}$} &
 \colhead{$\mu_{I}$} &
 \colhead{$\mu_{0}$} \\
 \colhead{}  & \colhead{(1)} & \colhead{(2)}  &
 \colhead{(3)}  & \colhead{(4)}  &
 \colhead{(5)}  & \colhead{(6)}  &
 \colhead{(7)}  & \colhead{(8)}  &
 \colhead{(9)}  & \colhead{(10)}} 

\startdata
\tableline
\noalign{\smallskip}
\multicolumn{11}{c}{\footnotesize NGC\,3627 \citep{Saha:etal:99}}\\
\noalign{\smallskip}
\tableline
        & C2-V4   &  1.623  &  24.55  &  23.53  &  30.48  &  30.40  &  30.27  &  30.35  &  30.20  &  29.96  \\
        & C2-V8   &  1.602  &  24.87  &  24.05  &  30.73  &  30.84  &  31.01  &  30.62  &  30.66  &  30.71  \\
        & C2-V10  &  1.342  &  24.71  &  24.23  &  29.77  &  30.16  &  30.74  &  29.79  &  30.10  &  30.58  \\
        & C2-V12  &  1.415  &  25.03  &  24.05  &  30.32  &  30.22  &  30.07  &  30.30  &  30.13  &  29.87  \\
 $\!$ * & C2-V13  &  1.260  &  25.83  &  24.99  &  30.64  &  30.64  &  30.64  &  30.70  &  30.63  &  30.52  \\
        & C2-V15  &  1.431  &  25.71  &  24.70  &  31.05  &  30.92  &  30.74  &  31.02  &  30.83  &  30.53  \\
        & C2-V17  &  1.613  &  24.24  &  23.20  &  30.14  &  30.03  &  29.87  &  30.02  &  29.84  &  29.56  \\
        & C2-V19  &  1.763  &  24.80  &  23.86  &  31.16  &  31.19  &  31.25  &  30.96  &  30.92  &  30.86  \\
        & C2-V20  &  1.415  &  24.84  &  24.03  &  30.13  &  30.20  &  30.31  &  30.11  &  30.11  &  30.11  \\
 $\!$ * & C2-V22  &  1.255  &  25.19  &  24.48  &  29.98  &  30.11  &  30.31  &  30.05  &  30.11  &  30.20  \\
        & C2-V29  &  1.272  &  25.58  &  24.61  &  30.42  &  30.30  &  30.11  &  30.48  &  30.29  &  29.98  \\
        & C2-V32  &  1.407  &  25.29  &  24.26  &  30.55  &  30.40  &  30.17  &  30.54  &  30.32  &  29.98  \\
        & C2-V33  &  1.505  &  24.64  &  23.68  &  30.20  &  30.15  &  30.07  &  30.14  &  30.01  &  29.82  \\
        & C2-V34  &  1.681  &  24.35  &  23.28  &  30.46  &  30.34  &  30.16  &  30.30  &  30.11  &  29.82  \\
        & C2-V35  &  1.452  &  25.24  &  24.35  &  30.64  &  30.64  &  30.64  &  30.60  &  30.53  &  30.42  \\
        & C3-V1   &  1.288  &  25.77  &  24.98  &  30.66  &  30.72  &  30.81  &  30.71  &  30.70  &  30.67  \\
        & C3-V3   &  1.477  &  24.99  &  23.98  &  30.47  &  30.35  &  30.18  &  30.42  &  30.23  &  29.95  \\
        & C3-V4   &  1.431  &  25.22  &  24.35  &  30.56  &  30.57  &  30.59  &  30.53  &  30.47  &  30.38  \\
        & C3-V5   &  1.288  &  25.01  &  24.32  &  29.90  &  30.06  &  30.30  &  29.95  &  30.04  &  30.17  \\
        & C3-V6   &  1.342  &  25.14  &  24.24  &  30.20  &  30.16  &  30.10  &  30.22  &  30.11  &  29.94  \\
        & C3-V8   &  1.288  &  25.38  &  24.38  &  30.27  &  30.12  &  29.89  &  30.32  &  30.10  &  29.76  \\
        & C3-V10  &  1.613  &  24.21  &  23.35  &  30.11  &  30.18  &  30.29  &  29.99  &  29.99  &  29.98  \\
 $\!$ **& C4-V2   &  1.272  &  25.76  &  25.17  &  30.60  &  30.85  &  31.25  &  30.65  &  30.84  &  31.12  \\
        & C4-V4   &  1.415  &  25.42  &  24.76  &  30.70  &  30.92  &  31.27  &  30.68  &  30.83  &  31.07  \\
        & C4-V6   &  1.283  &  25.47  &  24.58  &  30.34  &  30.30  &  30.24  &  30.39  &  30.28  &  30.11  \\
\tableline
               \multicolumn{5}{c}{mean (all):}  &  30.42  &  30.43  &  30.45  &  30.40  &  30.33  &  30.24  \\
\tableline
        \multicolumn{5}{c}{mean (restricted):}  &  30.43  &  30.44  &  30.45  &  30.40  &  30.33  &  30.23  \\
        \multicolumn{5}{c}{}                    &         &     23  &   0.87  &         &     23  &   0.87  \\
\multicolumn{5}{c}{mean ($2\sigma$ - adopted):} &  30.42  &  30.42  &  30.41  &  30.39  &  30.31  &  30.19  \\
        \multicolumn{5}{c}{}                    &         &     22  &   0.09  &         &     22  &   0.09  \\
\tableline
\noalign{\smallskip}
\multicolumn{11}{c}{\footnotesize NGC\,3982 \citep{Saha:etal:01b}}\\
\noalign{\smallskip}
\tableline
        & C1-V1   &  1.468  &  26.60  &  25.58  &  32.04  &  31.92  &  31.73  &  32.00  &  31.80  &  31.50  \\
        & C1-V2   &  1.386  &  26.72  &  25.89  &  31.91  &  31.96  &  32.03  &  31.91  &  31.88  &  31.84  \\
        & C1-V4   &  1.633  &  26.86  &  25.78  &  32.81  &  32.67  &  32.46  &  32.69  &  32.47  &  32.14  \\
 $\!$ * & C1-V5   &  1.328  &  27.08  &  26.03  &  32.09  &  31.90  &  31.61  &  32.12  &  31.86  &  31.46  \\
        & C2-V1   &  1.687  &  26.22  &  25.38  &  32.34  &  32.46  &  32.63  &  32.19  &  32.23  &  32.28  \\
        & C2-V2   &  1.449  &  26.35  &  25.65  &  31.74  &  31.93  &  32.22  &  31.71  &  31.82  &  32.00  \\
        & C2-V3   &  1.607  &  26.01  &  25.51  &  31.89  &  32.32  &  32.98  &  31.77  &  32.13  &  32.68  \\
        & C2-V5   &  1.572  &  26.71  &  25.87  &  32.48  &  32.56  &  32.69  &  32.38  &  32.39  &  32.40  \\
        & C2-V6   &  1.400  &  26.77  &  25.58  &  32.01  &  31.70  &  31.22  &  32.00  &  31.62  &  31.03  \\
        & C2-V10  &  1.439  &  26.89  &  25.96  &  32.25  &  32.21  &  32.15  &  32.22  &  32.11  &  31.93  \\
        & C3-V1   &  1.613  &  24.85  &  24.08  &  30.75  &  30.91  &  31.16  &  30.63  &  30.72  &  30.85  \\
        & C3-V3   &  1.330  &  26.94  &  26.11  &  31.96  &  31.99  &  32.04  &  31.99  &  31.95  &  31.88  \\
        & C4-V1   &  1.484  &  26.95  &  25.50  &  32.44  &  31.90  &  31.07  &  32.39  &  31.77  &  30.83  \\
        & C4-V2   &  1.535  &  26.61  &  25.48  &  32.26  &  32.04  &  31.72  &  32.18  &  31.89  &  31.46  \\
 $\!$ * & C4-V3   &  1.320  &  26.86  &  25.74  &  31.84  &  31.58  &  31.19  &  31.88  &  31.55  &  31.04  \\
        & C4-V4   &  1.402  &  25.57  &  25.03  &  30.81  &  31.15  &  31.67  &  30.80  &  31.07  &  31.48  \\
        & C4-V6   &  1.525  &  25.72  &  24.81  &  31.34  &  31.34  &  31.35  &  31.26  &  31.20  &  31.09  \\
\tableline
               \multicolumn{5}{c}{mean (all):}  &  31.94  &  31.91  &  31.88  &  31.89  &  31.79  &  31.64  \\
\tableline
        \multicolumn{5}{c}{mean (restricted):}  &  31.94  &  31.94  &  31.94  &  31.87  &  31.80  &  31.69  \\
        \multicolumn{5}{c}{}                    &         &     15  &   1.29  &         &     15  &   1.26  \\
\multicolumn{5}{c}{mean ($2\sigma$ - adopted):} &  31.94  &  31.94  &  31.94  &  31.87  &  31.80  &  31.69  \\
        \multicolumn{5}{c}{}                    &         &     15  &   0.15  &         &     15  &   0.15  \\
\tableline
\noalign{\smallskip}
\multicolumn{11}{c}{\footnotesize NGC\,4496A \citep{Saha:etal:96b}}\\
\noalign{\smallskip}
\tableline
        & C1-V1   &  1.511  &  25.57  &  24.71  &  31.15  &  31.19  &  31.27  &  31.08  &  31.05  &  31.01  \\
        & C1-V4   &  1.447  &  25.86  &  25.06  &  31.24  &  31.33  &  31.47  &  31.21  &  31.22  &  31.25  \\
        & C1-V5   &  1.449  &  26.04  &  25.29  &  31.42  &  31.57  &  31.78  &  31.39  &  31.46  &  31.56  \\
        & C1-V6   &  1.294  &  26.27  &  25.61  &  31.18  &  31.37  &  31.66  &  31.22  &  31.34  &  31.52  \\
        & C1-V9   &  1.462  &  25.70  &  24.77  &  31.13  &  31.09  &  31.04  &  31.09  &  30.98  &  30.81  \\
        & C1-V10  &  1.301  &  26.33  &  25.41  &  31.26  &  31.19  &  31.09  &  31.30  &  31.16  &  30.95  \\
        & C1-V12  &  1.653  &  24.75  &  23.99  &  30.77  &  30.95  &  31.23  &  30.63  &  30.74  &  30.90  \\
        & C1-V13  &  1.462  &  25.42  &  24.48  &  30.85  &  30.80  &  30.73  &  30.81  &  30.69  &  30.51  \\
        & C1-V16  &  1.505  &  25.14  &  24.34  &  30.70  &  30.81  &  30.97  &  30.64  &  30.67  &  30.72  \\
 $\!$ **& C1-V17  &  1.352  &  26.29  &  25.69  &  31.38  &  31.64  &  32.05  &  31.39  &  31.59  &  31.88  \\
        & C2-V12  &  1.613  &  25.63  &  24.60  &  31.53  &  31.43  &  31.28  &  31.41  &  31.24  &  30.98  \\
        & C2-V14  &  1.724  &  24.47  &  23.60  &  30.71  &  30.80  &  30.95  &  30.53  &  30.55  &  30.58  \\
        & C2-V17  &  1.602  &  25.15  &  24.29  &  31.01  &  31.08  &  31.19  &  30.90  &  30.90  &  30.89  \\
        & C2-V18  &  1.447  &  25.39  &  24.64  &  30.78  &  30.92  &  31.13  &  30.74  &  30.81  &  30.91  \\
        & C2-V20  &  1.699  &  25.27  &  24.38  &  31.43  &  31.50  &  31.60  &  31.27  &  31.26  &  31.25  \\
        & C2-V21  &  1.716  &  25.23  &  24.32  &  31.45  &  31.50  &  31.57  &  31.27  &  31.25  &  31.21  \\
        & C2-V22  &  1.708  &  24.44  &  23.57  &  30.63  &  30.72  &  30.85  &  30.46  &  30.47  &  30.50  \\
        & C2-V24  &  1.328  &  26.22  &  25.56  &  31.24  &  31.44  &  31.74  &  31.27  &  31.39  &  31.59  \\
 $\!$ * & C2-V27  &  1.255  &  26.31  &  25.26  &  31.10  &  30.89  &  30.58  &  31.17  &  30.89  &  30.46  \\
        & C3-V2   &  1.519  &  25.26  &  24.61  &  30.87  &  31.12  &  31.51  &  30.80  &  30.98  &  31.26  \\
        & C3-V3   &  1.720  &  25.14  &  24.18  &  31.37  &  31.37  &  31.37  &  31.19  &  31.12  &  31.00  \\
        & C3-V4   &  1.720  &  24.37  &  23.47  &  30.60  &  30.66  &  30.75  &  30.42  &  30.41  &  30.38  \\
        & C3-V12  &  1.531  &  25.49  &  24.49  &  31.14  &  31.05  &  30.91  &  31.06  &  30.90  &  30.65  \\
        & C3-V17  &  1.398  &  25.79  &  24.94  &  31.02  &  31.05  &  31.09  &  31.02  &  30.97  &  30.90  \\
        & C3-V21  &  1.663  &  25.23  &  24.28  &  31.28  &  31.28  &  31.27  &  31.14  &  31.06  &  30.93  \\
        & C3-V28  &  1.591  &  25.45  &  24.62  &  31.28  &  31.38  &  31.52  &  31.17  &  31.19  &  31.23  \\
        & C3-V34  &  1.531  &  25.84  &  24.99  &  31.49  &  31.55  &  31.64  &  31.41  &  31.40  &  31.38  \\
        & C3-V36  &  1.283  &  26.39  &  25.64  &  31.27  &  31.37  &  31.51  &  31.32  &  31.35  &  31.38  \\
        & C3-V37  &  1.716  &  24.50  &  23.72  &  30.72  &  30.89  &  31.17  &  30.54  &  30.65  &  30.80  \\
 $\!$ * & C3-V39  &  1.255  &  26.30  &  25.46  &  31.09  &  31.09  &  31.09  &  31.16  &  31.09  &  30.97  \\
        & C4-V7   &  1.845  &  24.52  &  23.51  &  31.12  &  31.11  &  31.10  &  30.89  &  30.80  &  30.66  \\
        & C4-V10  &  1.386  &  26.07  &  25.38  &  31.26  &  31.45  &  31.73  &  31.26  &  31.37  &  31.55  \\
        & C4-V14  &  1.369  &  25.64  &  24.98  &  30.78  &  30.99  &  31.32  &  30.78  &  30.92  &  31.14  \\
        & C4-V16  &  1.281  &  26.18  &  25.20  &  31.04  &  30.91  &  30.72  &  31.10  &  30.90  &  30.59  \\
        & C4-V20  &  1.401  &  25.49  &  24.81  &  30.73  &  30.93  &  31.24  &  30.72  &  30.85  &  31.04  \\
        & C4-V22  &  1.484  &  25.19  &  24.47  &  30.68  &  30.87  &  31.15  &  30.63  &  30.74  &  30.91  \\
        & C4-V23  &  1.436  &  25.46  &  24.57  &  30.80  &  30.80  &  30.81  &  30.78  &  30.70  &  30.59  \\
 $\!$ **& C4-V27  &  1.375  &  25.58  &  24.62  &  30.73  &  30.65  &  30.52  &  30.74  &  30.58  &  30.34  \\
        & C4-V28  &  1.267  &  26.50  &  25.58  &  31.32  &  31.25  &  31.14  &  31.38  &  31.24  &  31.02  \\
        & C4-V32  &  1.412  &  25.62  &  24.87  &  30.89  &  31.02  &  31.23  &  30.87  &  30.93  &  31.03  \\
        & C4-V35  &  1.281  &  26.42  &  25.67  &  31.28  &  31.38  &  31.54  &  31.34  &  31.37  &  31.41  \\
        & C4-V37  &  1.748  &  25.11  &  24.06  &  31.42  &  31.34  &  31.22  &  31.23  &  31.07  &  30.84  \\
        & C4-V40  &  1.556  &  25.43  &  24.41  &  31.14  &  31.05  &  30.90  &  31.05  &  30.88  &  30.62  \\
\tableline
               \multicolumn{5}{c}{mean (all):}  &  31.08  &  31.13  &  31.24  &  31.01  &  31.00  &  30.99  \\
\tableline
        \multicolumn{5}{c}{mean (restricted):}  &  31.07  &  31.14  &  31.24  &  31.01  &  31.00  &  30.99  \\
        \multicolumn{5}{c}{}                    &         &     41  &   0.68  &         &     41  &   0.72  \\
\multicolumn{5}{c}{mean ($2\sigma$ - adopted):} &  31.08  &  31.14  &  31.24  &  31.01  &  31.00  &  30.99  \\
        \multicolumn{5}{c}{}                    &         &     39  &   0.05  &         &     39  &   0.05  \\
\tableline
\noalign{\smallskip}
\multicolumn{11}{c}{\footnotesize NGC\,4527 \citep{Saha:etal:01a}}\\
\noalign{\smallskip}
\tableline
 s1,2   & C1-V2  &  1.318  &  26.28  &  25.44  &  31.26  &  31.28  &  31.31  &  31.30  &  31.24  &  31.16  \\
 s1,2   & C1-V4  &  1.312  &  25.96  &  25.14  &  30.92  &  30.96  &  31.01  &  30.96  &  30.92  &  30.87  \\
 s1     & C1-V5  &  1.405  &  25.61  &  24.45  &  30.86  &  30.58  &  30.16  &  30.85  &  30.50  &  29.96  \\
 s1     & C1-V7  &  1.712  &  25.07  &  23.68  &  31.26  &  30.84  &  30.18  &  31.09  &  30.59  &  29.82  \\
 s1     & C1-V8  &  1.310  &  26.03  &  24.95  &  30.98  &  30.76  &  30.43  &  31.02  &  30.73  &  30.29  \\
 s1,2   & C1-V10 &  1.653  &  25.04  &  24.02  &  31.06  &  30.98  &  30.88  &  30.92  &  30.77  &  30.55  \\
 s1     & C1-V11 &  1.525  &  26.78  &  25.15  &  32.40  &  31.69  &  30.59  &  32.33  &  31.54  &  30.33  \\
 s1,2   & C2-V1  &  1.590  &  25.34  &  24.24  &  31.17  &  30.99  &  30.73  &  31.06  &  30.81  &  30.43  \\
 s1,2   & C2-V2  &  1.444  &  25.88  &  24.94  &  31.25  &  31.20  &  31.13  &  31.22  &  31.10  &  30.91  \\
 s1,2   & C3-V2  &  1.593  &  25.42  &  24.40  &  31.25  &  31.16  &  31.02  &  31.14  &  30.98  &  30.72  \\
 s1     & C3-V7  &  1.763  &  24.91  &  24.26  &  31.26  &  31.59  &  32.09  &  31.07  &  31.32  &  31.70  \\
 s1     & C3-V8  &  1.749  &  25.18  &  23.93  &  31.50  &  31.22  &  30.79  &  31.31  &  30.95  &  30.41  \\
 s1,2   & C3-V9  &  1.686  &  24.72  &  23.67  &  30.84  &  30.74  &  30.60  &  30.68  &  30.51  &  30.26  \\
 s1,2   & C3-V14 &  1.377  &  26.03  &  25.09  &  31.19  &  31.13  &  31.03  &  31.19  &  31.06  &  30.85  \\
 s1,2   & C3-V16 &  1.377  &  25.83  &  24.89  &  30.99  &  30.93  &  30.83  &  31.00  &  30.86  &  30.65  \\
 s2     & C4-V3  &  1.430  &  25.63  &  24.67  &  30.96  &  30.88  &  30.77  &  30.93  &  30.79  &  30.56  \\
 s2     & C4-V15 &  1.346  &  25.86  &  25.04  &  30.93  &  30.97  &  31.03  &  30.95  &  30.92  &  30.87  \\
        & C4-V16 &  1.511  &  25.50  &  24.17  &  31.07  &  30.65  &  30.01  &  31.01  &  30.51  &  29.76  \\
 s1     & C4-V18 &  1.465  &  25.98  &  24.62  &  31.42  &  30.95  &  30.25  &  31.37  &  30.84  &  30.02  \\
 s1,2   & C4-V21 &  1.430  &  25.65  &  24.66  &  30.98  &  30.87  &  30.71  &  30.96  &  30.78  &  30.50  \\
 s1,2   & C4-V22 &  1.610  &  24.88  &  23.85  &  30.76  &  30.67  &  30.52  &  30.64  &  30.48  &  30.22  \\
 s1     & C4-V26 &  1.346  &  26.26  &  25.10  &  31.33  &  31.04  &  30.60  &  31.35  &  30.98  &  30.43  \\
\tableline
               \multicolumn{5}{c}{sample1}  &  31.19  &  31.03  &  30.78  &  31.13  &  30.89  &  30.53  \\
               \multicolumn{5}{c}{sample2}  &  31.04  &  30.98  &  30.89  &  31.00  &  30.86  &  30.66  \\
\tableline
                  \multicolumn{5}{c}{mean}  &  31.12  &  31.01  &  30.84  &  31.06  &  30.88  &  30.59  \\
\tableline
\noalign{\smallskip}
\multicolumn{11}{c}{\footnotesize NGC\,4536 \citep{Saha:etal:96a}}\\
\noalign{\smallskip}
\tableline
        & C1-V3   &  1.480  &  26.04  &  24.91  &  31.52  &  31.29  &  30.94  &  31.47  &  31.17  &  30.71  \\
 $\!$ * & C1-V4   &  1.312  &  26.05  &  25.35  &  31.01  &  31.17  &  31.41  &  31.05  &  31.13  &  31.26  \\
        & C1-V5   &  1.505  &  25.65  &  24.69  &  31.21  &  31.16  &  31.07  &  31.15  &  31.02  &  30.83  \\
        & C1-V6   &  1.551  &  25.75  &  24.45  &  31.45  &  31.07  &  30.49  &  31.36  &  30.91  &  30.22  \\
        & C1-V10  &  1.580  &  25.31  &  24.30  &  31.10  &  31.02  &  30.89  &  31.00  &  30.84  &  30.60  \\
        & C2-V4   &  1.458  &  25.81  &  25.09  &  31.23  &  31.40  &  31.67  &  31.19  &  31.29  &  31.44  \\
        & C2-V6   &  1.435  &  26.00  &  25.42  &  31.35  &  31.65  &  32.12  &  31.32  &  31.55  &  31.91  \\
        & C2-V8   &  1.771  &  24.74  &  23.90  &  31.12  &  31.26  &  31.47  &  30.92  &  30.98  &  31.07  \\
        & C2-V9   &  1.763  &  25.39  &  24.48  &  31.75  &  31.81  &  31.91  &  31.55  &  31.54  &  31.52  \\
        & C2-V14  &  1.726  &  25.12  &  24.20  &  31.37  &  31.41  &  31.48  &  31.19  &  31.16  &  31.11  \\
        & C2-V16  &  1.519  &  25.31  &  24.46  &  30.92  &  30.97  &  31.07  &  30.85  &  30.83  &  30.81  \\
 $\!$ * & C2-V20  &  1.327  &  26.21  &  25.21  &  31.23  &  31.08  &  30.87  &  31.26  &  31.04  &  30.72  \\
        & C2-V22  &  1.623  &  25.61  &  24.88  &  31.54  &  31.75  &  32.06  &  31.41  &  31.55  &  31.75  \\
        & C2-V29  &  1.477  &  26.14  &  25.34  &  31.62  &  31.72  &  31.87  &  31.57  &  31.59  &  31.63  \\
        & C2-V30  &  1.490  &  25.75  &  24.88  &  31.27  &  31.30  &  31.35  &  31.21  &  31.17  &  31.11  \\
        & C2-V34  &  1.491  &  25.93  &  25.11  &  31.45  &  31.53  &  31.66  &  31.40  &  31.40  &  31.42  \\
        & C2-V36  &  1.535  &  25.24  &  24.57  &  30.90  &  31.14  &  31.51  &  30.82  &  30.99  &  31.25  \\
        & C3-V3   &  1.813  &  24.80  &  24.06  &  31.31  &  31.56  &  31.93  &  31.09  &  31.26  &  31.52  \\
        & C3-V7   &  1.348  &  26.06  &  24.95  &  31.14  &  30.89  &  30.52  &  31.16  &  30.84  &  30.35  \\
        & C3-V11  &  1.407  &  25.53  &  24.77  &  30.79  &  30.91  &  31.09  &  30.78  &  30.82  &  30.89  \\
        & C3-V12  &  1.364  &  25.18  &  24.39  &  30.31  &  30.38  &  30.50  &  30.32  &  30.32  &  30.33  \\
        & C3-V18  &  1.386  &  26.00  &  25.17  &  31.20  &  31.24  &  31.30  &  31.19  &  31.16  &  31.12  \\
        & C3-V21  &  1.799  &  25.13  &  23.85  &  31.60  &  31.30  &  30.85  &  31.39  &  31.01  &  30.44  \\
        & C3-V22  &  1.562  &  25.33  &  24.50  &  31.07  &  31.16  &  31.29  &  30.98  &  30.99  &  31.01  \\
 $\!$ **& C3-V24  &  1.447  &  25.55  &  24.23  &  30.94  &  30.50  &  29.85  &  30.90  &  30.40  &  29.63  \\
        & C3-V25  &  1.771  &  24.49  &  23.43  &  30.87  &  30.79  &  30.66  &  30.67  &  30.51  &  30.26  \\
 $\!$ **& C3-V31  &  1.458  &  25.62  &  25.34  &  31.04  &  31.65  &  32.58  &  31.00  &  31.54  &  32.36  \\
        & C3-V32  &  1.597  &  25.60  &  24.47  &  31.45  &  31.24  &  30.94  &  31.34  &  31.06  &  30.64  \\
        & C4-V5   &  1.387  &  25.53  &  24.88  &  30.72  &  30.95  &  31.30  &  30.72  &  30.88  &  31.11  \\
        & C4-V8   &  1.699  &  25.35  &  24.28  &  31.50  &  31.39  &  31.23  &  31.34  &  31.16  &  30.87  \\
        & C4-V13  &  1.740  &  25.44  &  24.13  &  31.72  &  31.38  &  30.87  &  31.54  &  31.12  &  30.49  \\
\tableline
               \multicolumn{5}{c}{mean (all):}  &  31.22  &  31.23  &  31.25  &  31.13  &  31.07  &  30.98  \\
\tableline
        \multicolumn{5}{c}{mean (restricted):}  &  31.22  &  31.24  &  31.26  &  31.13  &  31.07  &  30.98  \\
        \multicolumn{5}{c}{}                    &         &     29  &   1.18  &         &     29  &   1.17  \\
\multicolumn{5}{c}{mean ($2\sigma$ - adopted):} &  31.24  &  31.25  &  31.26  &  31.15  &  31.08  &  30.98  \\
        \multicolumn{5}{c}{}                    &         &     27  &   0.09  &         &     27  &   0.09  \\
\tableline
\noalign{\smallskip}
\multicolumn{11}{c}{\footnotesize NGC\,4639 \citep{Saha:etal:97}}\\
\noalign{\smallskip}
\tableline
        & C1-V1   &  1.473  &  27.03  &  25.94  &  32.49  &  32.30  &  32.01  &  32.44  &  32.18  &  31.77  \\
        & C1-V2   &  1.336  &  26.78  &  26.09  &  31.82  &  31.99  &  32.25  &  31.84  &  31.94  &  32.09  \\
        & C1-V5   &  1.505  &  26.14  &  25.31  &  31.70  &  31.78  &  31.89  &  31.64  &  31.64  &  31.64  \\
        & C2-V1   &  1.613  &  25.79  &  24.95  &  31.69  &  31.78  &  31.92  &  31.57  &  31.59  &  31.62  \\
        & C2-V3   &  1.681  &  25.24  &  24.48  &  31.35  &  31.54  &  31.83  &  31.19  &  31.31  &  31.49  \\
        & C2-V4   &  1.415  &  26.67  &  25.72  &  31.96  &  31.89  &  31.79  &  31.94  &  31.80  &  31.58  \\
        & C2-V6   &  1.477  &  26.18  &  25.31  &  31.66  &  31.69  &  31.73  &  31.61  &  31.56  &  31.50  \\
 $\!$ * & C2-V7   &  1.322  &  26.96  &  26.02  &  31.96  &  31.88  &  31.75  &  31.99  &  31.84  &  31.60  \\
        & C3-V1   &  1.538  &  26.72  &  25.90  &  32.39  &  32.48  &  32.62  &  32.31  &  32.32  &  32.35  \\
 $\!$ **& C3-V6   &  1.531  &  25.87  &  25.52  &  31.52  &  32.08  &  32.93  &  31.44  &  31.93  &  32.67  \\
        & C3-V7   &  1.505  &  26.26  &  25.52  &  31.82  &  31.99  &  32.24  &  31.76  &  31.85  &  31.99  \\
 $\!$ * & C3-V8   &  1.322  &  26.28  &  25.17  &  31.28  &  31.03  &  30.64  &  31.31  &  30.99  &  30.49  \\
        & C3-V10  &  1.602  &  25.65  &  25.08  &  31.51  &  31.87  &  32.42  &  31.40  &  31.69  &  32.12  \\
        & C3-V11  &  1.763  &  26.39  &  25.36  &  32.75  &  32.69  &  32.60  &  32.55  &  32.42  &  32.22  \\
        & C4-V1   &  1.716  &  26.01  &  25.07  &  32.22  &  32.24  &  32.28  &  32.04  &  31.99  &  31.92  \\
\tableline
               \multicolumn{5}{c}{mean (all):}  &  31.87  &  31.95  &  32.06  &  31.80  &  31.80  &  31.80  \\
\tableline
        \multicolumn{5}{c}{mean (restricted):}  &  31.91  &  32.02  &  32.19  &  31.83  &  31.86  &  31.92  \\
        \multicolumn{5}{c}{}                    &         &     13  &   0.74  &         &     13  &   0.72  \\
\multicolumn{5}{c}{mean ($2\sigma$ - adopted):} &  31.95  &  32.02  &  32.13  &  31.86  &  31.86  &  31.86  \\
        \multicolumn{5}{c}{}                    &         &     12  &   0.09  &         &     12  &   0.09  \\
\tableline
\noalign{\smallskip}
\multicolumn{11}{c}{\footnotesize NGC\,5253 \citep{Saha:etal:95}} \\
\noalign{\smallskip}
\tableline
 $\!$ * &  C1-V2  &  0.497  &  25.25  &  24.32  &  27.70  &  27.41  &  26.98  &  28.06  &  27.71  &  27.17  \\
 $\!$ * &  C2-V1  &  0.431  &  26.01  &  25.00  &  28.26  &  27.87  &  27.29  &  28.62  &  28.18  &  27.51  \\
        &  C2-V3  &  0.951  &  24.22  &  23.31  &  28.07  &  27.92  &  27.70  &  28.37  &  28.10  &  27.69  \\
 $\!$ * &  C3-V1  &  0.899  &  24.39  &  22.76  &  28.08  &  27.20  &  25.86  &  28.39  &  27.39  &  25.88  \\
        &  C3-V2  &  1.066  &  23.61  &  23.24  &  27.81  &  28.24  &  28.88  &  27.98  &  28.33  &  28.87  \\
 $\!$ * &  C3-V3  &  0.746  &  24.45  &  23.47  &  27.67  &  27.40  &  26.99  &  27.99  &  27.63  &  27.07  \\
 $\!$ * &  C3-V4  &  1.099  &  23.65  &  22.80  &  27.96  &  27.91  &  27.83  &  28.10  &  27.98  &  27.80  \\
 $\!$ * &  C3-V5  &  0.983  &  23.86  &  23.10  &  27.81  &  27.82  &  27.84  &  28.11  &  27.99  &  27.82  \\
        &  C3-V6  &  0.786  &  24.80  &  24.06  &  28.14  &  28.12  &  28.09  &  28.46  &  28.34  &  28.16  \\
 $\!$ * &  C4-V1  &  0.641  &  24.68  &  23.96  &  27.57  &  27.54  &  27.48  &  27.92  &  27.79  &  27.61  \\
        &  C4-V2  &  1.137  &  23.54  &  22.50  &  27.96  &  27.73  &  27.39  &  28.09  &  27.79  &  27.34  \\
        &  C4-V3  &  1.207  &  23.08  &  22.55  &  27.72  &  28.02  &  28.48  &  27.81  &  28.04  &  28.39  \\
\tableline
               \multicolumn{5}{c}{mean (all):}  &  27.90  &  27.76  &  27.57  &  28.16  &  27.94  &  27.61  \\
\tableline
        \multicolumn{5}{c}{mean (restricted):}  &  27.94  &  28.01  &  28.11  &  28.14  &  28.12  &  28.09  \\
        \multicolumn{5}{c}{}                    &         &      5  &   0.27  &         &      5  &   0.27  \\
\tableline
\noalign{\smallskip}
\multicolumn{11}{c}{\footnotesize IC\,4182 \citep{Saha:etal:94}}\\
\noalign{\smallskip}
\hline
 $\!$ * &  C1-V1  &  0.842  &  24.47  &  23.71  &  27.98  &  27.96  &  27.92  &  28.30  &  28.17  &  27.96  \\
 $\!$ * &  C1-V2  &  0.863  &  24.54  &  23.75  &  28.12  &  28.07  &  27.99  &  28.43  &  28.27  &  28.03  \\
        &  C1-V4  &  1.393  &  23.29  &  22.45  &  28.50  &  28.54  &  28.60  &  28.50  &  28.46  &  28.41  \\
 $\!$ * &  C1-V5  &  0.964  &  24.10  &  23.55  &  27.99  &  28.21  &  28.53  &  28.29  &  28.38  &  28.52  \\
        &  C1-V6  &  1.623  &  21.87  &  21.19  &  27.79  &  28.05  &  28.45  &  27.67  &  27.85  &  28.13  \\
 $\!$ * &  C2-V1  &  0.760  &  24.65  &  23.86  &  27.91  &  27.83  &  27.72  &  28.24  &  28.06  &  27.79  \\
        &  C2-V2  &  1.574  &  22.86  &  21.85  &  28.63  &  28.55  &  28.42  &  28.53  &  28.38  &  28.13  \\
 $\!$ * &  C2-V3  &  0.838  &  24.57  &  24.00  &  28.07  &  28.23  &  28.48  &  28.39  &  28.44  &  28.53  \\
 $\!$ * &  C3-V1  &  0.629  &  24.98  &  24.59  &  27.84  &  28.13  &  28.57  &  28.18  &  28.39  &  28.70  \\
 $\!$ * &  C3-V7  &  0.790  &  25.13  &  24.63  &  28.48  &  28.70  &  29.04  &  28.80  &  28.92  &  29.10  \\
 $\!$ **&  C3-V9  &  1.332  &  22.80  &  22.60  &  27.83  &  28.49  &  29.50  &  27.85  &  28.44  &  29.34  \\
        &  C3-V10 &  1.021  &  24.56  &  23.76  &  28.63  &  28.61  &  28.58  &  28.82  &  28.73  &  28.59  \\
        &  C3-V11 &  1.124  &  24.04  &  23.28  &  28.42  &  28.47  &  28.55  &  28.56  &  28.54  &  28.50  \\
        &  C3-V12 &  1.560  &  22.36  &  21.57  &  28.09  &  28.22  &  28.42  &  28.00  &  28.06  &  28.14  \\
 $\!$ * &  C4-V1  &  0.565  &  24.83  &  24.74  &  27.49  &  28.06  &  28.93  &  27.84  &  28.34  &  29.09  \\
 $\!$ * &  C4-V2  &  0.766  &  24.80  &  24.43  &  28.08  &  28.42  &  28.95  &  28.41  &  28.65  &  29.02  \\
 $\!$ * &  C4-V4  &  0.714  &  24.85  &  24.51  &  27.97  &  28.33  &  28.88  &  28.30  &  28.57  &  28.98  \\
 $\!$ * &  C4-V5  &  0.638  &  25.10  &  99.99  &  27.99  & \nodata & \nodata &  28.33  & \nodata & \nodata \\
        &  C4-V7  &  1.568  &  22.70  &  21.79  &  28.46  &  28.47  &  28.49  &  28.36  &  28.30  &  28.21  \\
 $\!$ **&  C4-V8  &  1.547  &  22.72  &  22.17  &  28.41  &  28.78  &  29.34  &  28.32  &  28.62  &  29.07  \\
 $\!$ * &  C4-V9  &  0.852  &  24.65  &  24.29  &  28.20  &  28.57  &  29.15  &  28.51  &  28.78  &  29.18  \\
        &  C4-V10 &  1.255  &  23.14  &  22.77  &  27.93  &  28.40  &  29.12  &  28.00  &  28.40  &  29.00  \\
        &  C4-V11 &  1.623  &  22.33  &  21.40  &  28.25  &  28.26  &  28.28  &  28.13  &  28.06  &  27.96  \\
        &  C4-V14 &  1.342  &  23.42  &  22.62  &  28.48  &  28.54  &  28.64  &  28.50  &  28.49  &  28.48  \\
        &  C4-V15 &  1.314  &  23.36  &  22.31  &  28.33  &  28.14  &  27.85  &  28.37  &  28.10  &  27.70  \\
        &  C4-V16 &  1.204  &  23.76  &  23.11  &  28.39  &  28.57  &  28.84  &  28.48  &  28.59  &  28.75  \\
 $\!$ * &  C4-V17 &  0.866  &  24.85  &  24.34  &  28.44  &  28.67  &  29.02  &  28.75  &  28.87  &  29.05  \\
        &  C4-V18 &  1.428  &  23.20  &  22.25  &  28.52  &  28.46  &  28.37  &  28.50  &  28.36  &  28.16  \\
\tableline
               \multicolumn{5}{c}{mean (all):}  &  28.19  &  28.36  &  28.62  &  28.33  &  28.42  &  28.54  \\
\tableline
        \multicolumn{5}{c}{mean (restricted):}  &  28.31  &  28.44  &  28.63  &  28.31  &  28.36  &  28.44  \\
        \multicolumn{5}{c}{}                    &         &     15  &   0.84  &         &     15  &   0.90  \\
\multicolumn{5}{c}{mean ($2\sigma$ - adopted):} &  28.34  &  28.41  &  28.51  &  28.34  &  28.35  &  28.37  \\
        \multicolumn{5}{c}{}                    &         &     13  &   0.08  &         &     13  &   0.10  \\
\enddata                          
\end{deluxetable}

\subsection{The Correction for Absorption}
\label{sec:distance:absorption}
The true distance moduli $\mu^{0}$(Gal) and $\mu^{0}$(LMC) of the
individual Cepheids, i.e.\ after correction for absorption, are given
by 
\begin{equation}
   \mu^{0} \, = \, \frac{{\cal R}_V}{{\cal R}_V - {\cal R}_I} \, \mu_I
   \, - \, \frac{{\cal R}_I}{{\cal R}_V  - {\cal R}_I} \, \mu_V.
\label{eq:mu_R}
\end{equation}
The absorption ratios with respect to $E(B\!-\!V)$ have
been determined for Galactic Cepheids in \citeauthor{Sandage:etal:04}
to be ${\cal R}_{V}=3.23$ and ${\cal R}_{I}=1.95$. 
(The corresponding numbers for $E(V\!-\!I)$ become then 
${\cal R}_{V,V-I}=2.52$ and ${\cal R}_{I,V-I}=1.52$).
It is assumed that the same absorption law holds also for
extragalactic Cepheids. In that case it follows 
\begin{equation}
   \mu^{0} = 2.52\mu_I - 1.52\mu_V. 
\label{eq:mu_0}
\end{equation}
The true distance moduli $\mu^{0}$(Gal) and $\mu^{0}$(LMC) of
individual Cepheids are listed in Table~\ref{tab:cep:complete},
column~7 and 10. 
Equation~(\ref{eq:mu_0}) is not strictly correct because it treats all
deviations from the ridge line of the P-L relation as if absorption
were the only reason for such deviations. 
However, the intrinsic half-width of the instability strip of
$\Delta(V\!-\!I)=0.13$ (\citeauthor{Sandage:etal:04}, Fig.~8) 
causes a Cepheid of fixed period, but at the red (blue) boundary of
the strip to be fainter (brighter) by $\Delta
M_{V}=\beta_{V,V\!-\!I}\times0.13$ and  $\Delta
M_{I}=\beta_{I,V\!-\!I}\times0.13$, where $\beta$ is the slope 
of the constant-period lines. For LMC-like galaxies
$\beta_{V,V\!-\!I}=2.43$ and $\beta_{I,V\!-\!I}=1.43$
(\citeauthor{Sandage:etal:04}, equations 29\,\&\,44). 
Hence a Cepheid is at the red (blue) edge fainter (brighter) by
$\Delta M_{V}=0.32$ and $\Delta M_{I}=0.19$ than a Cepheid with the
same period  lying at the ridge line of the P-L relation. 
These absolute-magnitude offsets at $P=\mbox{const}$ cause
coincidentally through equation~(\ref{eq:mu_0}) a distance error of
only $\Delta\mu^{0}=\mp0.008$.  
The consequence of the finite width of the instability strip is shown
by the true envelope lines to the central $\mu_{V}$
vs. $\mu_{V}-\mu_{I}$ diagnostic diagram in, for example, Figure~11 of 
\citet{Saha:etal:96a}. 

     The situation is less favorable for Cepheids with a P-L relation
like Galactic Cepheids, whose constant-period lines have slopes of
$\beta_{V,V\!-\!I}=0.66$ and $\beta_{I,V\!-\!I}=-0.34$
(\citeauthor{Sandage:etal:04}, \S~4.2.1). This makes 
Cepheids at the red edge of the instability strip fainter by $\Delta
M_{V}=0.09$ and {\em brighter\/} by $\Delta M_{I}=-0.04$ and decreases
the distance modulus by $\Delta\mu^{0}=0.24$ through 
equation~(\ref{eq:mu_0}). The error is statistically compensated by
Cepheids at the blue edge, provided that absorption is treated
strictly algebraically, even if the measurements report $\mu_I>\mu_V$
which is, physically speaking, unrealistic.

     The possibility that the Galactic reddening factors 
${\cal R}_{V,V-I}$ and ${\cal R}_{I,V-I}$ vary from galaxy to galaxy,
affecting thus equation~(\ref{eq:mu_0}), has only a minor effect on the
individual galaxy distances because the mean reddening of the Cepheids
of the galaxies in Table~\ref{tab:cep:additional} amounts to only
$\langle E(V\!-\!I)\rangle=0.105$. Even a drastic change of ${\cal R}$
of $\pm0.5$ introduces therefore an average change of the distance
moduli of only $0.05\mag$ (2.5\% in distance).

\subsection{The Mean Cepheid Distances}
\label{sec:distance:mean}
The mean apparent distance moduli $\langle\mu_V\rangle$ and
$\langle\mu_I\rangle$ as well as the mean true distance modulus
$\langle\mu^0\rangle$ (i.e.\ corrected for absorption) from
equation~(\ref{eq:mu_0}) are given in Table~\ref{tab:cep:complete} at
the bottom of the tabulation for each galaxy, considering all Cepheids
listed. In an additional line the same values are shown for a
restricted Cepheid sample, where Cepheids with $P<10\;$days were
omitted. In some galaxies all known Cepheids have $P>10\;$days, but at
the shortest available periods they are still suspiciously bright due
to the selection effect described by \citet{Sandage:88}. 
In these cases one to three, in rare cases up to five Cepheids with the
shortest periods were also excluded. This additional period cut is
efficient in eliminating biased Cepheids without introducing any new
distance bias because period cuts, while reducing the sample size, are
in principle statistically permissible with no effect on the mean
distance. In the present case the anti-bias cut has a very modest
effect on the adopted distances of the relevant galaxies increasing
them by less than $0.02\mag$ on average. The excluded Cepheids are
marked with an asterisk in Table~\ref{tab:cep:complete}.
The number of remaining Cepheids and the dispersion in $\mu^0$ is
given in an auxiliary line. The next line gives the adopted distance
moduli $\langle\mu_V\rangle$, $\langle\mu_I\rangle$, and
$\langle\mu^0\rangle$ after $2\sigma$-clipping. The Cepheids excluded
through clipping are marked with two asterisks. The last auxiliary
line shows the final number of Cepheids considered as well as the mean
error of $\langle\mu^0\rangle$. 

     In some cases (NGC\,4496A, 4639, 5253, IC\,4182) the apparent
modulus $\langle\mu_I\rangle$ is larger than $\langle\mu_V\rangle$ due
to random observational errors. This implies negative, and hence
unphysical absorption. 
However, like in \S~\ref{sec:distance:individual} for individual Cepheids, 
the negative absorption values should be formally carried through to
give the most probable mean distance of an {\em ensemble\/} of galaxies. 
Negative absorption values, which yield too large distances, are to
compensate too small distances obtained from an overestimate of the
absorption. An overestimate is not as conspicuous as an underestimate,
but occurs with the same likelihood as the latter due to random
observational errors. The proper allowance for negative absorption
increases the adopted distance $\langle\mu^0\rangle$ of the four
galaxies involved by $0.06-0.09\mag$. The {\em mean\/} distance of the
8 galaxies in Table~\ref{tab:cep:correct} (below) is increased by
$0.04\mag$. 

     The true modulus $\langle\mu^0\rangle$ can be determined in two
ways, either \\
1) one averages the $\mu_V$ and $\mu_I$
and inserts the means $\langle\mu_V\rangle$ and $\langle\mu_I\rangle$
into equation~(\ref{eq:mu_0}), or \\ 
2) one calculates the individual $\mu^0$ from equation~(\ref{eq:mu_0})
and averages over $\mu^0$ to obtain $\langle\mu^0\rangle$. \\
In either case the true modulus $\langle\mu^0\rangle$ is the same, but
the apparent mean error $\epsilon(\langle\mu^0\rangle)$ becomes about
three times larger in case 1) than in case 2). 
The reason is that the individual $\mu_V$ and $\mu_I$ are correlated,
mainly because of the intrinsic width of the instability strip
(Cepheids bright in $V$ are also bright in $I$), and because of
individual absorption. The statistics of the correlation effect on
$\epsilon(\langle\mu^0\rangle)$ has been worked out by 
\citet{Ngeow:Kanbur:05}; the smaller error
$\epsilon(\langle\mu^0\rangle)$ from route 2) is more realistic.

\subsection{Comparison of the Distance Moduli from the Galactic and
            LMC P-L Relations} 
\label{sec:distance:comparison}
\begin{figure*}[t]
   \epsscale{0.65}
   \plotone{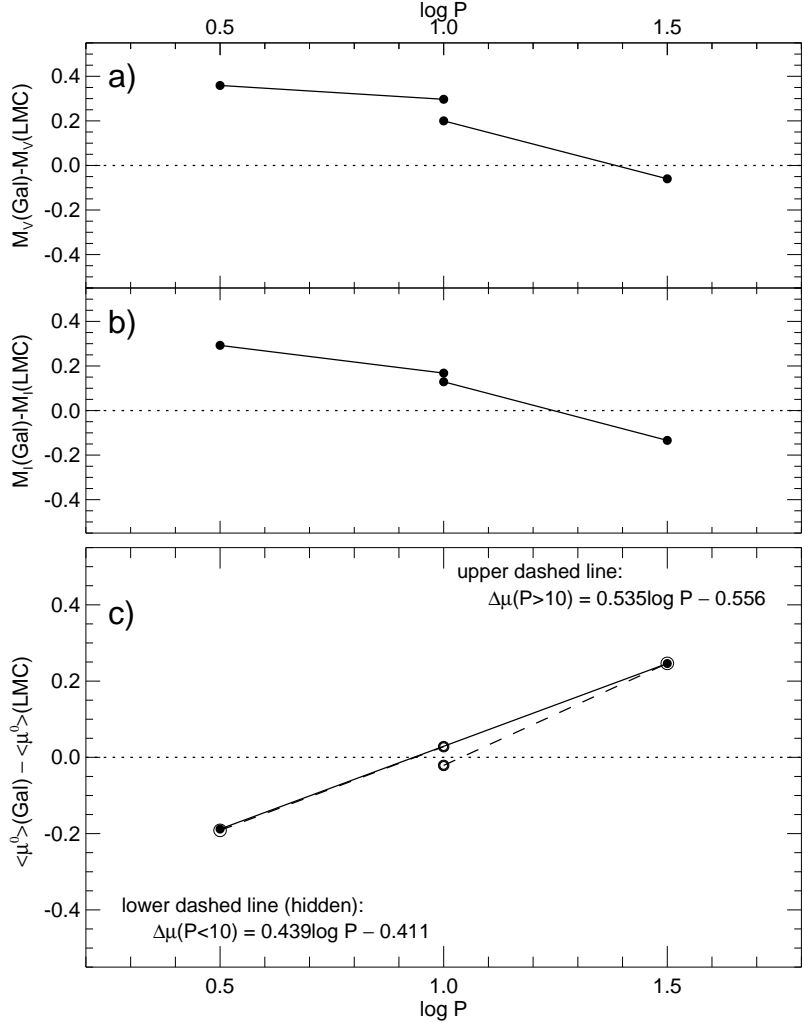}
   \caption{a) The difference in $M_{V}$ of Galactic and LMC
   Cepheids as a function of $\log P$. b) same for $M_{I}$. c) The
   distance differences $\Delta\mu$ from equation~(\ref{eq:mu_0}) of
   Galactic and LMC Cepheids of given apparent magnitude $m_{V}$ and
   $m_{I}$ as a function of $\log P$ (dashed lines, the lower one is
   hidden). The adopted, approximate relation in
   equation~(\ref{eq:Dmu_0}) is shown as a full line. Note that near
   $\log P=1$ $\Delta\mu$ becomes small in any case.}
\label{fig:distance:diff}
\end{figure*}

The different P-L relations of the Galaxy and LMC, derived in
\citeauthor{Sandage:etal:04}, cause Cepheids of given period to have
different luminosities. The differences of $M$(Gal) and $M$(LMC) in
$V$ and $I$ as a function of $\log P$ are shown in
Figure~\ref{fig:distance:diff}a,b. Inserting the different $M_{V}$ and
$M_{I}$ into equation~(\ref{eq:mu_0}) yields also different true
distance moduli $\langle\mu^0\rangle$ for Cepheids with the same
period. The resulting differences $\Delta\mu^{0}$ as a function of
$\log P$ are shown in Figure~\ref{fig:distance:diff}c. The relation
has a discontinuity at $\log P=1$, because it was not attempted in
\citeauthor{Sandage:etal:04} to join the LMC P-L relations for
Cepheids with $\log P\lessgtr1$ by force. Yet here the two segments of
the $\Delta\mu$-$\log P$ relation for short- and long-period Cepheids
are well approximated by a single straight line (to within $0.03\mag$)
of the form
\begin{equation}
   \Delta\mu = \mu(\mbox{Gal}) - \mu(\mbox{LMC}) = 0.434\log P - 0.405, 
\label{eq:Dmu_0}
\end{equation}
which we adopt in the following. It can be seen that LMC-like Cepheids
yield larger distances at short periods (up to $\log P=0.93$), at
longer periods Galaxy-like Cepheids yield larger distances.

     The distance coincidence at $\log P=0.93$ ($P=8.5\;$days) has no
{\em physical\/} significance. Neither the absolute magnitudes $M_{V}$
nor $M_{I}$ of Galactic and LMC Cepheids are the same at that
period. (Equal $M_{V}$ occurs at $\log P=1.38$, equal $M_{I}$ at $\log
P=1.25$). The agreement in distance at $\log P=0.93$ depends on the
slopes and zero-points of the $V$ and $I$ P-L relations of the Galaxy
and LMC (equations~\ref{eq:PLgal:V}\,-\,\ref{eq:PLlmc:I:lt1}) and
holds through equation~(\ref{eq:mu_0}) for unreddened
Cepheids which follow exactly either of these P-L relations. It holds
also for Cepheids with other metallicities, provided that the slopes
and zero-points of the P-L relations vary linearly with metallicity
between the Galaxy and LMC.
However, for Cepheids which follow different P-L relations in $V$
and/or $I$ for intrinsic or observational reasons the coincidence
point at $\log P=0.93$ has no significance at all. This becomes
important in \S~\ref{sec:analysis} where it is shown that the galaxy
distances derived from actual Cepheid data do depend on period in
general. This is due to slope and zero-point variations of the 
{\em observed\/} P-L relations. {\em In these cases there is no reason
  to give the distance at $\log P=0.93$ any preference.} 
-- The coincidence period of $\log P=0.93$ is also shifted in case of
reddened Cepheids because equation~(\ref{eq:mu_0}) yields somewhat
different color excesses $E(V\!-\!I)$ depending whether one uses the
Galactic or LMC P-L relations (see \S~\ref{sec:analysis}).

\section{CEPHEID DISTANCES AND METALLICITY CORRECTIONS}
\label{sec:metal}
%
\subsection{The Determination of the Metallicity Correction}
\label{sec:metal:det}
It was shown in \citeauthor{Sandage:etal:04} that Cepheids in the
Galaxy and in LMC occupy different places in the $\log L$-$\log T_{\rm e}$ 
plane and that they define different slopes. The lower-metallicity
Cepheids in LMC are 80-350$\;$K warmer at constant luminosity,
depending on period.  
It was also shown that this forces the period-color (P-C) relations
and hence the P-L relations to be different in the two galaxies. 
This effect is beyond the metal-dependent blanketing effect, which
affects the P-C and hence also the P-L relations.
Although it cannot be proved at present that the different
positions in the $\log L$-$\log T_{\rm e}$ diagram are as well the
result of the different metallicities, it is the most plausible
assumption. It is therefore assumed in the following that metallicity
is indeed the primary parameter that determines the shape and slope of
the P-L relation. 

     With these precepts the distance modulus differences $\Delta\mu$,
shown in Figure~\ref{fig:distance:diff}c and expressed in
equation~(\ref{eq:Dmu_0}), are the result of the metallicity difference
between the Galaxy and LMC, the implication being that the metallicity
corrections are a strong function of the mean period $\langle\log
P\rangle$ of the Cepheids. Galactic, high-metallicity Cepheids give
shorter distances at short periods, yet larger distances at long
periods, the transition being at $\langle\log P\rangle\approx0.93$. 
The exact position of the crossover point is somewhat dependent on the
adopted distances of the Galactic Cepheids and of LMC 
(see \S~\ref{sec:zeropoint}).   
Moreover, we assume that the metallicity-dependent distance correction
$\Delta\mu_{Z}$ is -- at least over a reasonable range of abundances
-- a linear function of the metallicity itself.

\begin{figure}[t]
   \epsscale{0.65}
   \plotone{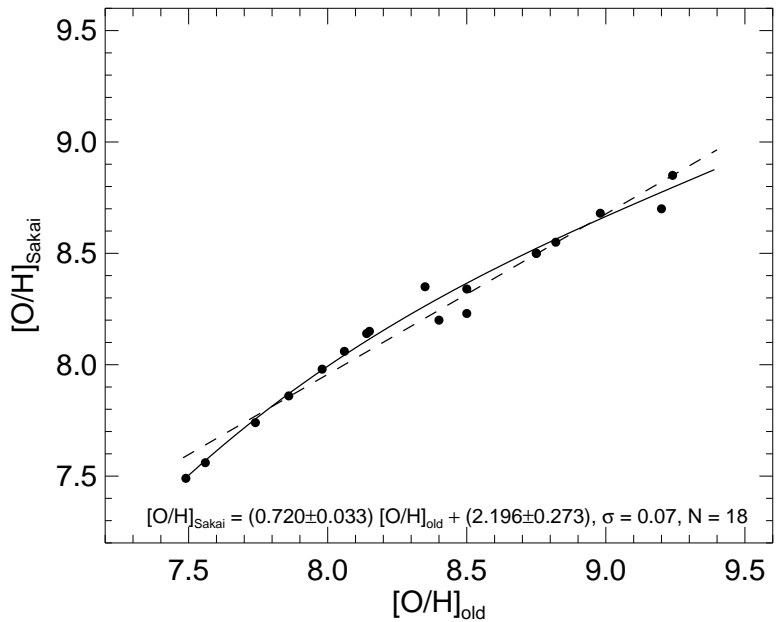}
   \caption{The relation between new, $T_{\rm e}$-based values of
   $[\frac{O}{H}]_{\rm Sakai}$ and the old values
   $[\frac{O}{H}]_{\rm old}$, both given by
   \citet{Sakai:etal:04}. The full line is the adopted polynomial
   fit; the dashed line a linear fit.}
\label{fig:OH:new}
\end{figure}

     As a measure of the metallicity the oxygen-to-hydrogen ratio has
been adopted. \citet{Kennicutt:etal:98} have determined
the ratios $[\frac{O}{H}]=12+\log(\frac{O}{H})$ and their radial
gradients over the face of the galaxy for many galaxies. 
\citet{Ferrarese:etal:00} have interpolated these values according to
the average position of the Cepheids in their parent galaxies and
added some galaxies in their list. Some additional values of
$[\frac{O}{H}]$ have been taken from the literature
\citep[e.g.][]{Riess:etal:05}. The 1998 scale has recently been 
revised by $T_{\rm e}$-based $[\frac{O}{H}]$ values which are
significantly smaller for the highest metallicities
\citep{Sakai:etal:04}. These authors give old and new $[\frac{O}{H}]$
values for 18 galaxies. Their data are plotted in
Figure~\ref{fig:OH:new} together with a polynomial regression. The 
regression has been used to convert all $[\frac{O}{H}]_{\rm old}$
values into new values $[\frac{O}{H}]_{\rm Sakai}$ of the Sakai
et~al. system. {\em All metallicities quoted in this paper are in the
  new system.}  

     Concurrently with the compression of the $[\frac{O}{H}]$ scale
the classical solar value of $[\frac{O}{H}]_{\sun}=8.89$
\citep{Grevesse:Anders:89} has been lowered to 
$[\frac{O}{H}]_{\sun}=8.7$ \citep{AllendePrieto:etal:01,Holweger:01}. 
The mean oxygen abundance of 68 Galactic Cepheids is $0.08$ sub-solar
\citep{Andrievsky:etal:02}. We therefore adopt for the Cepheids which
define the Galactic P-L relation $[\frac{O}{H}]=8.6$.
The value for LMC in the new (Sakai et~al.) scale is
$[\frac{O}{H}]_{\rm LMC}=8.34$ \citep{Sakai:etal:04}. 

     The implication here that Galactic Cepheids have sub-solar
[O/H] seems in contradiction with
\S~\ref{sec:distance:individual}, where it was argued that the
calibrating Cepheids in clusters have solar [Fe/H] on
average. Solar [Fe/H] holds also for the Cepheids with BBW
distances. Nine of them have a mean $[\mbox{Fe/H}]=-0.02\pm0.03$, and
they as well lie close to the solar circle ($\langle
R\rangle=8.0\;$kpc). However, the puzzling discrepancy is just what is
expected from present abundance determinations. 
\citet{Kovtyukh:etal:05} find for Cepheids on the solar circle a mean
value of $[\mbox{O/Fe}]=-0.1$.

\begin{figure*}[t]
   \epsscale{1.1}
   \plottwo{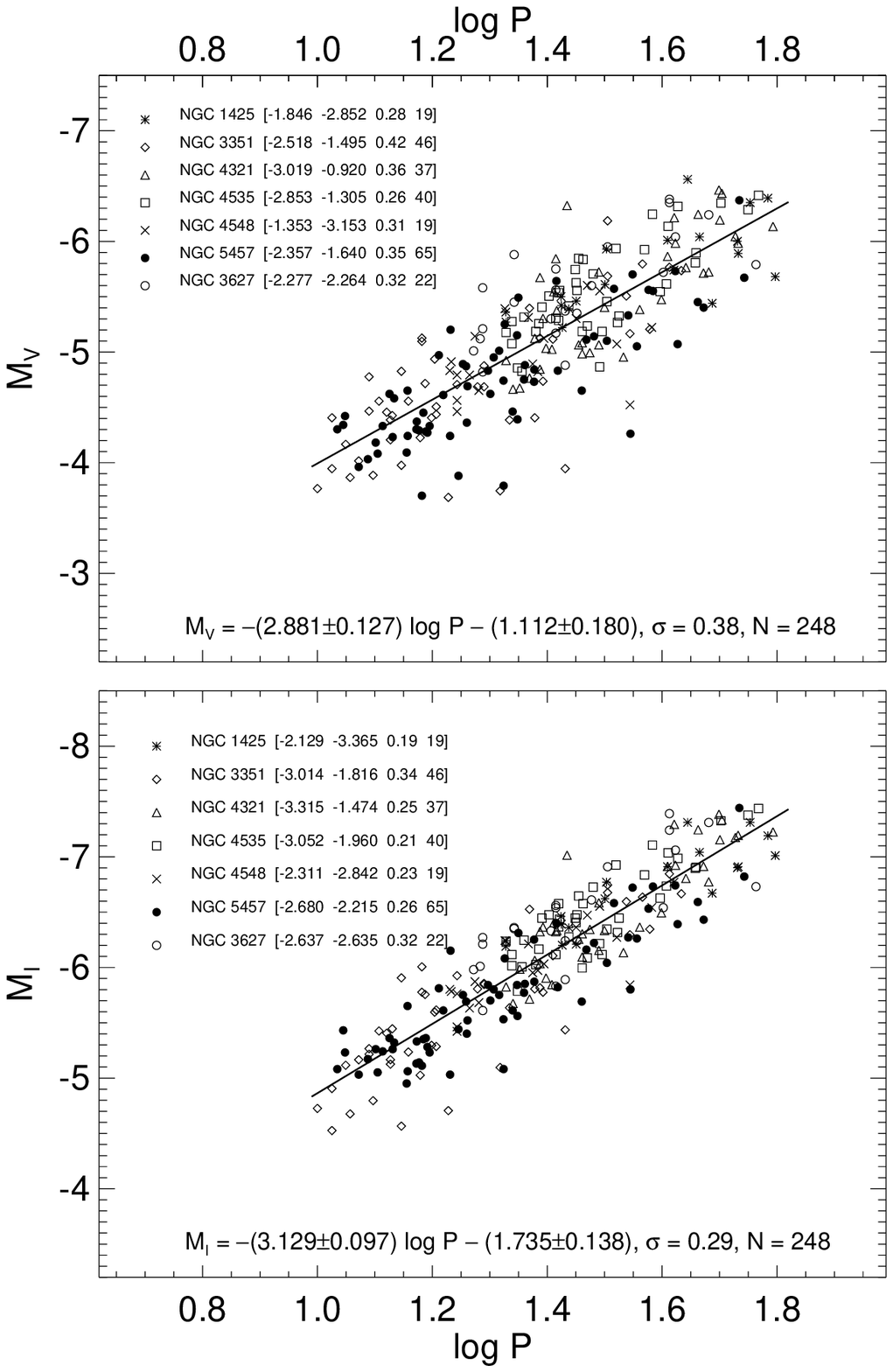}{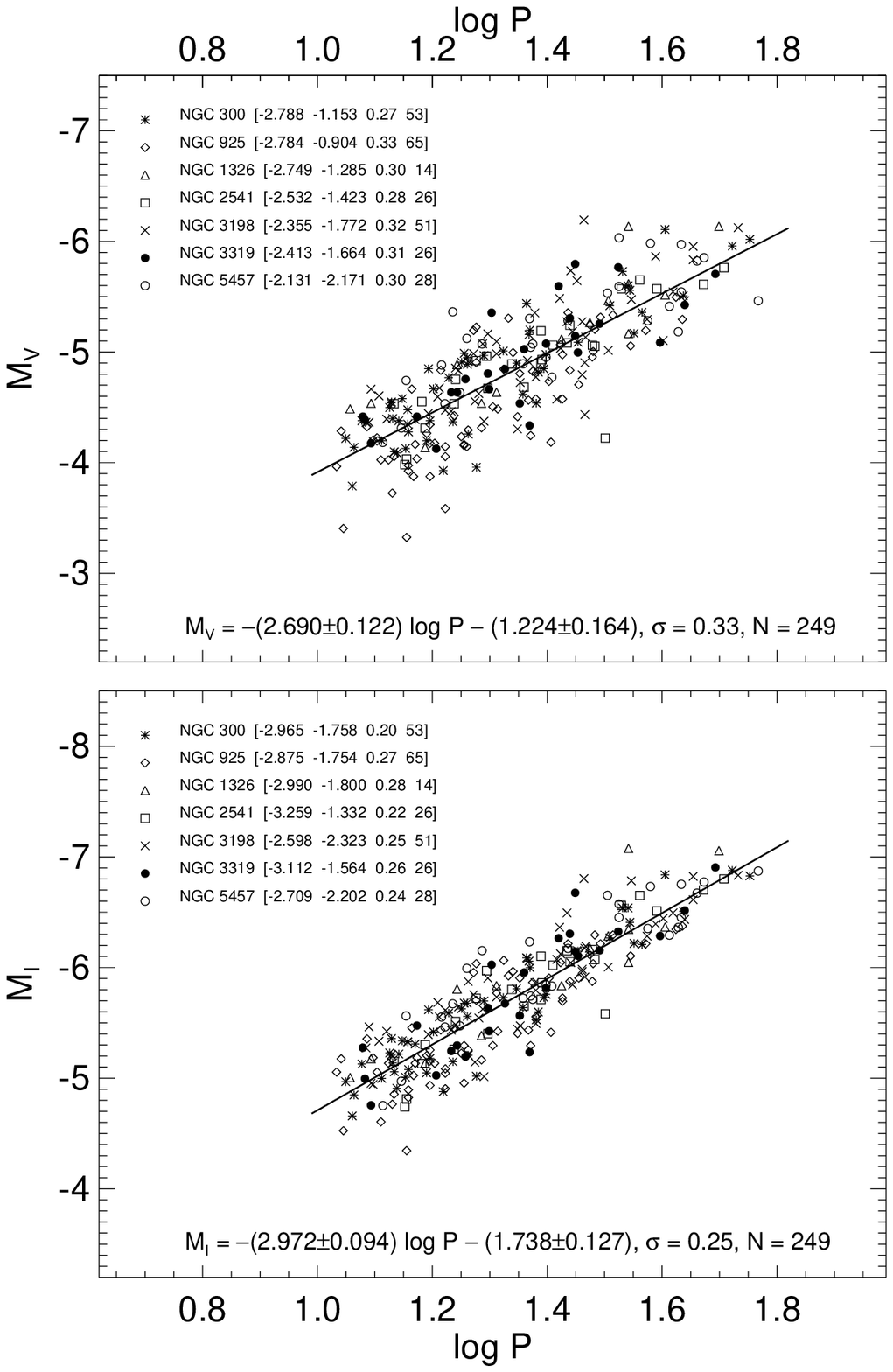}
   \caption{{\em Left panel:} The composite P-L relation in $V$ and $I$
   of the Cepheids in 7 {\em metal-rich\/} galaxies. 
   {\em Right panel:} The composite P-L relation in $V$ and $I$ of the
   Cepheids in 7 {\em metal-poor\/} galaxies.} 
\label{fig:rpMetal}
\end{figure*}

     The conclusion, that the {\em slope\/} of the P-L relation
increases with the metallicity, finds support from external data. 
In Figure~\ref{fig:rpMetal} the absolute magnitudes in $V$ and $I$,
based on the respective adopted distances $\mu^{0}_{Z}$ from
Table~\ref{tab:cep:additional}, are plotted against $\log P$ for all
Cepheids in the 7 most metal-rich ($[\frac{O}{H}]_{\rm Sakai}>8.65$)
and in 7 metal-poor galaxies with $8.20<[\frac{O}{H}]_{\rm Sakai}<8.45$.
Galaxies with less than 15 Cepheids are not shown. 
The slope of the metal-rich Cepheids is somewhat shallower than the
Galactic P-L relation (cf.\ Fig.~\ref{fig:rpMetal}), although it
should be somewhat steeper because their mean metallicity is higher 
than that of Galactic Cepheids ($\langle[\frac{O}{H}]\rangle=8.76$ 
compared to 8.6). The mean slope of the metal-poor Cepheids is
marginally steeper than that of the LMC Cepheids although their mean
metallicity is nearly identical. 
Obviously the expected dependencies do not work out
exactly, but it must be considered that the test is very exacting on
the data, because the interval in $\log P$ of the Cepheids
in most galaxies is even narrower than the narrow interval considered,
and relative distance errors affect the slope determination.
Moreover the observational magnitude errors of the
individual Cepheids are large. Yet the main point here is that the
slope {\em difference\/} between metal-poor and metal-rich Cepheids is
at least suggestive at a level of $\sim1\sigma$ ($0.19\pm0.16$ in $V$
and $0.16\pm0.13$ in $I$), the latter defining a {\em steeper\/} P-L
relation. 

     From the present precepts it follows that the distance difference
$\Delta\mu=\mu(\mbox{Gal})-\mu(\mbox{LMC})$ in
equation~(\ref{eq:Dmu_0}) is caused by a metallicity difference of
$[\frac{O}{H}]_{\rm Gal}-[\frac{O}{H}]_{\rm LMC}=8.6-8.34=0.26$. By
linear extrapolation the period-dependent distance correction for
$\Delta[\frac{O}{H}]$ becomes 
\begin{equation}
   \Delta\mu_{Z} = 1.67(\log P - 0.933)([\frac{O}{H}]-A), 
   \label{eq:OH:correct}
\end{equation}
(note that 0.434 in equation~(\ref{eq:Dmu_0}) divided by
$0.26=1.67$). $A=8.60$ for the correction of $\mu(\mbox{Gal})$ and
$A=8.34$ for $\mu(\mbox{LMC})$. In case of multiple Cepheids per 
galaxy, $\log P$ may be replaced by the mean value 
$\langle\log P\rangle$. 

     Equation~(\ref{eq:OH:correct}) has been somewhat extrapolated to
hold for the range $8.2<[\frac{O}{H}]<8.7$. For the few galaxies
considered here, whose values lie below or above this range, the
limiting values of 8.2 and 8.7, respectively, have been adopted. The
lower limit is chosen, because it is known that the P-L relations of
SMC with $[\frac{O}{H}]=7.98$ lie {\em between\/} those of LMC and the
Galaxy (\citeauthor{Tammann:etal:03}), thus providing a warning
against exaggerated extrapolation. An extrapolation to values above
$[\frac{O}{H}]=8.7$ is questionable because none of the even
metal-richest galaxies has steeper P-L relations in $V$ and $I$ (as
poorly as the slopes may be determined) than the Galaxy with an
adopted value of $[\frac{O}{H}]=8.62$ 
(see Figure~\ref{fig:rpMetal}).

     The galaxies considered in Table~\ref{tab:cep:correct} and in the
Appendix require from equation~(\ref{eq:OH:correct}), allowing for
their respective values of $[\frac{O}{H}]$ and $\langle\log P\rangle$,
modulus correction of $\Delta\mu_{Z}=-0.26$ to $+0.10\mag$ for
$\mu$(Gal), and $\Delta\mu_{Z}=-0.09$ to $+0.36\mag$ for $\mu$(LMC).

\begin{deluxetable}{lcccccccccc}
\tablewidth{0pt}
\tabletypesize{\scriptsize}
\tablecaption{Metallicity-corrected distance moduli of the 8 program
  galaxies.\label{tab:cep:correct}} 
\tablehead{
 \colhead{Galaxy} & 
 \colhead{$[\frac{O}{H}]_{\rm Sakai}$} &
 \colhead{$\mu^{0}(\mbox{Gal})$} &
 \colhead{$\mu^{0}(\mbox{LMC})$} &
 \colhead{$\langle\log P\rangle$} &
 \colhead{$\Delta\mu_{Z}(\mbox{Gal})$} &
 \colhead{$\Delta\mu_{Z}(\mbox{LMC})$} &
 \colhead{$\mu^{0}_{Z}(\mbox{Gal})$} &
 \colhead{$\mu^{0}_{Z}(\mbox{LMC})$} &
 \colhead{$\mu^{0}_{Z}$} &
 \colhead{$\epsilon(\mu^{0}_{Z})$} \\
 \colhead{(1)} & \colhead{(2)}  &
 \colhead{(3)} & \colhead{(4)}  &
 \colhead{(5)} & \colhead{(6)}  &
 \colhead{(7)} & \colhead{(8)}  &
 \colhead{(9)} & \colhead{(10)} &
 \colhead{(11)}
} 
\startdata
NGC\,3627  & 8.80 & 30.41 & 30.19 & 1.452 & $+$0.09 & $+$0.31 & 30.50 & 30.50 & 30.50 & 0.09 \\
NGC\,3982  & 8.52 & 31.94 & 31.69 & 1.502 & $-$0.07 & $+$0.17 & 31.87 & 31.87 & 31.87 & 0.15 \\
NGC\,4496A & 8.53 & 31.24 & 30.99 & 1.514 & $-$0.06 & $+$0.19 & 31.18 & 31.17 & 31.18 & 0.05 \\
NGC\,4527  & 8.52 & 30.84 & 30.59 & 1.498 & $-$0.07 & $+$0.17 & 30.76 & 30.77 & 30.76 & 0.20 \\
NGC\,4536  & 8.58 & 31.26 & 30.98 & 1.566 & $-$0.02 & $+$0.25 & 31.24 & 31.23 & 31.24 & 0.09 \\
NGC\,4639  & 8.67 & 32.13 & 31.86 & 1.552 & $+$0.07 & $+$0.34 & 32.20 & 32.19 & 32.20 & 0.09 \\
NGC\,5253  & 8.15 & 28.11 & 28.09 & 1.029 & $-$0.06 & $-$0.02 & 28.04 & 28.06 & 28.05 & 0.12 \\
IC\,4182   & 8.20 & 28.51 & 28.32 & 1.387 & $-$0.30 & $-$0.11 & 28.20 & 28.22 & 28.21 & 0.08 \\
\enddata                          
\end{deluxetable}
     After application of equation~(\ref{eq:OH:correct}) the corrected
moduli $\mu_{Z}(\mbox{Gal})$ and $\mu_{Z}(\mbox{LMC})$ become
nearly identical by construction. The agreement does therefore not
provide an independent check of the zero-point of the distance
scale. On the contrary, the corrections $\Delta\mu_{Z}$ in
equation~(\ref{eq:OH:correct}) do depend on the adopted distances of
the Galactic Cepheids and of LMC (see below).

     The calculations of the metallicity-corrected distances
$\mu^{0}_{Z}$ of the eight program galaxies in
Table~\ref{tab:cep:complete} are shown
in Table~\ref{tab:cep:correct}.

     In Table~\ref{tab:cep:correct} column~2 gives the values of
$[\frac{O}{H}]$ in the new \citep{Sakai:etal:04} scale. 
Column~3-4 repeat the adopted moduli $\mu^{0}(\mbox{Gal})$ and
$\mu^{0}(\mbox{LMC})$ from Table~\ref{tab:cep:complete}. 
Column~5 gives the mean value $\langle\log P\rangle$ of the Cepheids
used for the solution. 
The metallicity correction for $\mu^{0}$(Gal) and $\mu^{0}$(LMC),
respectively, from equation~(\ref{eq:OH:correct}) are shown in
column~6 \& 7. 
The resulting, almost identical moduli are shown in column~8 \&
9. Column~10 gives the adopted, mean metal-corrected  modulus
$\mu^{0}_{Z}$; their random errors are in column 11.

\subsection{Tests of the Adopted Metallicity Correction}
\label{sec:metal:test}
To test the metallicity-corrected distances $\mu^{0}_{Z}$ in
Table~\ref{tab:cep:correct} additional Cepheid distances are
needed. In the Appendix we have applied the new P-L relations in
equations~(\ref{eq:PLgal:V}-\ref{eq:PLlmc:I:lt1}) to all available
Cepheids in galaxies with $[\frac{O}{H}]>8.2$ and have corrected their
distances by means of equation~(\ref{eq:OH:correct}). Among the 37
galaxies are 21 galaxies which have been observed with HST by other
authors. Their photometric zero-point is not necessarily the same as
derived in the present paper. Possible differences are, however,
negligible in the present context.

     A first test of the validity of the adopted metallicity
corrections is provided by M\,101 (NGC\,5457). While the inner,
metal-rich Cepheids of this galaxy yield uncorrected moduli which are
$0.3-0.4\mag$ smaller than from outer, metal-poor Cepheids, the
adopted, metallicity-corrected moduli agree within $0.02\mag$.

\begin{deluxetable}{lcccccccccc}
\tablewidth{0pt}
\tabletypesize{\scriptsize}
\tablecaption{TRGB distances and metallicity-corrected Cepheid 
distances.\label{tab:TRGB}}
\tablehead{
 \colhead{Galaxy} & 
 \colhead{$[\frac{O}{H}]_{\rm old}$} &
 \colhead{$[\frac{O}{H}]_{\rm Sakai}$} &
 \colhead{$\mu(\mbox{TRGB})$} &
 \colhead{$\mu^{0}_{Z}(\mbox{Cepheids})$} &
 \colhead{$\Delta\mu$} \\
 \colhead{(1)} & \colhead{(2)} &
 \colhead{(3)} & \colhead{(4)} &
 \colhead{(5)} & \colhead{(6)} 
} 
\startdata
NGC\,224   & 8.98 & 8.68 & 24.47$\pm$0.11 & 24.54 & $+$0.07 \\
NGC\,300   & 8.35 & 8.35 & 26.65$\pm$0.07 & 26.48 & $-$0.17 \\
NGC\,598   & 8.82 & 8.55 & 24.81$\pm$0.04 & 24.64 & $-$0.17 \\
NGC\,3031  & 8.75 & 8.50 & 28.03$\pm$0.12 & 27.80 & $-$0.23 \\
NGC\,3351  & 9.24 & 8.85 & 30.39$\pm$0.13 & 30.10 & $-$0.29 \\
NGC\,3621  & 8.75 & 8.50 & 29.36$\pm$0.11 & 29.30 & $-$0.06 \\
NGC\,5253  & 8.15 & 8.15 & 27.88$\pm$0.11 & 28.05 & $+$0.17 \\
NGC\,5457i & 9.20 & 8.70 & 29.42$\pm$0.11 & 29.16 & $-$0.26 \\
NGC\,5457o & 8.50 & 8.23 & 29.42$\pm$0.11 & 29.18 & $-$0.24 \\
NGC\,6822  & 8.14 & 8.14 & 23.37$\pm$0.07 & 23.31 & $-$0.06 \\
IC\,4182   & 8.40 & 8.20 & 28.25$\pm$0.06 & 28.21 & $-$0.04 \\
\enddata                          
\end{deluxetable}
\subsubsection{New Cepheid distances vs. TRGB distances}
\label{sec:metal:test:TRGB}
\begin{figure}[t]
   \epsscale{0.60}
   \plotone{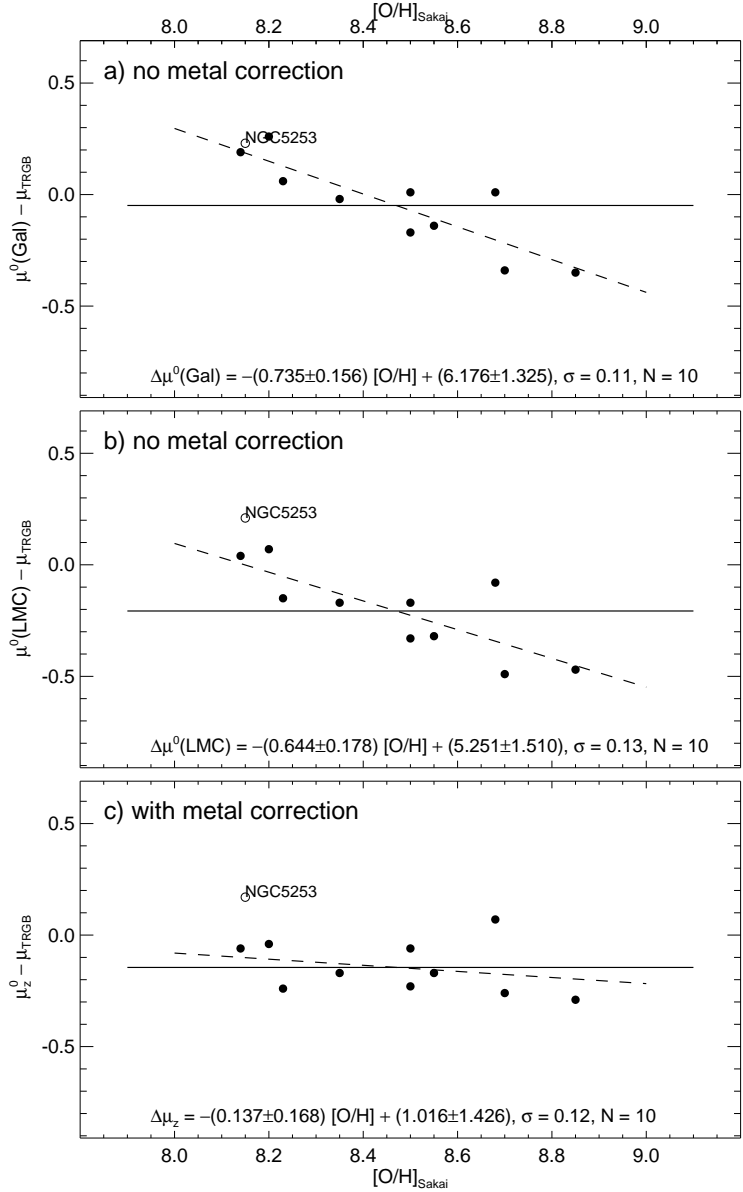}
   \caption{The difference of TRGB distances and Cepheid distances
   from Table~A1 as a function of the metallicity of the parent galaxy.  
   a) Using distances from the Galactic P-L relation; 
   b) using distances from the LMC P-L relation; 
   c) using metallicity-corrected Cepheid distances; the remaining
   slope is insignificant. The dashed lines are fits to the data; the
   horizontal lines show the {\em mean\/} distance difference. 
   NGC\,5253 is not considered because its distance is insensitive to
   metallicity due to the short period of its Cepheids (see
   equation~\ref{eq:OH:correct}). The small scatter may be noted.} 
\label{fig:TRGB}
\end{figure}

\citet{Sakai:etal:04} have published tip-of-the-red-giant branch
(TRGB) distances of nine galaxies 
(i.e. based on the brightest stars in what was earlier 
called the Baade sheet which he discussed in his 1944 resolution 
studies of M\,31 and its companions [\citealt{Baade:44a,Baade:44b}]).
They have also published $[\frac{O}{H}]$ values in the new system
of these nine galaxies, whose corrected Cepheid distances can be found
in the Appendix. The independent distances are compared in
Table~\ref{tab:TRGB} and plotted in Figure~\ref{fig:TRGB}.
It is obvious that after the metallicity correction is applied the
difference of the two sets of distance determinations show hardly any
dependence on metallicity.

     The systematic difference between the corrected Cepheid distances
and the TRGB distances is $0.12\pm0.04\mag$, the latter being 
{\em larger}, which is taken up again in \S~\ref{sec:zeropoint:add}.

\subsubsection{New Cepheid distances vs. SN\,Ia luminosities}
\label{sec:metal:test:SNeIa}
\begin{figure}[t]
   \epsscale{0.62}
   \plotone{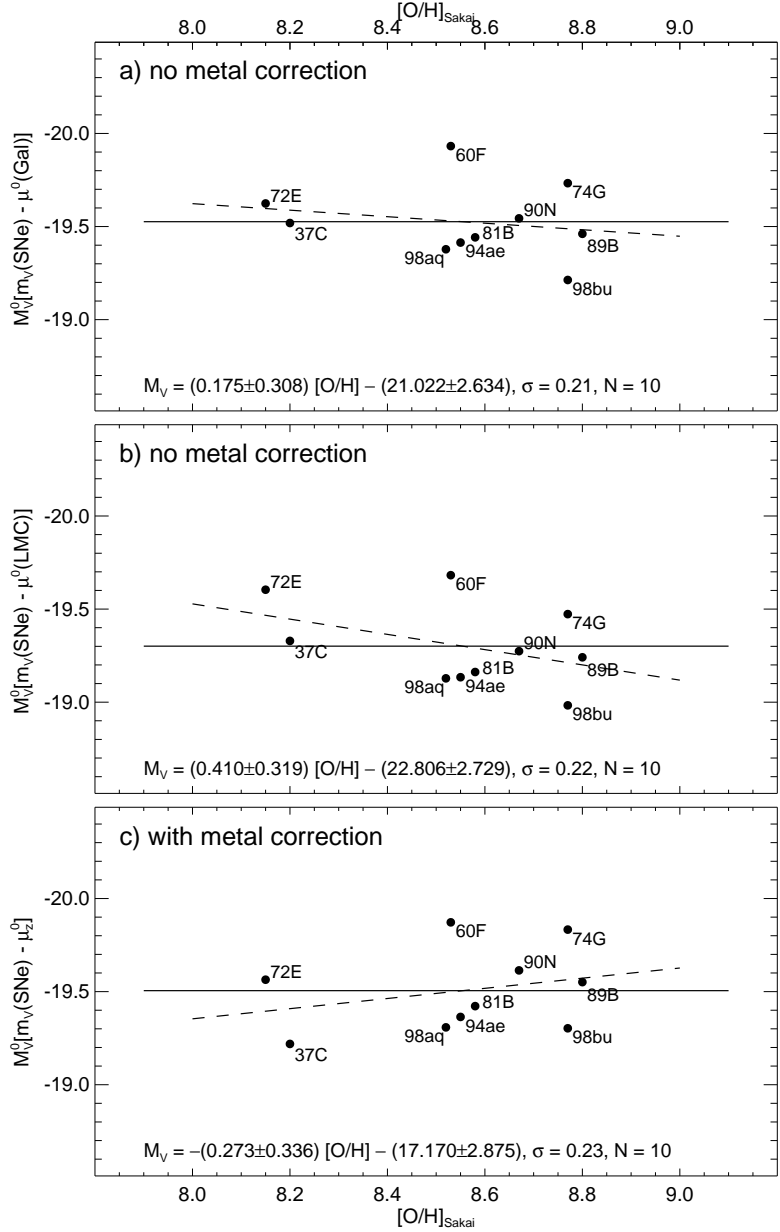}
   \caption{The absolute SNe\,Ia magnitude $M_{V}$ using the host
   galaxy distances from Table~A1 as a function of the metallicity of
   the parent galaxy. 
   a) Using distances from the Galactic P-L relation; 
   b) using distances from the LMC P-L relation; 
   c) using metallicity-corrected Cepheid distances; 
   the significant metallicity dependence in panel b) is removed. 
   The dashed lines are fits to the data; the horizontal lines show
   the {\em mean\/} absolute magnitude. The individual SNe\,Ia are
   identified.}  
\label{fig:SNe}
\end{figure}

Another test of the corrected Cepheid moduli $\mu^{0}_{Z}$ is provided
by SNe\,Ia taken as standard candles. Their luminosity depends on the
type of the host galaxy, and hence on the metallicity which, however,
is expected to be fully compensated by the normalization. If their
magnitudes are normalized to a fixed value of the decline rate $\Delta
m_{15}$, the corrected magnitudes $m^{\rm corr}_{V}$ exhibit a scatter
of only $\sigma_{V}=0.14$\mag, which suggest that their metallicity
dependence has been successfully compensated for
(\citeauthor{Reindl:etal:05}).  Therefore their absolute magnitudes
$M_{V}(\mbox{SNe\,Ia})=m^{\rm corr}_{V}-\mu^{0}_{Z}$ should show no
correlation with $[\frac{O}{H}]$. The data are shown in
Figure~\ref{fig:SNe} and show, if $\mu^{0}_{Z}$ is used,
indeed no significant slope with $[\frac{O}{H}]$.

\subsubsection{New Cepheid distances vs. velocity distances}
\label{sec:metal:test:velocity}
It may be noted that the remaining marginal dependencies on
$[\frac{O}{H}]$ in Figure~\ref{fig:TRGB}c \& \ref{fig:SNe}c have
opposite signs. 
Increasing the present metallicity corrections to better satisfy the
TRGB data would worsen the metal independence of the SN\,Ia luminosities. 

     It may also be noted in passing that, if we had assumed
$[\frac{O}{H}]_{\rm Gal}=8.7$ instead of 8.6 for the Galactic
Cepheids, the resulting dependencies in Figures~\ref{fig:TRGB} \&
\ref{fig:SNe} would be considerably steeper.

\begin{figure}[t]
   \epsscale{0.60}
   \plotone{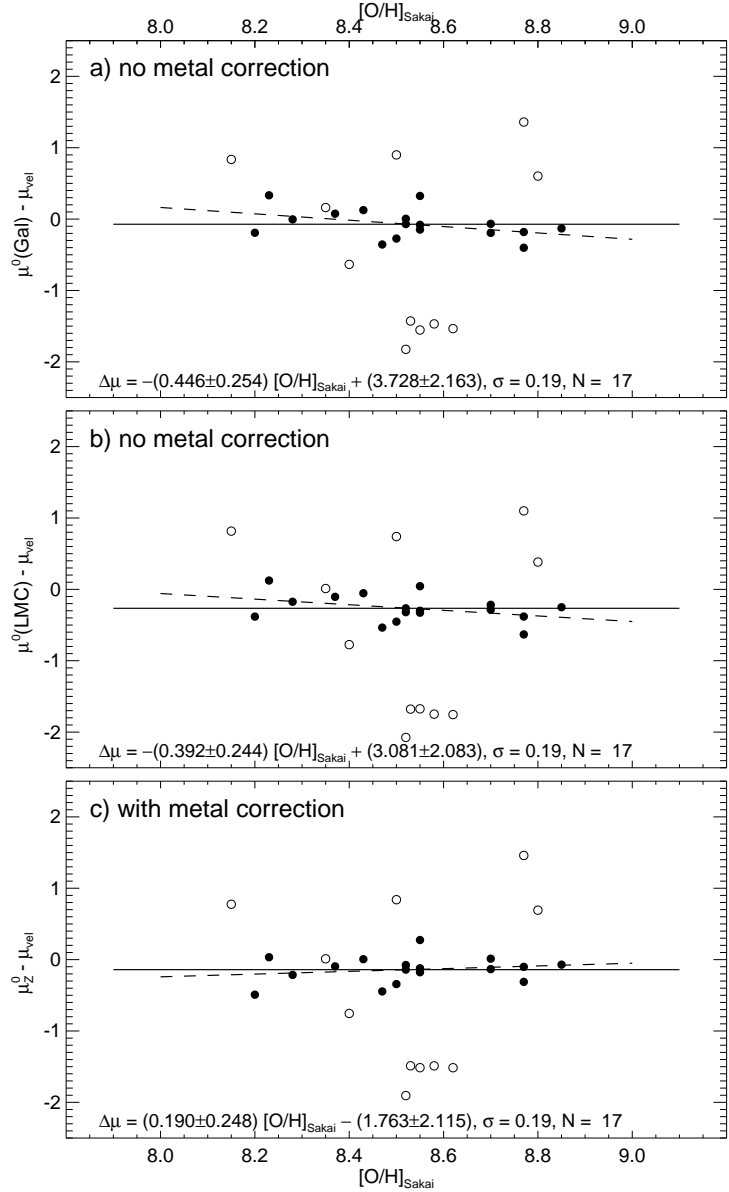}
   \caption{The difference of velocity distances and Cepheid distances
   from Table~\ref{tab:cep:additional} as a function of the
   metallicity of the parent galaxy. 
   a) Using distances from the Galactic P-L relation; 
   b) using distances from the LMC P-L relation; 
   c) using metallicity corrected Cepheid distances (see text). 
   The dashed lines are fits to the data; the horizontal lines show
   the {\em mean\/} distance difference. Members of the Virgo and
   Fornax clusters and galaxies within the noisy $25^{\circ}$ circle
   about the Virgo cluster center \citep{Tammann:etal:02} are shown as
   open symbols.}
\label{fig:velocity}
\end{figure}

   The velocity distances $\mu_{\rm vel}=5(\log\frac{v_{220}}{60})+25$
(with an arbitrary value of $H_{0}=60$) offer a final test for the
adopted Cepheid distances. The test is precarious because the
recession velocities $v_{220}$ (corrected for Virgo-centric infall) are
small and the influence of peculiar velocities on the velocity
distances is correspondingly large. To minimize their effect 
we consider only the 19 galaxies from Table~\ref{tab:cep:additional}
which have $\mu^{0}_{Z}>28.2$ and which lie outside of the
particularly noisy $25^{\circ}$ region around the Virgo cluster
\citep{Tammann:etal:02}; also Fornax cluster members are
excluded. While the differences $\mu^{0}({\rm Gal}) - 
\mu_{\rm vel}$ and $\mu^{0}({\rm LMC}) - \mu_{\rm vel}$ show a
significant dependence on the metallicity $[\frac{O}{H}]_{\rm Sakai}$
(Fig.~\ref{fig:velocity}a,b), the signal is effectively removed
through the adopted metallicity corrections (Fig.~\ref{fig:velocity}c). 

\subsubsection{Comparison with the metallicity corrections of 
               \citet{Sakai:etal:04}}
\label{sec:metal:test:sakai}
An alternative way to estimate the metallicity dependence of Cepheid
distances is to compare them with independent distance determinations
of galaxies. \citet{Kennicutt:etal:98} and \citet{Sakai:etal:04} have
used TRGB distances in the $I$ band which are assumed to be free of
metallicity effects. The number of available galaxies with both kinds
of determinations is too small to study the metallicity dependence of
Cepheids as a function of period. The comparison therefore provides only
an average correction for the mean period of all Cepheids in the
galaxies considered.

     \citet{Sakai:etal:04} compared the Cepheid and TRGB distances of
17 galaxies. Their Cepheid distances were based on the P-L relations
in $V$ and $I$ by M/F (among others). Using the old $[\frac{O}{H}]$
scale they found a mean correction of
$\Delta\mu_{Z}=(-0.20\pm0.09)\Delta[\frac{O}{H}]_{\rm old}$. This value
may be too low for Cepheids with $P>10\;$days as considered here,
because seven of their galaxies have $\langle\log P\rangle<0.93$, in
which cases we believe the metallicity correction to be negative
(\S~\ref{sec:metal:det}). 
Repeating their determination with only the galaxies with $\langle\log
P\rangle>0.93$ their data yield indeed a steeper metal dependence of 
$\Delta\mu_{Z}=-(0.27\pm0.11)\Delta[\frac{O}{H}]_{\rm old}$, or if
based on the new $[\frac{O}{H}]$ scale, 
$\Delta\mu_{Z}=-(0.43\pm0.18)\Delta[\frac{O}{H}]_{\rm Sakai}$.

     In order to compare the present metallicity corrections with
those of \citet{Sakai:etal:04} we have calculated the $\mu^{0}$(M/F)
for all galaxies in Table~\ref{tab:cep:additional} using the same
Cepheids as there and the P-L relations of M/F. (The zero-point is
taken at $(m-M)^{0}_{\rm LMC}=18.54$ which, however, is irrelevant for
the slope of the metallicity corrections). The results are shown in
Table~\ref{tab:07}, which is organized as follows:\\
Column~1: name of galaxy\\
Column~2 and 3: the $[\frac{O}{H}]$ abundance ratios in the old and
Sakai scale\\
Column~4: the metallicity-corrected Cepheid modulus as adopted here
(taken from Table~\ref{tab:cep:additional})\\
Column~5: the distance modulus $\mu^{0}$(M/F) \\
Column~6: the metallicity correction from
Equation~(\ref{eq:metal})\\
Column~7: the metallicity-corrected distance modulus
$\mu^{0}_{Z}$(M/F)\\
Column~8: The modulus difference between column~4 and 6.
\begin{deluxetable}{lcccccccccc}
\tablewidth{0pt}
\tabletypesize{\scriptsize}
\tablecaption{Cepheid distances from the P-L relations in $V$ and $I$
  of M/F and comparison with the distances adopted in the
  Appendix.\label{tab:07}} 
\tablehead{
 \colhead{Galaxy} & 
 \colhead{$[\frac{O}{H}]_{\rm old}$} &
 \colhead{$[\frac{O}{H}]_{\rm Sakai}$} &
 \colhead{$\mu^{0}_{Z}$} &
 \colhead{$\mu^{0}$(M/F)} &
 \colhead{$\Delta\mu^{0}$(M/F)} &
 \colhead{$\mu^{0}_{Z}$(M/F)} &
 \colhead{$\Delta\mu_{Z}$} \\
 \colhead{(1)} & \colhead{(2)} &
 \colhead{(3)} & \colhead{(4)} &
 \colhead{(5)} & \colhead{(6)} &
 \colhead{(7)} & \colhead{(8)} 
} 
\startdata
NGC224   & 8.98 & 8.68 & 24.54 & 24.45 & $-$0.12 & 24.57 & $-$0.03 \\ 
NGC300   & 8.35 & 8.35 & 26.48 & 26.58 & $+$0.09 & 26.49 & $-$0.01 \\ 
NGC598   & 8.82 & 8.55 & 24.64 & 24.60 & $-$0.04 & 24.64 & $+$0.00 \\ 
NGC925   & 8.55 & 8.40 & 29.84 & 29.91 & $+$0.06 & 29.85 & $-$0.01 \\ 
NGC1326A & 8.50 & 8.37 & 31.17 & 31.28 & $+$0.08 & 31.20 & $-$0.03 \\ 
NGC1365  & 8.96 & 8.64 & 31.46 & 31.33 & $-$0.10 & 31.43 & $+$0.03 \\ 
NGC1425  & 9.00 & 8.67 & 31.96 & 31.78 & $-$0.11 & 31.89 & $+$0.07 \\ 
NGC1637  & 8.75 & 8.52 & 30.40 & 30.37 & $-$0.02 & 30.39 & $+$0.01 \\  
NGC2090  & 8.80 & 8.55 & 30.48 & 30.44 & $-$0.04 & 30.48 & $+$0.00 \\  
NGC2403  & 8.80 & 8.55 & 27.43 & 27.42 & $-$0.04 & 27.46 & $-$0.03 \\  
NGC2541  & 8.50 & 8.37 & 30.50 & 30.60 & $+$0.08 & 30.52 & $-$0.02 \\  
NGC2841  & 8.80 & 8.55 & 30.75 & 30.71 & $-$0.04 & 30.75 & $+$0.00 \\
NGC3031  & 8.75 & 8.50 & 27.80 & 27.80 & $+$0.00 & 27.80 & $+$0.00 \\ 
NGC3198  & 8.60 & 8.43 & 30.80 & 30.85 & $+$0.04 & 30.81 & $-$0.01 \\  
NGC3319  & 8.38 & 8.28 & 30.74 & 30.89 & $+$0.14 & 30.75 & $-$0.01 \\  
NGC3351  & 9.24 & 8.85 & 30.10 & 29.99 & $-$0.23 & 30.22 & $-$0.12 \\  
NGC3368  & 9.20 & 8.77 & 30.34 & 30.16 & $-$0.18 & 30.34 & $+$0.00 \\  
NGC3370  & 8.80 & 8.55 & 32.37 & 32.31 & $-$0.04 & 32.35 & $+$0.02 \\  
NGC3621  & 8.75 & 8.50 & 29.30 & 29.30 & $+$0.00 & 29.30 & $+$0.00 \\  
NGC3627  & 9.25 & 8.80 & 30.50 & 30.33 & $-$0.20 & 30.53 & $-$0.03 \\  
NGC3982  & 8.75 & 8.52 & 31.87 & 31.84 & $-$0.02 & 31.86 & $+$0.01 \\  
NGC4258  & 9.06 & 8.70 & 29.63 & 29.54 & $-$0.13 & 29.67 & $-$0.04 \\  
NGC4321  & 9.13 & 8.74 & 31.18 & 30.97 & $-$0.16 & 31.13 & $+$0.05 \\  
NGC4414  & 9.20 & 8.77 & 31.65 & 31.44 & $-$0.18 & 31.62 & $+$0.03 \\  
NGC4496A & 8.77 & 8.53 & 31.18 & 31.14 & $-$0.03 & 31.17 & $+$0.01 \\  
NGC4527  & 8.75 & 8.52 & 30.76 & 30.74 & $-$0.02 & 30.76 & $+$0.00 \\  
NGC4535  & 9.20 & 8.77 & 31.25 & 31.05 & $-$0.18 & 31.23 & $+$0.02 \\  
NGC4536  & 8.85 & 8.58 & 31.24 & 31.15 & $-$0.06 & 31.21 & $+$0.03 \\  
NGC4548  & 9.34 & 8.85 & 30.99 & 30.85 & $-$0.23 & 31.08 & $-$0.09 \\  
NGC4639  & 9.00 & 8.67 & 32.20 & 32.02 & $-$0.11 & 32.13 & $+$0.07 \\  
NGC4725  & 8.92 & 8.62 & 30.65 & 30.55 & $-$0.08 & 30.63 & $+$0.02 \\  
NGC5236  & 9.19 & 8.77 & 28.32 & 28.08 & $-$0.18 & 28.26 & $-$0.06 \\  
NGC5253  & 8.15 & 8.15 & 28.05 & 28.12 & $+$0.22 & 27.90 & $+$0.15 \\  
NGC5457i & 9.20 & 8.70 & 29.16 & 29.03 & $-$0.13 & 29.16 & $+$0.00 \\  
NGC5457o & 8.50 & 8.23 & 29.18 & 29.40 & $+$0.17 & 29.23 & $-$0.05 \\  
NGC6822  & 8.14 & 8.14 & 23.31 & 23.50 & $+$0.23 & 23.27 & $+$0.04 \\  
NGC7331  & 8.67 & 8.47 & 30.89 & 30.91 & $+$0.01 & 30.90 & $-$0.01 \\  
IC4182   & 8.40 & 8.20 & 28.21 & 28.44 & $+$0.19 & 28.25 & $-$0.04 \\ 
\enddata                          
   \vspace*{-0.2in} 
\end{deluxetable}

\begin{figure}[t]
   \epsscale{0.63}
   \plotone{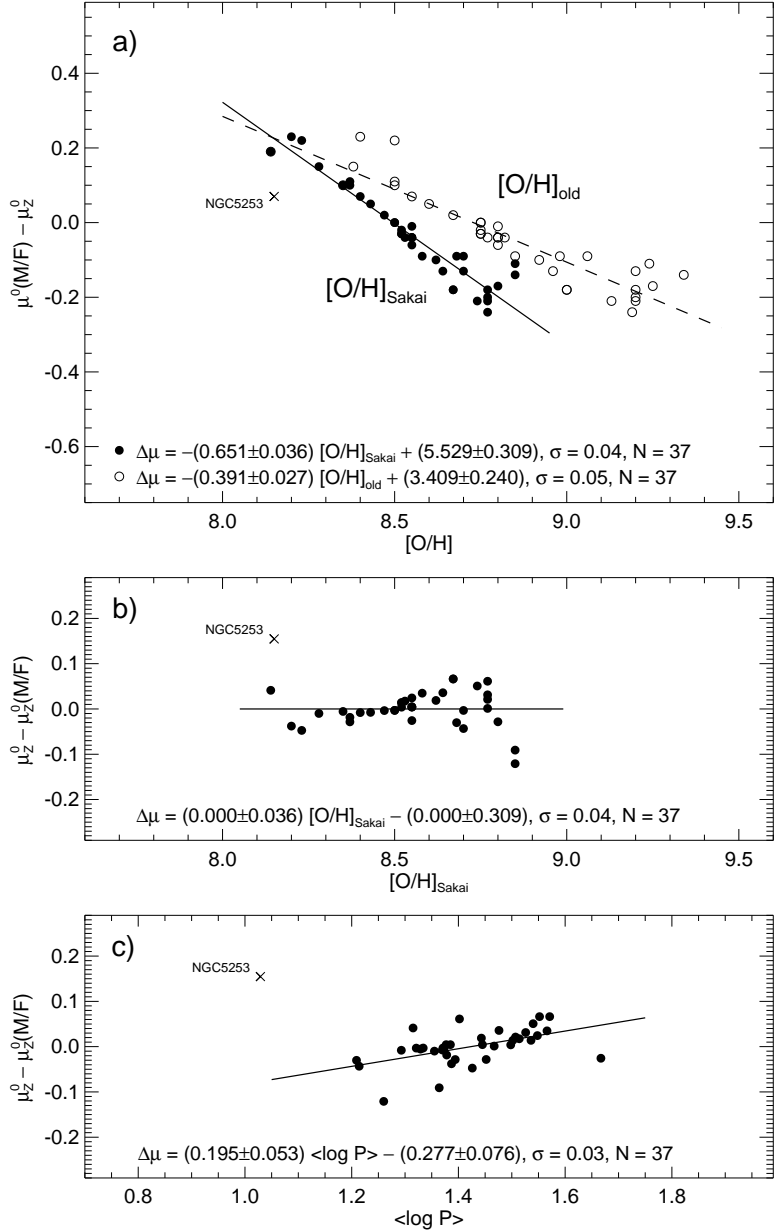}
   \caption{a) The difference of the uncorrected $\mu^{0}$(M/F) and the
   adopted metallicity-corrected $\mu^{0}_{Z}$ of
   Table~\ref{tab:cep:additional} against the metallicity
   $[\frac{O}{H}]$ in the old (open symbols) and new (closed symbols)
   system. 
   b) The difference of the metallicity corrected moduli 
   $\mu^{0}_{Z} - \mu^{0}_{Z}$(M/F) as a function of 
   $[\frac{O}{H}]_{\rm Sakai}$. 
   c) Same as b), but plotted against the mean period 
   $\langle \log P \rangle$ of the Cepheids.}  
\label{fig:14}
\end{figure}

     The difference between the $\mu^{0}$(M/F) (without metallicity
corrections) and the corrected $\mu^{0}_{Z}$ (column~4) are plotted
against the values of $[\frac{O}{H}]$ (in the old and Sakai systems)
in Figure~\ref{fig:14}a.

     The Figure shows little scatter and the expected trend. The
$\mu^{0}$(M/F) fall progressively short of the $\mu^{0}_{Z}$ as
$[\frac{O}{H}]$ increases. To bring the $\mu^{0}$(M/F) into the
$\mu^{0}_{Z}$ system, a metallicity correction must be applied of
\begin{equation}
   \Delta\mu_{Z}(\mbox{M/F})=\mu^{0}(\mbox{M/F})-\mu^{0}_{Z} =
   -(0.39\pm0.03)\Delta[\frac{O}{H}]_{\rm old} =
   -(0.65\pm0.04)\Delta[\frac{O}{H}]_{\rm Sakai}.
\label{eq:metal}
\end{equation} 
The relevance of this result is that the coefficient $-0.39$, which is
independent of any TRGB distances, is only marginally larger than the
TRGB-based coefficients $-0.20\pm0.09$ (for all periods) and 
$-0.27\pm0.11$ (for only $\langle\log P\rangle >0.93$) of
\citet{Sakai:etal:04}. With other words if we wanted to replace our
period-dependent metallicity correction by a single correction for all
periods, we would derive in an independent way a metallicity
correction only marginally larger than that of \citet{Sakai:etal:04}.
We take the reasonable agreement of the two independent metallicity
corrections as a confirmation that the present corrections are sound
on average.

     If the metallicity corrections $\Delta\mu_{Z}$ of
equation~(\ref{eq:metal}) are added to the $\mu^{0}$(M/F) (column~5 of
Table~\ref{tab:07}) one obtains the corrected values
$\mu^{0}_{Z}$(M/F) in column~7. The difference between our moduli
$\mu^{0}_{Z}$ and $\mu^{0}_{Z}$(M/F) is shown in column~8. The average
difference is zero by construction, but the small scatter of
$0.04\mag$ about zero and the independence of
$[\frac{O}{H}]_{\rm Sakai}$ (Fig.~\ref{fig:14}b) are 
most remarkable since it must be recalled
that the moduli $\mu^{0}_{Z}$ are based on the new P-L relations of
the Galaxy and LMC (equations~\ref{eq:PLgal:V}-\ref{eq:PLlmc:I:lt1}),
while the $\mu^{0}_{Z}$(M/F) rest on the old P-L relation of
\citet{Madore:Freedman:91}. The unexpected agreement is
explained by the fact that the slopes of the M/F P-L relations in $V$
and $I$ happen to lie {\em between\/} the corresponding slopes of the
Galaxy and LMC. The difference $\mu^{0}_{Z} - \mu^{0}_{Z}(\mbox{M/F})$
does, however, show a significant dependence on $\langle\log P\rangle$ 
(Fig.~\ref{fig:14}c), as must be expected from the period term in 
equation~(\ref{eq:OH:correct}).

     If we are correct that the metallicity correction changes sign
below $\log P=0.93$, the close agreement of $\mu^{0}_{Z}$ and
$\mu^{0}_{Z}$(M/F) must break down for short-period Cepheids, because
the metallicity correction in equation~(\ref{eq:metal}) to derive
$\mu^{0}_{Z}$(M/F) is independent of period.

\section{AN ANALYSIS OF CEPHEID DISTANCES}
\label{sec:analysis}
\begin{figure}[t]
   \epsscale{0.84}
   \plotone{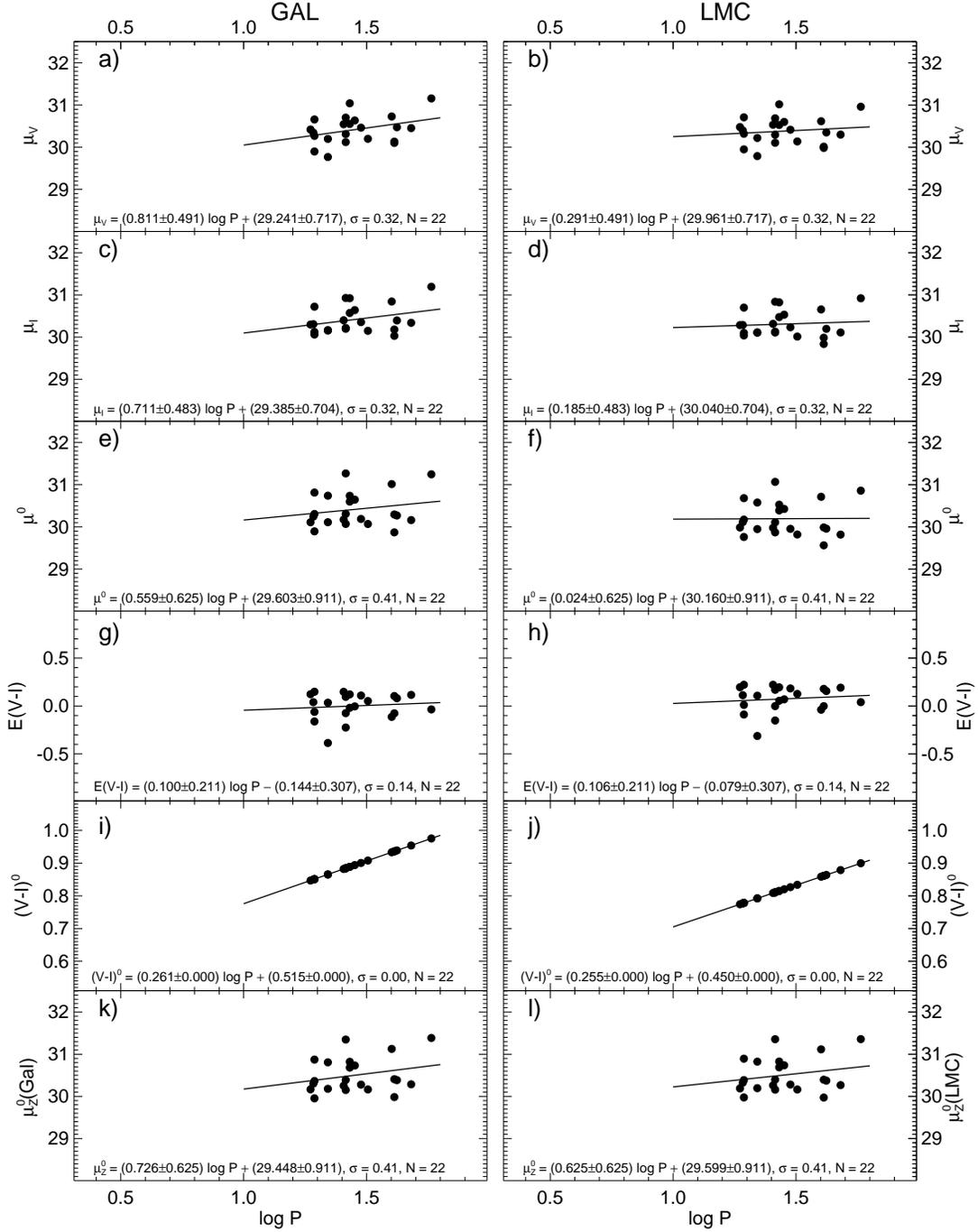}
   \caption{Apparent, true, and metallicity-corrected distance
   moduli as well as reddening $E(V\!-\!I)$ of {\em individual\/}
   Cepheids in NGC\,3627 as determined from the Galactic P-L
   relations (left panels) and LMC P-L relations (right panels) and
   equation~(\ref{eq:mu_0}) as a function of $\log P$.}  
\label{fig:15}
\end{figure}

It would be highly undesirable if distances from individual Cepheids are 
period-dependent. However this will always be the case if 
Cepheids of a given galaxy do not 
follow the slope of an adopted P-L relation:  
clearly, in such a case, the derived apparent distances of {\em individual\/} 
Cepheids become period-dependent. 
This is illustrated in Figure~\ref{fig:15}, a-d, with the Cepheids of
NGC\,3627, which suggests a P-L relation for $V$ and $I$ at 
$P > 10\;$days that is
even flatter than in the LMC. 
Consequentially the individual apparent moduli  
$\mu_{Vi}$(Gal) and $\mu_{Ii}$(Gal) (based on the Galactic P-L
relation) and the $\mu_{Vi}$(LMC) and $\mu_{Ii}$(LMC) (based on the
LMC P-L relation) increase with $\log P$.  
This behavior has here only weak statistical significance, 
but its principal nature is clear.

     As long as it was believed that the P-L relation of Cepheids is 
universal (or affected only by a zero-point shift due to metallicity 
differences) one could assume that the slope differences are purely 
statistical, caused by small-number  statistics, the intrinsic width 
of the instability strip, absorption variations, etc. 
Since it is known that there are physical differences of the P-L 
relation slope, caused by the blanketing effect and 
temperature differences of Galactic and LMC Cepheids, the period 
dependence  of individual Cepheid distances cannot be discarded as a 
statistical fluke, but it is a systematic effect. 
This is a {\em general\/} concern affecting {\em all\/} previous and 
present Cepheid distances if the observed slope of the P-L relations 
in $V$ and $I$ does not coincide with the slope of the adopted 
P-L relations used for calibration.  

     The concern is heavily accentuated if equation~(\ref{eq:mu_0}) is
used to {\em simultaneously\/} solve for distance {\em and\/}
reddening. This is shown by the individual ``true'' distance moduli
$\mu^{0}_{i}$ of the NGC\,3627 Cepheids in Figure~\ref{fig:15}\,e\&f. 
Particularly the moduli $\mu^{0}_{i}$(Gal) from the Galactic P-L
relation show an important increase of the distance with period,
i.e.\ $\mu^{0}_{i}(\mbox{Gal}) = 30.27$ to $30.61$ for $\log P = 1.2$ to
$\log P = 1.8$, while $\mu^{0}_{i}$(LMC) varies only from 30.19 to
30.20 over the same period interval. 
For still shorter periods the range of distances increases
even further, which makes the adopted mean distance dependent on the
period range considered. It had previously been assumed that the
difference between  $\mu^{0}$(Gal) and  $\mu^{0}$(LMC) was due to only
a metallicity effect. 
It is now clear that the difference and its period dependence
must be driven by still another effect.  

     The additional error source lies in the individual color excesses
$E(V\!-\!I)_{i}$. Since $E(V\!-\!I)_{i}$ is simply given by the
difference between  $\mu_{Vi}$ and  $\mu_{Ii}$, where both terms vary
(differently) with period, it is not surprising that $E(V\!-\!I)_{i}$
varies with period as shown in Figure~\ref{fig:15}\,g\&h, which,
of course, is unphysical. Since the biased values of $E(V\!-\!I)_{i}$,
multiplied with the reddening-to-absorption ratios ${\cal R}_{V,V-I} =
2.52$ and ${\cal R}_{I,V-I} = 1.52$ enter directly into the true
distances of the individual Cepheids, it is by necessity that these
distances carry a strong, but spurious dependence on period
(Figure~\ref{fig:15}\,e\&f). This dependence, be it positive or 
negative, becomes more positive in case of the finally adopted
metallicity-corrected $\mu^{0}_{Z}$(Gal) and $\mu^{0}_{Z}$(LMC)
(Figure~\ref{fig:15}\,k\&l) because of the period dependence of the
metallicity corrections in equation~(\ref{eq:OH:correct}). 

     The situation is even worse because the color excesses
$E(V\!-\!I)_{i}$ do not only show an unphysical period dependence, but
they differ also in the mean. 
The value of $E(V\!-\!I)$ is $0.01$ at the mean period $\log P = 1.5$
in case of the Galactic P-L relations, and $0.08$ in case of the LMC
P-L relations (see the equations in Figure~\ref{fig:15}\,g\&h). The
difference of $\Delta E(V\!-\!I) = 0.07$ remains almost constant over all
periods. The individual moduli $\mu^{0}_{i}$(Gal) and
$\mu^{0}_{i}$(LMC) as well as the mean moduli $\mu^{0}$(Gal) and 
$\mu^{0}$(LMC) as derived from equation~(\ref{eq:mu_0}) are therefore
-- although the distances  may nearly agree -- inconsistent implying
different reddenings, and hence different absorption corrections. 

     The reason for the unfortunate period dependence of the
$E(V\!-\!I)_{i}$ is that the difference between the adopted
calibrating P-L relations in $V$ and $I$ imply also a specific P-C
relation which is strictly recovered in the dispersionless 
$(V\!-\!I)$-period relation in Figure~\ref{fig:15}\,i\&j. 
The actual disagreement between the calibrating P-C relation and the
observed P-C relation is thus shifted upon the color excesses
$E(V\!-\!I)_{i}$. 

\begin{figure}[t]
   \epsscale{0.84}
   \plotone{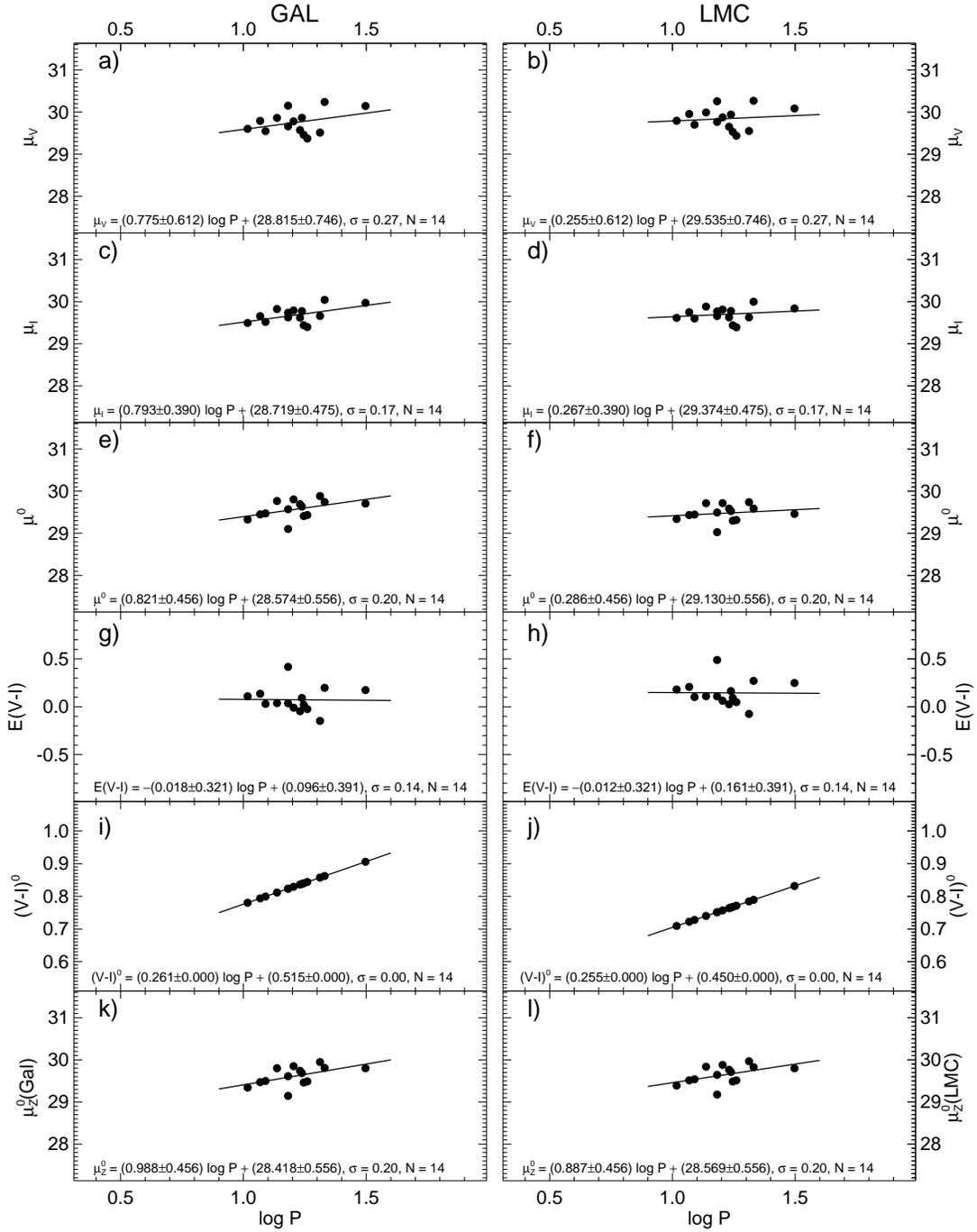}
   \caption{Same as Figure~\ref{fig:15} for the case of NGC\,4258.
   \vspace*{0.3in}} 
\label{fig:16}
\end{figure}

     The problem has been illustrated here with the Cepheids of
NGC\,3627 with $P > 10\;$days. 
Yet the situation is in no way particular for this galaxy. This is
illustrated in Figure~\ref{fig:16} for the additional example of
NGC\,4258. Unusual in the latter case is only that $E(V\!-\!I)$ is
nearly constant over all periods; this is because the observed P-C
relation happens to be very close to the P-C relations implied by the
calibrating P-L relations of the Galaxy and LMC. The character of the
period dependence of the parameters shown in Figures~\ref{fig:15} \&
\ref{fig:16} holds also for the short-period Cepheids with
$P<10\;$days with the exception that the LMC-based parameters may
change slope at $10\;$days, because of the break of the LMC P-L
relations at this period (\citeauthor{Sandage:etal:04}).
The problem of the period-dependent distances arises in similar form
whenever the Cepheids of a galaxy do not conform with the shape of the
P-L relations in $V$ and/or $I$ used for the calibration. 
Equation~(\ref{eq:mu_0}) elegantly hides the period dependence of
the individual moduli and the fact, that the reddening $E(V\!-\!I)$
depends on which P-L relations are used for the calibration.  

     Attempts to avoid the problem, for instance by imposing an
estimated mean color excess on all Cepheids, have not been able to
remove the period dependence. One is therefore left with the ambiguity
at which period the true distance should be read. -- 
It has been stated in \S~\ref{sec:distance:comparison} that the 
P-L relations of the Galaxy and LMC give identical distances at 
$\log P = 0.93$. This, however, holds only for the very specific
case of unreddened Cepheids which follow either
the $V$ and $I$ P-L relations of the Galaxy or LMC. Any other
slopes observed in other galaxies may have a quite different crossover
period. It would therefore be a mistake to read the best distance at
the period $\log P = 0.93$. 

     To estimate the remaining uncertainty of the Cepheid distances of
the galaxies in Table~\ref{tab:cep:additional}, we have determined a
factor $\pi$, which measures the variation of  $\mu^{0}_{Z}$ (being
the mean of $\mu^{0}_{Z}$(Gal) and $\mu^{0}_{Z}$(LMC)) with $\log P$, 
\begin{equation}
   \mu^{0}_{Z}(\mbox{corr}) - \mu^{0}_{Z}(\mbox{obs}) =  \pi\times\Delta\log P
\label{eq:pi}
\end{equation}
The listed distances in column~9 of Table~\ref{tab:cep:additional}
refer to the mean period in column~6. The negative and positive values
of $\pi$ are given in column~12 of Table~\ref{tab:cep:additional}. 
Since some of the $\pi$ values are absolutely even larger than 1, a
shift of the {\em mean\/} period of 0.3 of the Cepheids under
consideration can change the distance modulus by more than $0.3\mag$
or 15\% in distance. -- 
The mean value of $\pi$ of the 37 galaxies in
Table~\ref{tab:cep:additional} is $0.28\pm0.12$; therefore a 
shift again of $\log P = 0.3$ causes an {\em average\/} change of the
distance moduli of 0.094. The combined evidence of the 37 galaxies in
Table~\ref{tab:cep:additional} therefore defines the zero-point of the
distance scale to $\sim\!5\%$.

     The typical error of Cepheid distances of individual galaxies
derived from $V$ and $I$ observations has been frequently quoted to be
in the order of only $\la0.1\mag$. 
This is indeed the formal mean error of individual Cepheid distances
from equation~(\ref{eq:mu_0}).  
But it has always remained a puzzle how such an accuracy could be
achieved with only one or a few dozen Cepheids, where it is necessary
to solve for the distance {\em and\/} the absorption and in view of the
intrinsic width of the P-L relation, the possibly variable absorption
and the observational errors in magnitude and color. 
It is now clear that an additional uncertainty in the order of up to 
10-15\% is hidden in the period dependence of the distances of
individual Cepheids.

\section{THE ZERO-POINT OF THE CEPHEID DISTANCE SCALE}
\label{sec:zeropoint}
It is sometimes stated that the zero-point of extragalactic Cepheid
distances depended entirely on an {\em assumed\/} distance of
LMC. This is not anymore the case.

\subsection{The Zero-Point from Open Clusters, BBW Distances, and LMC}
\label{sec:zeropoint:GalLMC}
The Galactic P-L relations in equations~(\ref{eq:PLgal:V} \&
\ref{eq:PLgal:I}) rest on two independent zero-points (for details see
\citeauthor{Tammann:etal:03}, \citeauthor{Sandage:etal:04})\\
\noindent
(1) on the 37 Galactic Cepheids which are members of open
clusters. The cluster distances are obtained from main-sequence
fitting, with the Pleiades as a reference. The Pleiades modulus of
$5.61\mag$ comes from a variety of determinations, including now also
HIPPARCOS \citep{Makarov:02,Soderblom:etal:05}, and is secure to
better than $0.04\mag$. \\ 
\noindent
(2) on the physical Baade-Becker-Wesselink (BBW) distances of 32
Cepheids by \citet{Fouque:etal:03} and \citet{Barnes:etal:03}. These
distances are based on physical parameters of the stellar atmospheres
and are independent of any assumed astronomical distance.\footnote{%
    \citet{Gieren:etal:05} have revised the theoretically founded
    p-factor (to convert the observed radial velocities into
    pulsational velocities of the Cepheid atmospheres), the motive of
    the empirical correction being to force the cluster NGC\,1866 into
    the plane of LMC. The BBW distance of seven cluster Cepheids, of
    originally $18.36\pm0.06$, agrees {\em after\/} the revision with
    LMC at $18.56\pm0.04$. Yet the relative distance between NGC\,1866
    and LMC is open to debate. The P-L relations of 
    equations~(\ref{eq:PLgal:V} \& \ref{eq:PLgal:I}) and 
    (\ref{eq:PLlmc:V:lt1} \& \ref{eq:PLlmc:I:lt1}) 
    yield an ambiguous result for the same seven Cepheids of 
    $\mu^{0}(\mbox{LMC})=18.58\pm0.04$, $E(B\!-\!V)=0.10$,
    and $\mu^{0}(\mbox{Gal})=18.37\pm0.04$, $E(B\!-\!V)=0.05$; which
    of the two values is more correct depends on the
    (unknown) metallicity of the Cepheids proper. It is noted that the 
    {\em lower\/} distance agrees well with the main-sequence fitting
    of NGC\,1866 by 
    \citet[][$18.35\pm0.05, E(B\!-\!V)=0.06$]{Walker:etal:02}.
    The main objection against the p-factor revision, however, comes
    from the revised distances of Galactic Cepheids, which determine
    slopes of the $B,V,I$ P-L relations much flatter than the slopes
    from Galactic Cepheids in {\em open clusters\/}
    (cf. \citeauthor{Tammann:etal:03}). Since these slopes agreed
    well before the revision, 
    and since there is hardly any rational for the open-cluster
    distances (from zero-main-sequence fitting) to introduce a
    period-dependent error into the Galactic P-L relation,
    we do not consider here the distances from the revised p-factor.
    Moreover, Gieren's et~al. \citeyear{Gieren:etal:05} conclusion
    that the slopes of the Galactic and LMC P-L relations are the same
    is contradicted by the simple fact that Galactic Cepheids are so
    much redder than in LMC; this implies decisively different P-C
    relations, which in turn preclude identical P-L relations in the
    two galaxies.} 
The zero-point of the BBW distances, whose systematic error is
difficult to estimate, is in good agreement with the cluster Cepheids,
being fainter by only $0.05$, $0.07$, and $0.10\mag$ in $B$, $V$, and
$I$ at $P=10\;$days. Therefore the Cepheid samples under (1) and (2)
were combined to define the adopted P-L relations of
equations~(\ref{eq:PLgal:V} \& \ref{eq:PLgal:I}). 

     Parallaxes of Galactic Cepheids were determined by various
authors with HIPPARCOS and HST. Their results, discussed in
\citeauthor{Sandage:etal:04}, all suggest that the luminosities from
equations~(\ref{eq:PLgal:V} \& \ref{eq:PLgal:I}) are somewhat faint,
possibly by as much as $\sim0.1\mag$. If we had included these
measurement into the zero-point calibration, it would become brighter
by $\sim0.05\mag$. The systematic error of the adopted zero-point is
therefore estimated to be $\sim0.08\mag$ with a tendency to be
actually brighter.

     The P-L relations of LMC in equations~(\ref{eq:PLlmc:V} $-$
\ref{eq:PLlmc:I:lt1}) are based on an adopted LMC modulus of
$18.54\pm0.02$ (statistical) from \citeauthor{Sandage:etal:04}. 
This value comes from a compilation of various distance determinations
(\citeauthor{Tammann:etal:03}), {\em excluding\/} the P-L relation of
Cepheids because the different slopes of the Galactic 
and LMC P-L relations preclude a meaningful determination of the LMC
modulus by means of a Galactic P-L relation. 
The adopted value is supported by the most recent determinations,
i.e.\ $18.59\pm0.09$ from the TRGB distance \citep{Sakai:etal:04} and
$18.53\pm0.06$ from the BBW method \citep{Gieren:etal:05}. 
The systematic error of the LMC zero-point is unlikely to be larger
than $0.05\mag$.  

     The distance moduli in Table~\ref{tab:cep:correct} and
\ref{tab:cep:additional} of galaxies with Galactic metallicity depend
only on the Galactic zero-point, galaxies with the metallicity of LMC
depend only on the LMC zero-point. A mixed sample of metal-rich and
metal-poor galaxies depends on both zero-points in about equal
parts. In that case the zero-point error is likely to be smaller than
$0.08\mag$.

\subsection{Additional Evidence for the Zero-Point}
\label{sec:zeropoint:add}
The adopted distances can also be compared with external
data. For 10 galaxies of the present sample \citet{Sakai:etal:04} have
determined independent, metal-insensitive TRGB distances based on a
zero-point by \citet{Lee:etal:93} and \citet{DaCosta:Armandroff:90},
which in turn assume $M_{V}(\mbox{RR\,Lyr})=0.5-0.7$ depending on
[Fe/H]. As seen in Table~\ref{tab:TRGB} the TRGB distances 
are {\em larger\/} on average by $0.12\pm0.04\mag$.

     It is finally noted that the improved ``spectral-fitting
expanding atmosphere method'' (SEAM) yields for the Type IIP
SN\,1999em a distance of $\mu^{0}=30.48\pm0.29$ \citep{Baron:etal:04}
in excellent agreement with the Cepheid distance of the parent galaxy
NGC\,1637 in Table~A1 ($\mu^{0}_{Z}=30.40\pm0.07$), although the large
error does not yet allow a stringent test.

\subsection{The Case of NGC\,4258}
\label{sec:zeropoint:4258}
While the above arguments suggest a somewhat brighter zero-point,
an opposite signal comes from NGC\,4258. Its high-weight water maser
distance modulus of $(m-M)^{0} = 29.29\pm0.08$ (random) $\pm0.07$
(systematic) \citep{Herrnstein:etal:99} is $0.34$ smaller than
its Cepheid distance of $29.63\pm0.05$ (formal error; the realistic
error is rather $0.10$). Ongoing work on the Cepheids 
of NGC\,4258 and its water maser distance may bring a better agreement
\citep{Humphreys:etal:04,Macri:etal:04,Greenhill:etal:04}.
It must be stressed already here, however, that the strong period
dependence of the quoted Cepheid distance ($\pi=0.94$) makes the
discrepancy much less alarming as it may appear. 
If future discoveries of fainter Cepheids in NGC\,4258 with
correspondingly shorter periods shift the mean period of
$\langle\log P\rangle=1.21$ to say $\sim\!1.00$, it will lower the
Cepheid distance by $\sim\!0.2\mag$ as seen from
equation~(\ref{eq:pi}).  
The remaining difference between the maser and
Cepheid distance of $\sim\!0.14\pm0.15$ would lose its
significance in this case.  
The example may show that it would be unwise to set
the zero-point of the distance scale on a single Cepheid distance
because of the period dependence of the latter.

The water maser distance of NGC\,598 (M\,33) of
$(m-M)^{0}=24.31^{+0.51}_{-0.18}$ \citep{Brunthaler:etal:05} carries
still too large an error to be helpful for the zero-point
determination. It will take 5-10 good maser distances before the
method can provide a competitive zero-point for the P-L relations of
Cepheids. 

     In summary, it does not seem justified to change the adopted 
zero-point, which rests on the excellent Pleiades modulus, the quite
reliable LMC distance, and on the BBW method of Cepheids, only to
serve the contradictory evidence from HIPPARCOS/HST distances of
Cepheids and from TRGB distances on the one hand, and from a single
object like NGC\,4258 on the other hand. If one wanted to weight the
proposed changes of the distance scale zero-point, one would end up
with a near confirmation of the adopted zero-point, but in view of the
unknown systematic errors that would not be meaningful. 

     The conclusion is that it seems quite unlikely that the adopted
zero-point of the metallicity-corrected distance scale is off by more
than $0.10\mag$.   

\section{COMPARISON WITH EARLIER CEPHEID DISTANCES}
\label{sec:comparison}
The P-L relations and metallicity corrections adopted here lead
to Cepheid distances which can be compared in seven aspects with
earlier results by us and by others.

     (1) The revision of the photometric zero-point of the
WFPC2, which has affected six galaxies in Table~\ref{tab:cep:correct},
increases their mean distance by $0.04\pm0.02\mag$ as compared to the
distances given in our last summary paper \citep[][NGC\,4527 from
\citealt{Saha:etal:01a}]{Saha:etal:01b}, if both sets are reduced with
the $V$ \& $I$ P-L relations of \citet{Madore:Freedman:91}.

     (2) However, the distances $\mu^{0}$(Gal) of the eight galaxies
in Table~\ref{tab:cep:correct} are $0.16\pm0.02\mag$ {\em larger\/}
than the 2001 values (which are still on a LMC zero-point of $18.50$),
while the distances $\mu^{0}$(LMC) are $0.06\pm0.03\mag$ {\em
  smaller}. The difference between $\mu^{0}$(Gal) and $\mu^{0}$(LMC)
is reconciled by the metallicity corrections in
equation~(\ref{eq:OH:correct}), using the values of $[\frac{O}{H}]$
and $\langle\log P\rangle$ of the individual galaxies listed in
Table~\ref{tab:cep:correct}.  
The resulting distances, $\mu^{0}_{Z}$, 
(Table~\ref{tab:cep:correct}, column~10) are larger on average by
$0.10\pm0.05\mag$  than our 2001 values that were based on the P-L 
relations of \citet{Madore:Freedman:91} with an {\em assumed\/} 
zero-point at $(m-M)^{0}_{\rm LMC} = 18.50$ and no metallicity corrections.
Hence, our 2001 distance scale was within 
5\% of the new P-L relations of the Galaxy and LMC used here with
their new zero-points, the revised magnitudes from the WFPC2 data and
the present metallicity corrections. 

     (3) The re-discussed Cepheid distances of the parent galaxies of
eight SNe\,Ia by \citet{Gibson:etal:00}, based on the P-L relations of
\citet{Madore:Freedman:91} and with no metallicity corrections, are on
average smaller by $0.25\pm0.07$ than those in
Table~\ref{tab:cep:additional}. The difference would be reduced 
to $0.21\mag$ if the same zero-point from LMC (18.54) were used.

     (4) \citet{Ferrarese:etal:00} have compiled Cepheid distances of
28 galaxies contained also in Table~\ref{tab:cep:additional}. Their
distances are derived with the precepts as in (3). Yet they are on
average only $0.04\pm0.03\mag$ smaller than the present values,
provided they are based on a common LMC zero-point.

     (5) Also a favorable overall comparison is
obtained from the 32 galaxies whose distances were determined by
\citet{Tammann:etal:02} from the 1991 P-L relations without
metallicity corrections. They are smaller by only 
$0.03\pm0.02\mag$ ($\sigma=0.14$) than the new values
$\mu^{0}_{Z}$ in Table~\ref{tab:cep:additional}. Also the
\citet{Freedman:etal:01} distances derived with the same precepts,
that were published {\em but discarded\/} by \citet{Freedman:etal:01} 
(their Table~3 col.~2 replaced by their Table~4 col.~11:  see also 
item (6) below) for 30 galaxies in
Table~\ref{tab:cep:additional} are on average only 
$0.07\pm0.03\mag$ ($\sigma = 0.22$) smaller. 
If one applies to the 30 galaxies the metallicity corrections adopted
by these authors the difference would be reduced to a mere
$0.01\pm0.03$. (A common LMC zero-point is assumed).  

     (6) However, the distances that \citet[][their Table~4,
col. 11]{Freedman:etal:01} did adopt,
based on the P-L relations by \citet{Udalski:etal:99a}, which are a
first-order fit to the LMC OGLE data, a zero-point
with LMC at 18.50, and the metallicity correction of
\citet{Kennicutt:etal:98}, are $0.17\pm0.03\mag$
smaller on average than in Table~\ref{tab:cep:additional}.
In particular, the six metal-rich galaxies which
\citet{Freedman:etal:01} use for the luminosity calibration of the
SNe\,Ia are {\it too close} by 
$0.35\pm0.09\mag$ compared to Table~\ref{tab:cep:correct} (or
Table~\ref{tab:cep:additional}). Even if NGC\,5253, discussed in
\S~\ref{sec:newadopted:selection}, is excluded the mean difference
remains at $0.31\pm0.10\mag$.  

     The same difference is independently confirmed by 11 of
their galaxies for which \citet{Sakai:etal:04} have determined TRGB
(Baade sheet) distances. 

     (7) \citet{Riess:etal:05} have determined the Cepheid distances
of four SN\,Ia parent galaxies from the $V,I$ P-L relations as
defined by the LMC Cepheids with $P>10^{\rm d}$ \citep{Thim:etal:03},
a zero-point at $(m-M)_{\rm LMC}=18.50$, and metallicity corrections
from \citet{Sakai:etal:04}. Their mean distance is smaller than in 
Table~\ref{tab:cep:additional} by as much as $0.30\pm 0.07\mag$. 
The reason of the discrepancy is that the four galaxies are quite
metal-rich, i.e.\ their mean $[\frac{O}{H}]$ abundance is close to the
Galactic Cepheids. It would therefore be indicated to determine their
distances in first approximation by means of the {\em Galactic\/} P-L
relations (see Table~\ref{tab:cep:additional}, column~3). In that case
the authors would have recovered the adopted distances $\mu^{0}_{Z}$
in Table~\ref{tab:cep:additional} to within $0.01\pm0.03\mag$.
The reason for the good agreement of the metal-uncorrected $\mu$(Gal)
and the metal-corrected $\mu^{0}_{Z}$ is that the mean value of
$[\frac{O}{H}]=8.57$ of the four galaxies is so close to the adopted
value of the Galactic Cepheids ($[\frac{O}{H}]_{\rm Sakai}=8.60$),
that the metallicity corrections (nearly) cancel.

     The case illustrates that the period-independent bulk correction
for metallicity of \citet{Sakai:etal:04} cannot be applied for
long-period Cepheids and relatively large metallicity differences. A
minor point is that \citet{Sakai:etal:04} give the metallicity
correction for various P-L relations, but not for the flat P-L
relations of \citet{Thim:etal:03}. A comparison of the Cepheid
distances from the latter source and the TRGB distances of
\citet{Sakai:etal:04} gives for the relevant range
$8.2<[\frac{O}{H}]<8.8$ and $1.2<\log P<1.6$, an only slightly flatter 
slope of $\Delta\mu_{Z}= -0.36 \Delta [\frac{O}{H}]_{\rm old}$ than in
equation~(\ref{eq:metal}).

\section{CONCLUSIONS}
\label{sec:conclusions}
This paper up-dates the Cepheid distances of the eight principal
SNe\,Ia calibrating galaxies in the original program (plus 29
additional galaxies in the Appendix), based on a firm zero-point and
including metallicity corrections and realistic error estimates, which
together will be used in a forthcoming Paper~V summarizing our HST
program on the luminosity calibration of SNe\,Ia. These, in turn, when
combined with the Hubble diagram of SNe\,Ia in
\citeauthor{Reindl:etal:05}, will provide the global value of
$H_{0}$. The paper consists of two parts:  
(1) a re-calibration of the time-dependent photometric zero-point of
the WFPC2 camera on HST and corresponding magnitude corrections of the
relevant Cepheids, and 
(2) a discussion of Cepheid distances derived from the
different P-L relations in the Galaxy and in LMC, their implications
for the metallicity corrections, and a discussion of the
remaining open questions. 

The re-calibration of the WFPC2 zero-points, based on comparison of 
ground based photometry of selected objects that were also observed 
contemporaneously with the SNeIa Cepheid galaxies, has revealed that
small adjustments have to be made to the Cepheid magnitudes (and
colors) that were published previously. The corrections are listed by
chip and filter in Table~\ref{tab:final_delta_mags}. 
When these corrections are applied to the Cepheids that were used to
obtain distances in the previous papers, the dominant effect is to
reduce the reddening estimates in $E(V-I)$ by about 0.03 mag in the
mean (individual Cepheids in individual chips differ), with respect  
to the Holtzman zero-points.
When the {\it ad hoc} corrections used in the previous papers 
with respect to the Holtzman zero-points are accounted for, the 
inferred de-reddened distances for the six galaxies that were studied 
with WFPC2 increase by $ \sim 0.04 $  mag overall. Note again, that the 
two galaxies where Cepheids were found and measured with the older
WF/PC are not affected by this re-calibration. 

The 29 additional distance moduli presented in the appendix in
Table~\ref{tab:cep:additional} are not, in general on the same
photometric zero-point, since they come from different sources: most
other Cepheid work with WFPC2 is tied to the zero-point established by 
\citet{Hilletal:98}. To within 0.02 mag, the zero-points in this paper 
result in $V$ magnitudes very similar to the \citet{Hilletal:98} scale, 
but the $I$ magnitudes of the latter are brighter by $0.03$ mag 
systematically. However, these differences are small and within the 
envelope of the other uncertainties that affect the analyses in this
paper. 

In summation, we conclude that:

(1) The metal-poor Cepheids in LMC are bluer than their more
metal-rich counterparts in the Galaxy at fixed period. This is in part
a consequence of the metallicity-dependent line blanketing. Yet in
addition LMC Cepheids at fixed period have higher temperatures than
Galactic Cepheids as first shown by \citet{Laney:Stobie:86} and
confirmed in \citeauthor{Sandage:etal:04}; the same holds at fixed  
luminosity (\citeauthor{Sandage:etal:04}). The reason for this
additional temperature difference is not known at present, but we
assume as a working hypothesis in the present paper that the
temperature difference is also caused by metallicity variations.
 
     The color difference between Galactic and LMC Cepheids causes
also their P-L relations in $V$ and $I$ to be different at a high
level of significance (\citeauthor{Sandage:etal:04};
\citealt{Ngeow:Kanbur:04}). LMC Cepheids are brighter than their  
Galactic counterparts at short periods, but above $P=24\;$days (for
$V$) and $P=18\;$days (for $I$) Galactic Cepheids are brighter. 
The consequence is -- if the absorption-free (``true'') moduli
$\mu^{0}$(Gal) [from the Galactic P-L relations] and $\mu^{0}$(LMC)
[from the LMC P-L relations] are derived from the respective  
apparent moduli $\mu_{V}$ and $\mu_{I}$ through
equation~(\ref{eq:mu_0}) -- that the Galactic and LMC P-L relations
yield identical distances $\mu^{0}$ for unreddened Cepheids with 
$\log P\sim~0.93$ (8.5 days; see Fig~\ref{fig:distance:diff}c). (The
exact transition period depends on the adopted distances of the
Galactic Cepheids and of LMC). The cross-over period is shifted for
reddened Cepheids because the two different sets of P-L relations
yield different color excesses $E(V\!-\!I)$ from
equation~(\ref{eq:mu_0}), even if the true moduli $\mu^{0}$ are the
same. 

     (2) The fact, that the difference of the distance moduli as
derived from the Galactic or LMC P-L relations varies with period
implies that any metallicity corrections must depend on period. 
From the above the metallicity correction is zero for unreddened
Cepheids with $\log P = 0.93$. 
The correction changes sign at this value. 

     We adopt as a measure of the metallicity of Cepheids the 
oxygen-to-hydrogen ratio $[\frac{O}{H}] = 12 + \log(\frac{O}{H})$ in
the $T_{e}$-based scale of \citet{Sakai:etal:04}. All available values of
$[\frac{O}{H}]$ are transformed into this scale. The adopted value for
Galactic Cepheids is $[\frac{O}{H}] = 8.60$ and for LMC $8.34$. It  
follows from the present premises that the metallicity difference of
$\Delta[\frac{O}{H}]=0.26$ must be responsible for the
(period-dependent) difference between $\mu^{0}$(Gal) and
$\mu^{0}$(LMC). This allows, assuming that linear interpolation and
some extrapolation from $[\frac{O}{H}] = 8.60$ to $8.7$ are
permissible, to calculate the metallicity correction $\Delta\mu_{Z}$
for any values of $[\frac{O}{H}]$ and $\log P$
(equation~\ref{eq:OH:correct}). To avoid excessive extrapolation
$\Delta\mu_{Z}$ has been truncated at 8.2 and 8.7 for galaxies even  
less or more metal-rich. 

     The average correction applied to the galaxies in
Tables~\ref{tab:cep:correct} and \ref{tab:cep:additional} amounts to
$\Delta\mu_{Z}= -0.04\mag$ in case of the moduli $\mu^{0}$(Gal) based
on the Galactic P-L relations in equations~(\ref{eq:PLgal:V} \&
\ref{eq:PLgal:I}), and to $\Delta\mu_{Z} = +0.18\mag$ in case of
$\mu^{0}$(LMC) from the LMC P-L relations in equations
(\ref{eq:PLlmc:V} $-$ \ref{eq:PLlmc:I:lt1}). 
The main reason, why the correction to the $\mu^{0}$(LMC) 
is larger than to the $\mu^{0}$(Gal), is -- besides minor period
effects -- that the mean metallicity of $[\frac{O}{H}]=8.55$ for all
galaxies is close to the Galactic value.  

     (3) The present procedure finds validation in the fact, that the
resulting metallicity-corrected Cepheid distances $\mu^{0}_{Z}$, if
compared with independent TRGB distances by \citet{Sakai:etal:04}, do
not show any dependence on metallicity. Also the dependence of the
luminosity of SNe\,Ia on the metallicity of their parent galaxies
becomes nearly flat if their magnitudes are based on the
metallicity-corrected moduli $\mu^{0}_{Z}$.  Also a comparison of the
$\mu^{0}_{Z}$ with velocity distances $\mu_{\rm vel}$ shows no
significant metallicity dependence {\em after\/} the present
metallicity corrections are applied.

     \citet{Sakai:etal:04} have compared their TRGB distances with
Cepheid distances from the M/F P-L relations
\citep{Madore:Freedman:91} and concluded that the latter need an
over-all metallicity correction (excluding any period dependence) of
$\Delta\mu_{Z} = -(0.20\pm0.09) [\frac{O}{H}]_{\rm old}$; their
coefficient becomes $-(0.27\pm0.11)$ if only the galaxies with
$\langle\log P\rangle > 0.93$ (the cross-over period) are 
considered. If we follow their precepts by comparing the adopted 
metal-corrected moduli $\mu^{0}_{Z}$ in Table~\ref{tab:cep:additional}
(column 9), all of which have $\langle\log P\rangle > 0.93$,  
with the uncorrected M/F Cepheid moduli (Table~\ref{tab:07}), we find
an over-all correction of $\Delta\mu_{Z} =
-(0.39\pm0.03)[\frac{O}{H}]_{\rm old}$. 
The good statistical agreement lends further support to the adopted
metallicity corrections.  

     (4) The present method to correct for metallicity is based on the
{\em assumption\/} that the slopes of the P-L relations in $V$ and $I$
change continuously with increasing $[\frac{O}{H}]$ from LMC to the
Galaxy. The actual correlation between $[\frac{O}{H}]$ with the 
observed P-L relation slope is rather unsatisfactory 
(see e.g.\ Figure~\ref{fig:rpMetal}). 
This does not necessarily contradict the basic
assumption, because the observed slopes carry errors which are very
large in comparison with the effect sought. It is a notorious problem
to fix a reliable slope of the ridge line in view of the finite width
of the P-L relation, the occupation of which may in addition be biased
by magnitude and other effects. Moreover, the listed values of
$[\frac{O}{H}]$ are not error-free. They refer in most cases to the
mean radial distance of the Cepheids, but possible azimuthal
variations are not accounted for. Quoted errors range from 0.05 to
0.20.

     The disagreement between the observed and expected slopes of the
P-L relations has unpleasant consequences. It makes the apparent
moduli $\mu_{V}$ and $\mu_{I}$ of individual Cepheids in a given
galaxy to depend (slightly) on period. If the apparent moduli are
combined in equation~(\ref{eq:mu_0}) to yield the individual
true moduli $\mu^{0}$, the latter -- and the color excesses
$E(V\!-\!I)$ -- become frequently significantly dependent on
period. This period dependence is then somewhat modified by the
subsequent (period-dependent) metallicity correction. 
The situation is illustrated for two typical galaxies, NGC\,3627 and
NGC\,4258, in Figures~\ref{fig:15} \& \ref{fig:16}. It is here
clear that the final distance $\mu^{0}_{Z}$ depends on the period 
where the distance is read. If the mean period changes as more
Cepheids will be discovered (preferentially of shorter period), the
most probable distance will change with $\langle\log P\rangle$.

     As a measure of the sensitivity of the distances of individual
galaxies on the mean period a $\pi$-factor has been introduced in
equation~(\ref{eq:pi}). The individual values of $\pi$, which may be
positive or negative, are listed in Table~\ref{tab:cep:additional},
column 12. The absolute $\pi$-values of some galaxies exceed even 1,
which means that the distance changes by $0.2\mag$ or more if the mean
period changes by 0.2. This is an inherent problem of Cepheid
distances if it is attempted to solve for the distance and the
reddening from only two colors. The ambiguity can only be solved if
independent determinations of the color excesses of individual
Cepheids will become available. -- The mean value of $\pi$ of the  
galaxies in Table~\ref{tab:cep:additional} is 0.28;  variations of
$\langle\log P\rangle$  by 0.2 will therefore affect their mean
distance by only $0.056\mag$ (3\%). The zero-point of the distance  
scale defined by the combined evidence of the 37 galaxies in
Table~\ref{tab:cep:additional} is therefore quite secure. 

     The disagreement between the slopes of the calibrating P-L
relation(s) and of the observed P-L relation(s) is a general
problem. Hardly ever will the Cepheid observations follow exactly the
prescribed slope, which always results in a variation of their moduli
with period. Individual Cepheid distances of galaxies carry therefore
a larger uncertainty than frequently quoted.

     (5) The zero-point of the adopted distance scale rests on three
independent pillars, i.e.\ on Galactic Cepheids in open clusters and
hence on the Pleiades at $(m-M)^{0} = 5.61$, on the physical
Baade-Becker-Wesselink (BBW) method of moving atmospheres, and on an
adopted modulus of LMC of $18.54$. 
The weight with which the Galactic and LMC zero-points enter into the
Cepheid distance of a galaxy depends on its metallicity. A galaxy with
Galactic metallicity depends only on the Galactic zero-point, and a
galaxy like LMC depends only on the LMC zero-point. The weights shift
gradually for galaxies with intermediate metallicities.
 -- Although TRGB and HIPPARCOS distances suggest somewhat larger 
distances, and the water maser distance of NGC\,4258 smaller distances 
(see \S~\ref{sec:zeropoint:4258}), the adopted zero-point is believed
to be secure to within $0.10\mag$. 

     (6) The P-L relations and metallicity corrections adopted here
lead to Cepheid distances which may be compared with earlier results. 

     We first note that the revision of the photometric zero-point of
the WFPC2, which has affected six galaxies in
Table~\ref{tab:cep:correct}, increases their mean distance by
$0.04\pm0.02\mag$ as compared to the distances given in our last
summary paper (\citealt{Saha:etal:01b}; NGC\,4527 from
\citealt{Saha:etal:01a}), if both sets are reduced with the same P-L
relations of \citet{Madore:Freedman:91}.  

     A comparison of the mean difference between the adopted distances 
and several previous determinations is given in
\S~\ref{sec:comparison}. All previous determinations are based on some
form of the P-L relation of LMC with or without (period-independent) 
metallicity corrections. The average difference (in the sense present
$-$ previous) is only $+0.03$ to $0.07\mag$ for some earlier lists  
of Cepheid distances, if all are based on a common LMC distance as
zero-point (\citealt{Saha:etal:01a,Saha:etal:01b};
\citealt{Freedman:etal:01} [which would support our long distance
scale here; nevertheless, {\em this long distance scale was discarded
by these authors}]; 
\citealt{Ferrarese:etal:00}; 
\citealt{Tammann:etal:02}). 
The differences between previous values and our new values in this
paper would further decrease by roughly $0.06\mag$ if the
period-independent metallicity correction of \citet{Sakai:etal:04}
was applied. Somewhat larger average differences come from a
comparison with seven galaxies of \citet[][$0.15\mag$]{Gibson:etal:00}
and with the {\em adopted\/} values of 30 galaxies by
\citet[][$0.13\mag$]{Freedman:etal:01}, even after setting the
distances on a common zero-point and correcting them for metallicity
following \citet{Kennicutt:etal:98} or \citet{Sakai:etal:04}.  

     Important mean differences between our values in this paper and
those of previous studies are found 
-- again on a common zero-point -- for six galaxies reduced by
\citet[][$0.31\pm0.03\mag$]{Freedman:etal:01} and four galaxies 
reduced by \citet[][$0.26\pm0.07\mag$]{Riess:etal:05} using the P-L  
relations of the metal-poor LMC by \citet{Udalski:etal:99a} and by 
\citet{Thim:etal:03}, respectively. These galaxies, which all have
produced a SNe\,Ia, are on average almost as metal-rich as the
Galactic Cepheids and they have above average periods. Both facts
require according to equation~(\ref{eq:OH:correct}) considerably
larger metallicity corrections than applied by these authors. 

     The long distance scale of the present paper is not primarily
caused by the adopted metallicity corrections, but by the fact that
the Galactic P-L relations in $V$ and $I$ -- so far not used in
extragalactic work -- lead to significantly larger distances than
those of LMC (Figure~\ref{fig:distance:diff}c). Since the average  
metallicity of the galaxies in Table~\ref{tab:cep:additional} is close
to the Galactic value, it is to be expected that their distances are
larger than frequently anticipated. 

     (7) Present evidence suggests that the adopted metallicity
correction should not be extrapolated to still lower metallicities
($[\frac{O}{H}]\la8.1$). Somewhat preliminary evidence seems to
indicate that the SMC Cepheids with $[\frac{O}{H}]=7.98$ have
temperatures at fixed luminosity that lie {\em between\/} LMC and the
Galaxy and hence that their P-L relations are steeper than those of
LMC. With the precepts developed here this would lead to different
corrections for low-metallicity galaxies.

     (8) The present paper establishes a rather wide array of
distances for a total of 37 galaxies, i.e.\ absorption-corrected
moduli from the Galactic and LMC P-L relations
(Table~\ref{tab:cep:additional}, columns 3 \& 4) and their metallicity
corrections (columns 7 \& 8) as well as the adopted metal-corrected
mean from both relations (column 9). In addition Table~\ref{tab:07}
lists the uncorrected and metal-corrected moduli from the old M/F P-L
relation. It will therefore be possible in the forthcoming  
Paper~V to investigate the influence of these different distance
scales on the cosmic value of $H_{0}$. 

     (9) Proof is not yet available for our assumption used throughout 
the analysis that metallicity differences between the Galaxy and the
LMC is the only cause of the P-L difference. Until such proof is
available, our analysis here must remain provisional. Clarification
can be expected when the P-L relations are analyzed (in the same
manner as done in Papers I and II of this series) for the low
metallicity galaxies SMC, IC~1613, WLM  and others of intermediate
metallicity.

\acknowledgments

We are indebted to Dr. B.~Madore for the unpublished
photometry of the M\,31 Cepheids.
Dr.~A.~Riess has kindly communicated data on the Cepheids in NGC\,3370
and on SN\,1998ae prior to publication. 
We thank the many individuals at STScI who over time, made the
original observations with HST possible. We thanks the referees for
their careful comments, especially those which have resulted in a more
complete discussion of the photometry problems and their mitigation.  
A.~Saha and F.~Thim thank for support provided by NASA of the
retro-active analysis WFPC2 photometry zero-points through grant
HST-AR-09216.01A from the Space Telescope Science Institute, which is
operated by the Association of Universities for Research in Astronomy,
Inc., under NASA contract  NAS 5-26555.  
B.~Reindl thanks the Swiss National Science Foundation for financial
support.

\clearpage
\appendix
\section*{APPENDIX}
\section{Cepheid distances of 37 Galaxies from new P-L relations
  including metallicity corrections}
\label{app:cepheids}
Table~\ref{tab:cep:additional} lists 37 galaxies whose Cepheids have
been observed in $V$ and $I$ with HST and other telescopes by various
authors. The eight galaxies from Table~\ref{tab:cep:correct} have been
included for convenience. Not considered are galaxies with
$[\frac{O}{H}]<8.1$ for reasons stated in the text. The selection of
Cepheids follows the original authors, except that Cepheids with
$P<10\;$days are excluded (except for the case of NGC~5253). 
In cases where magnitudes were derived from
ALLFRAME and DoPHOT packages the former were preferred. The Galactic
and LMC P-L relations in 
equation~(\ref{eq:PLgal:V}\,\&\,\ref{eq:PLgal:I}) and 
(\ref{eq:PLlmc:V}\,\&\,\ref{eq:PLlmc:I}), respectively, were used to
derive the apparent 
moduli $\mu_{VI}$(Gal) and $\mu_{VI}$(LMC), which in turn were
inserted into equation~(\ref{eq:mu_0}) to yield the true moduli
$\mu^{0}$(Gal) and $\mu^{0}$(LMC) for each Cepheid. The mean moduli of
all Cepheids in a given galaxy, excluding $2\sigma$ deviations, lead
to the ``true'' moduli $\langle\mu^{0}\rangle$(Gal) and
$\langle\mu^{0}\rangle$(LMC) 
in columns 3 \& 4. Finally, the new abundances $[\frac{O}{H}]$ in the
$T_{e}$-based system of \citet{Sakai:etal:04} (cf.\
Fig.~\ref{fig:OH:new}) in column~2 and the mean period 
$\langle\log P\rangle$ were used to derive through
equation~(\ref{eq:OH:correct}) the metallicity corrections
$\Delta\mu_{Z}$(Gal) and $\Delta\mu_{Z}$(LMC) (columns 7 \& 8) and
hence the fully corrected moduli $\mu^{0}_{Z}$ and their {\em
formal\/} errors in columns 9 \& 10. Since the moduli
$\mu^{0}_{Z}$(Gal) and $\mu^{0}_{Z}$(LMC) are nearly identical by
construction only the mean value $\mu^{0}_{Z}$ is shown. 
Column~11 gives the decrease of the distance modulus in case the
formally negative absorption is set to zero. 
The period dependence $\pi$ (in equation~\ref{eq:pi}) of the mean
moduli $\mu^{0}_{Z}$ is listed in column~12. The references in Column
13 refer to the observations used for the distance determination. 

     The distances $\mu^{0}_{Z}$ are on the zero-point discussed in
Section~\ref{sec:zeropoint}. The distance of the LMC cannot be determined
from the P-L relation of Cepheids; its adopted distance of $18.54$ had 
to be used as one
of the ingredients of the zero-point definition. Adopting the Galactic
P-L relation for the LMC Cepheids and going through the above
procedure would just recover the value of $18.54$.

\makeatletter
\def\thetable{A\@arabic\c@table}%
\makeatother

 \setcounter{table}{0}

\setlength\textheight{9.3in}%
\clearpage

\setlength\oddsidemargin{-0.50in}%
\setlength\evensidemargin{-0.50in}%

\clearpage
\setlength\textheight{9.3in}%

\begin{deluxetable}{lccccccccccccc}
\tablewidth{0pt}
\tabletypesize{\scriptsize}
\tablecaption{Metallicity-corrected distance moduli of 37
  galaxies.\label{tab:cep:additional}}  
\tablehead{
 \colhead{Galaxy} & 
 \colhead{[O/H]} &
 \colhead{$\mu^{0}(\mbox{Gal})$} &
 \colhead{$\mu^{0}(\mbox{LMC})$} &
 \colhead{$N$} &
 \colhead{$\langle\log P\rangle$} &
 \colhead{$\Delta\mu_{Z}(\mbox{Gal})$} &
 \colhead{$\Delta\mu_{Z}(\mbox{LMC})$} &
 \colhead{$\mu^{0}_{Z}$} &
 \colhead{$\epsilon(\mu^{0}_{Z})$} &
 \colhead{$\delta\mu^{0}$} &
 \colhead{$\pi$} &
 \colhead{Ref.} \\ 
 \colhead{(1)}  & \colhead{(2)}  &
 \colhead{(3)}  & \colhead{(4)}  &
 \colhead{(5)}  & \colhead{(6)}  &
 \colhead{(7)}  & \colhead{(8)}  &
 \colhead{(9)}  & \colhead{(10)} &
 \colhead{(11)} & \colhead{(12)} &
 \colhead{(13)}
} 
\startdata
 NGC\,224   & 8.68 & 24.48 & 24.39 & 25 & 1.209 & $+$0.04 & $+$0.16 & 24.54 & 0.07   &      & $+$0.96 & (1)  \\
 NGC\,300   & 8.35 & 26.63 & 26.48 & 56 & 1.329 & $-$0.17 & $-$0.01 & 26.48 & 0.03   & 0.03 & $-$0.03 & (2)  \\
 NGC\,598   & 8.55 & 24.67 & 24.49 & 10 & 1.385 & $-$0.04 & $+$0.16 & 24.64 & 0.06   &      & $+$0.63 & (3)  \\
 NGC\,925   & 8.40 & 29.96 & 29.82 & 65 & 1.293 & $-$0.12 & $+$0.04 & 29.84 & 0.04   &      & $+$0.35 & (4)  \\
 NGC\,1326A & 8.37 & 31.35 & 31.16 & 14 & 1.394 & $-$0.18 & $+$0.02 & 31.17 & 0.10   & 0.03 & $-$0.04 & (5)  \\
 NGC\,1365  & 8.64 & 31.42 & 31.19 & 32 & 1.476 & $+$0.04 & $+$0.27 & 31.46 & 0.06   &      & $+$0.84 & (6)  \\
 NGC\,1425  & 8.67 & 31.89 & 31.61 & 19 & 1.571 & $+$0.07 & $+$0.35 & 31.96 & 0.11   &      & $+$1.25 & (7)  \\
 NGC\,1637  & 8.52 & 30.48 & 30.21 & 18 & 1.536 & $-$0.08 & $+$0.18 & 30.40 & 0.07   &      & $+$0.48 & (8)  \\
 NGC\,2090  & 8.55 & 30.51 & 30.33 & 28 & 1.377 & $-$0.04 & $+$0.16 & 30.48 & 0.04   &      & $+$0.16 & (9)  \\
 NGC\,2403  & 8.55 & 27.39 & 27.27 & 9  & 1.667 & $-$0.06 & $+$0.26 & 27.43 & 0.15   &      & \nodata & (10) \\   
 NGC\,2541  & 8.37 & 30.67 & 30.49 & 26 & 1.378 & $-$0.17 & $+$0.02 & 30.50 & 0.06   &      & $-$1.05 & (11) \\
 NGC\,2841  & 8.80 & 30.79 & 30.57 & 18 & 1.445 & $-$0.04 & $+$0.18 & 30.75 & 0.06   &      & $-$0.13 & (12) \\
 NGC\,3031  & 8.50 & 27.86 & 27.70 & 24 & 1.334 & $-$0.07 & $+$0.11 & 27.80 & 0.09   &      & $+$0.27 & (13) \\
 NGC\,3198  & 8.43 & 30.92 & 30.74 & 51 & 1.370 & $-$0.12 & $+$0.07 & 30.80 & 0.08   &      & $+$0.44 & (14) \\
 NGC\,3319  & 8.28 & 30.95 & 30.78 & 26 & 1.355 & $-$0.22 & $-$0.04 & 30.74 & 0.08   & 0.02 & $-$1.01 & (15) \\
 NGC\,3351  & 8.85 & 30.04 & 29.92 & 46 & 1.260 & $+$0.05 & $+$0.20 & 30.10 & 0.07   &      & $+$0.09 & (16) \\
 NGC\,3368  & 8.77 & 30.25 & 30.02 & 7  & 1.467 & $+$0.09 & $+$0.32 & 30.34 & 0.11   &      & $+$1.37 & (17) \\
 NGC\,3370  & 8.55 & 32.42 & 32.14 & 64 & 1.548 & $-$0.05 & $+$0.22 & 32.37 & 0.03   &      & $+$0.36 & (18) \\
 NGC\,3621  & 8.50 & 29.37 & 29.19 & 31 & 1.371 & $-$0.07 & $+$0.12 & 29.30 & 0.06   &      & $+$0.36 & (19) \\
 NGC\,3627  & 8.80 & 30.41 & 30.19 & 22 & 1.452 & $+$0.09 & $+$0.31 & 30.50 & 0.09   &      & $+$0.68 & (T5) \\
 NGC\,3982  & 8.52 & 31.94 & 31.69 & 15 & 1.502 & $-$0.07 & $+$0.17 & 31.87 & 0.15   &      & $+$1.89 & (20) \\
 NGC\,4258  & 8.70 & 29.57 & 29.48 & 14 & 1.214 & $+$0.05 & $+$0.17 & 29.63 & 0.05   &      & $+$0.94 & (21) \\
 NGC\,4321  & 8.74 & 31.08 & 30.81 & 39 & 1.540 & $+$0.10 & $+$0.36 & 31.18 & 0.05   &      & $+$0.25 & (22) \\ 
 NGC\,4414  & 8.77 & 31.55 & 31.29 & 10 & 1.526 & $+$0.10 & $+$0.36 & 31.65 & 0.17   & 0.03 & $+$0.39 & (23) \\
 NGC\,4496A & 8.53 & 31.24 & 30.99 & 39 & 1.514 & $-$0.06 & $+$0.19 & 31.18 & 0.05   & 0.07 & $-$0.47 & (T5) \\
 NGC\,4527  & 8.52 & 30.84 & 30.59 & 19 & 1.498 & $-$0.07 & $+$0.17 & 30.76 & 0.20   &      & $-$1.24 & (T5) \\
 NGC\,4535  & 8.77 & 31.15 & 30.90 & 42 & 1.507 & $+$0.10 & $+$0.34 & 31.25 & 0.04   &      & $+$0.47 & (24) \\
 NGC\,4536  & 8.58 & 31.26 & 30.98 & 27 & 1.566 & $-$0.02 & $+$0.25 & 31.24 & 0.09   & 0.01 & $+$0.35 & (T5) \\
 NGC\,4548  & 8.85 & 30.92 & 30.74 & 19 & 1.364 & $+$0.07 & $+$0.26 & 30.99 & 0.04   &      & $+$0.09 & (25) \\
 NGC\,4639  & 8.67 & 32.13 & 31.86 & 12 & 1.552 & $+$0.07 & $+$0.34 & 32.20 & 0.09   & 0.07 & $+$0.90 & (T5) \\
 NGC\,4725  & 8.62 & 30.63 & 30.41 & 19 & 1.443 & $+$0.02 & $+$0.24 & 30.65 & 0.06   &      & $+$0.73 & (26) \\
 NGC\,5236  & 8.77 & 28.24 & 28.04 & 9  & 1.402 & $+$0.08 & $+$0.28 & 28.32 & 0.13   &      & $+$0.89 & (27) \\
 NGC\,5253  & 8.15 & 28.11 & 28.09 & 5  & 1.029 & $-$0.07 & $-$0.03 & 28.05 & 0.27   & 0.07 & $-$0.23 & (T5) \\
 NGC\,5457i & 8.70 & 29.08 & 28.93 & 65 & 1.321 & $+$0.06 & $+$0.23 & 29.16 & 0.04   &      & $+$0.69 & (28) \\
 NGC\,5457o & 8.23 & 29.48 & 29.27 & 28 & 1.426 & $-$0.30 & $-$0.09 & 29.18 & 0.08   & 0.03 & $-$0.51 & (29) \\
 NGC\,6822  & 8.14 & 23.56 & 23.41 & 21 & 1.315 & $-$0.26 & $-$0.09 & 23.31 & 0.03   &      & $-$0.49 & (30) \\
 NGC\,7331  & 8.47 & 30.98 & 30.80 & 13 & 1.373 & $-$0.09 & $+$0.10 & 30.89 & 0.10   &      & $+$0.18 & (31) \\
 IC\,4182   & 8.20 & 28.51 & 28.32 & 13 & 1.387 & $-$0.30 & $-$0.11 & 28.21 & 0.09   & 0.07 & $-$1.24 & (T5) \\
\enddata                          
\tablecomments{If the negative absorption of $\mu^{0}$(Gal) or
  $\mu^{0}$(LMC) [only for IC\,4182] is set to zero, column~11 shows
  the amount by which $\mu^{0}_{Z}$ becomes smaller.  \vspace*{-0.5cm}}
\vspace*{-0.5cm}

\tablerefs{(T5) repeated from Table~\ref{tab:cep:correct}, 
            (1) Madore \& Freedman (2005, private communication),
            (2) \citet{Gieren:etal:04},
            (3) \citet{Freedman:etal:91},
            (4) \citet{Silbermann:etal:96},
            (5) \citet{Prosser:etal:99},
            (6) \citet{Silbermann:etal:99},
            (7) \citet{Mould:etal:00},
            (8) \citet{Leonard:etal:03},
            (9) \citet{Phelps:etal:98},
            10) NGC\,2403. Periods and $B$ magnitudes from
            \citet{Tammann:Sandage:68}. $I$ magnitudes from
            \citet{Freedman:Madore:88}. The Galactic and LMC P-L
            relations and ${\cal R}_{B}=4.23$ are from
            \citeauthor{Sandage:etal:04}. Then, in
            analogy to equation~\ref{eq:mu_0}, $\mu^{0} = 1.86\mu_I -
            0.86\mu_B$.
           (11) \citet{Ferrarese:etal:98},
           (12) \citet{Macri:etal:01},
           (13) \citet{Freedman:etal:94},
           (14) \citet{Kelson:etal:99},
           (15) \citet{Sakai:etal:99},
           (16) \citet{Graham:etal:97},
           (17) \citet{Tanvir:etal:99},
           (18) \citet{Riess:etal:05},
           (19) \citet{Rawson:etal:97},
           (20) T5. A re-analysis of the HST magnitudes by
           \citet{Stetson:Gibson:01} yields $\mu^{0}_{Z}=31.66$ with
           the present precepts. 
           (21) \citet{Newman:etal:01},
           (22) \citet{Ferrarese:etal:96},
           (23) \citet{Turner:etal:98},
           (24) \citet{Macri:etal:99},
           (25) \citet{Graham:etal:99},
           (26) \citet{Gibson:etal:99},
           (27) \citet{Thim:etal:03},
           (28) \citet{Stetson:etal:98} (Cepheids in an inner,
           metal-rich region of M\,101),
           (29) \citet{Kelson:etal:96}, (Cepheids in an outer,
           metal-poor region of M\,101),
           (30) \citet{Pietrzynski:etal:04},
           (31) \citet{Hughes:etal:98}.
}
\end{deluxetable}

 \setlength\textheight{8.4in}%
\clearpage
 \setlength\oddsidemargin{0in}%
 \setlength\evensidemargin{0in}%
 \setlength\textheight{8.4in}%
 \setlength\topmargin{0in}%



\end{document}